\documentclass[a4paper,twoside,final,fleqn,usenatbib]{mnras}

\usepackage{newtxtext,newtxmath}

\usepackage[T1]{fontenc}

\DeclareRobustCommand{\VAN}[3]{#2}
\let\VANthebibliography\thebibliography
\def\thebibliography{\DeclareRobustCommand{\VAN}[3]{##3}\VANthebibliography}

\newcommand{\jk}[1]{{{{#1}}}}
\newcommand{\mz}[1]{{#1}}
\newcommand{\mztwo}[1]{{#1}}
\newcommand{\mzrevi}[1]{{#1}}

\usepackage{graphicx}	
\usepackage{amsmath}
\usepackage{amssymb}	
\usepackage{caption}
\usepackage{rotating}

\title[]{Dust extinction map of the Galactic plane based on the VVV survey data}

\author[M. Zhang et al.]{
M. Zhang,$^{1,2}$\thanks{E-mail: miaomiao@pmo.ac.cn}
and J. Kainulainen$^{3,2}$
\\
$^{1}$Purple Mountain Observatory, and Key Laboratory for Radio Astronomy, Chinese Academy of Sciences, Nanjing 210023, China\\
$^{2}$Max-Planck-Institut f\"{u}r Astronomie, K\"{o}nigstuhl 17, D-69117 Heidelberg, Germany\\
$^{3}$Chalmers University of Technology, Department of Space, Earth and Environment, SE-412 93 Gothenburg, Sweden
}

\date{Accepted XXX. Received YYY; in original form ZZZ}

\pubyear{2022}

\begin{document}
\label{firstpage}
\pagerange{\pageref{firstpage}--\pageref{lastpage}}
\maketitle

\begin{abstract}
\jk{Dust extinction is one of the most reliable tracers of the gas distribution in the Milky Way. The near-infrared (NIR) Vista Variables in the V\'ia L\'actea (VVV) survey enables extinction mapping based on stellar photometry over a large area in the Galactic plane. We devise a novel extinction mapping approach, XPNICER, by bringing together VVV photometric catalogs, stellar parameter data from StarHorse catalogs, and previously published Xpercentile and PNICER extinction mapping techniques. We apply the approach to the VVV survey area, resulting in an extinction map that covers the Galactic disk between 295\degr~$\lesssim$ $l$ $\lesssim$ 350\degr~and -2\degr~$\lesssim$ $b$ $\lesssim$~2\degr, and the Galactic bulge between -10\degr~$\lesssim$ $b$ $\lesssim$ 5\degr. The map has 30\arcsec spatial resolution and it traces extinctions typically up to $A_V\sim$~10-20\,mag and maximally up to $A_V\sim$~30\,mag. We compare our map to previous dust based maps, concluding that it provides \mzrevi{a high}-fidelity extinction-based map, especially in its ability to recover both the diffuse dust component of the Galaxy and moderately extincted giant molecular cloud regions. The map is especially useful as independent, extinction-based data on the Galactic dust distribution and applicable for a wide range of studies from individual molecular clouds to the studies of the Galactic stellar populations.}
\end{abstract}



\section{INTRODUCTION}

\jk{Dust is a crucial component of the interstellar medium (ISM), important as a tracer of the Galactic ISM, a diagnostic tool of the physical properties of the ISM, and as a foreground for extragalactic observations.} 
%
%
The spatial distribution of the Galactic dust has been mapped using thermal dust emission with many datasets \citep{sfd1998,planck-dust-2014,ppmap2017}. For example, \citet[][SFD hereafter]{sfd1998} presented a full-sky reddening map with the spatial resolution of $\sim$6\arcmin~using the the Cosmic Background Explorer \citep[COBE,][]{cobe1992} and the Infrared Astronomical Satellite \citep[IRAS,][]{iras1984} emission maps. \mz{The SFD map} is currently the most widely-used large-scale dust reddening map. However, converting  thermal dust emission to dust column density requires assumptions about both the dust emissivity and the line-of-sight dust temperature distribution \citep{draine2009,padoan-pp6} that 
usually \mz{have} large uncertainties due to the lack of detailed information about dust properties \citep{ossenkopf1994}.

Another commonly-used dust mapping technique is measuring the reddening and extinction towards \mz{a large number of} stellar objects, which is then treated as \mz{a} discrete sampling of a continuous dust distribution. The extinction towards an individual star can be obtained by comparing its observed spectral energy distribution (SED) to intrinsic SED \citep{lada1994}, which, in principle, requires detailed stellar spectroscopic information, but does not depend on the assumptions about dust temperature and properties. Therefore, dust extinction can be \mz{more reliable} probe of column density compared to thermal dust emission and gas tracers such as $^{13}$CO \citep{goodman2009}. 

In the optical regime, a large \mz{number} of works \citep{green2014,andrae2018,starhorse2019,bai2020,starhorse2020} have attempted to determine the stellar parameters, distance, and reddening based on the modern wide-field optical photometric and spectroscopic survey data \citep{sdss2000,skymapper2007,lamost2012,pan-starrs2016,gaia2018} and the stellar evolutionary models \citep{parsec2012,basti2018}. The obtained stellar parameter catalogs can be naturally used to map the three-dimensional (3D) dust distribution \citep{sale2014,rezaei2018,chen2019,green2019,lallement2019,leike2019,leike2020}. However, the dynamical range of the dust maps based on optical surveys is limited, only up to a few to $\sim$10 magnitudes of visual extinction, and thus not enough to trace the high column density regions closely \mz{connected to} the star formation process \citep{gao2004,lada2010,mypub2019}.

Since extinction decreases with wavelength, we can expect to detect more background stars at longer wavelength. \mz{Indeed}, extinction mapping with the NIR multi-band photometric survey data \citep{2mass2006,ukidss2007,vvv2010} can measure the dust column density with a dynamic range $\sim$5-10 times larger than the optical extinction maps \citep{lada1994}.
However, due to the lack of large NIR spectroscopic and astrometry surveys, it is difficult to derive accurate stellar parameters for NIR sources. The NIR extinction towards individual stars are usually obtained by simply comparing the observed NIR colors and the \mz{average} intrinsic colors. \mz{This is facilitated by two observationally confirmed facts}: 1) the dispersion of the stellar intrinsic colors is relatively small at NIR wavelength \citep{nicer2001,girardi2005,davenport2014}; and 2) the variation of the NIR extinction law is small \citep{cardelli1989,wangjiang2014,meingast2018,wangchen2019}. In practice, the intrinsic colors of stars are 
usually inferred with the nearby, supposedly extinction-free reference fields. 

Based on above well-established approach, several two-dimensional (2D) extinction mapping techniques have been developed, each employing a somewhat different detailed formalism (NICE: \citealt{lada1994}; NICER: \citealt{nicer2001}; NICEST: \citealt{nicest2009}; PNICER: \citealt{pnicer2017}; XNICER: \citealt{xnicer2018}; see also, e.g., \citealt{cambresy2002,gutermuth2009} for further variants). These techniques have been used extensively to investigate column density structures of nearby star forming regions \citep[e.g.,][]{kai2006,kai2007,kai2009,lombardi06,lombardi08ext,lombardi10,gutermuth11,schneider2011,alves14}, but \mz{they are rarely applied} to study the Galactic plane. This is because these techniques are based on observing stars that are behind the dust clouds, and when peering through the Galactic plane, the relative configuration of dust and stars can be complex, which makes extinction mapping prohibitively difficult \citep[e.g.,][]{lombardi05, kainulainen11alves,juvela16}. 
\mz{To improve on this point,} \citet{dobashi2008} \mz{developed} a technique named "X percentile method" as an extension of the NICE method. The X percentile method selects the X percentile reddest stars along the line of sight as the background stars and thus can efficiently remove the contamination by foreground sources, which makes it suitable to estimate the extinction towards distant and/or dense dust clouds located behind a large number of foreground stars. \citet{dobashi2011} applied the X percentile method to the Two Micron All-sky Survey catalogs \citep[2MASS,][]{2mass2006} and obtained an all-sky extinction map that had a \mz{variable} spatial resolution of 1\arcmin$-$12\arcmin~\mz{and} a dynamical range of up to $A_V\sim$~20\,mag, \mz{limited by the relatively poor} resolution and sensitivity of 2MASS survey. \mz{In general, the map is still not refined enough} to trace the fine dust structures in the Galactic plane. 
\mz{In summary}, the complementary view of Galactic dust provided by NIR extinction mapping is largely missing towards the Galactic plane.

In this paper, \mz{we make progress by deriving a new extinction-based dust map that can trace the dust distribution in the Galactic plane better than its predecessors. To do this, we develop a new} NIR extinction mapping technique, XPNICER, \mz{that is based on} PNICER \citep{pnicer2017} and Xpercentile \citep{dobashi2008} methods. We \mz{will apply} XPNICER to the VVV survey \citep{vvv2010} photometric catalogs \citep{vvvdaophotmypub} and obtained a 2D extinction map with the spatial resolution of 30\arcsec~and the dynamical range up to $A_V\sim$~30\,mag, covering the whole VVV survey area in the Galactic plane. In Sect.~\ref{sect:data} we described the source catalogs that we used, including the VVV DaoPHOT catalog \citep{vvvdaophotmypub} that has been recalibrated with the aim to decrease the zero magnitude bias. We described the XPNICER technique in Sect.~\ref{sect:method} and presented the extinction map in Sect.~\ref{sect:results}. In Sect.~\ref{sect:discussion} we  compared our extinction maps with several previous dust based maps and gave some suggestions and caveats to the usage of our map. Finally, we summarized our results and conclusions in Sect.~\ref{sect:summary}.

\section{DATA}\label{sect:data}

\subsection{VVV photometric catalogs}\label{sect:vvv-daophot-catalog}

\mz{We use in this paper photometric data from the VVV survey as the main data to derive our extinction map. We describe below the survey and the data to the degree relevant to this work. For more detailed descriptions we refer to the respective survey and instrument papers.}

The VVV survey uses VIRCAM \citep[VISTA InfraRed CAMera;][]{vircam1,vircam2} equipped on the VISTA telescope \citep{vista2015} to map $\sim$562 square degrees in the inner Galactic plane, including the Galactic bulge (-10 $\leq$ $l$ $\leq$ 10, -10 $\leq$ $b$ $\leq$ 5) and part of the adjacent Galactic plane (-65 $\leq$ $l$ $\leq$ -10, -2 $\leq$ $b$ $\leq$ 2). The survey covers five bands, $Z, Y, J, H, K_s$, and it extends over the time period of five years \citep{vvv2010, vvvdr1}.

VIRCAM is equipped with a 4$\times$4 detector array and each detector has 2048$\times$2048 pixels$^2$ in size with a pixel scale of 0.339\arcsec. A single pointing with the detector array is called a pawprint and six pawprints can be combined to construct a contiguous field of $\sim$1.687 deg$^2$ named a tile. The whole VVV survey consists of 348 tiles, including 196 bulge tiles (names start with "b"), and 152 disk tiles (names start with "d"). The tile names and coordinates can be found in \citet{vvvdr1}. 

\mz{Several works have used the VVV data to produce photometric catalogs.} \citet{vvvdr1} released a catalog based on aperture photometry with the 5$\sigma$ limiting magnitude of $K_s\sim$~17-18 mag for most tiles and $K_s\sim$~15-16 mag in the crowded fields. \citet{vvvdophot2018} presented a PSF photometric catalog based on DoPHOT algorithm \citep{dophot1989,dophot1993} with the limiting magnitude (5$\sigma$) of $K_s\sim$~17-18 mag. \citet{vvvdaophotmypub} performed the PSF photometry on the stacked VVV survey images using DaoPHOT algorithm \citep{daophot1987} and presented a catalog that was about one magnitude deeper than the DoPHOT photometric catalog released by \citet{vvvdophot2018}. To detect more faint background sources, we decided to use the PSF photometric catalog presented by \citet{vvvdaophotmypub}.

We note that \citet{vvvdaophotmypub}'s DaoPHOT photometric catalog was calibrated with \citet{vvvdophot2018}'s DoPHOT catalog, which \mz{in turn} was calibrated with \mz{\citet{vvvdr1}'s aperture photometric catalog}. The flux of sources in the aperture photometric catalog was calibrated with the 2MASS sources \citep{vistacali2018}. Specifically, \citet{vistacali2018} constructed the empirical transformations between 2MASS and VISTA photometric systems, considering the effect of interstellar reddening. Then the 2MASS sources with VISTA counterparts were transformed to the VISTA system and treated as the local secondary standard stars to calibrate the VVV aperture photometric catalog tile by tile. However, \citet{hajdu2020} has pointed out that there is bias in the photometric zero-points of the calibrated VVV aperture photometric catalog. The sources of the bias were identified as 1) artifacts in some detector images due to the high level of sky background; 2) high blending effect of 2MASS sources in the crowed fields. Given the calibration chain, the DaoPHOT catalog released by \citet{vvvdaophotmypub} should also have the similar zero magnitude bias. 

\begin{figure*}
    \centering
    \includegraphics[width=1.0\linewidth]{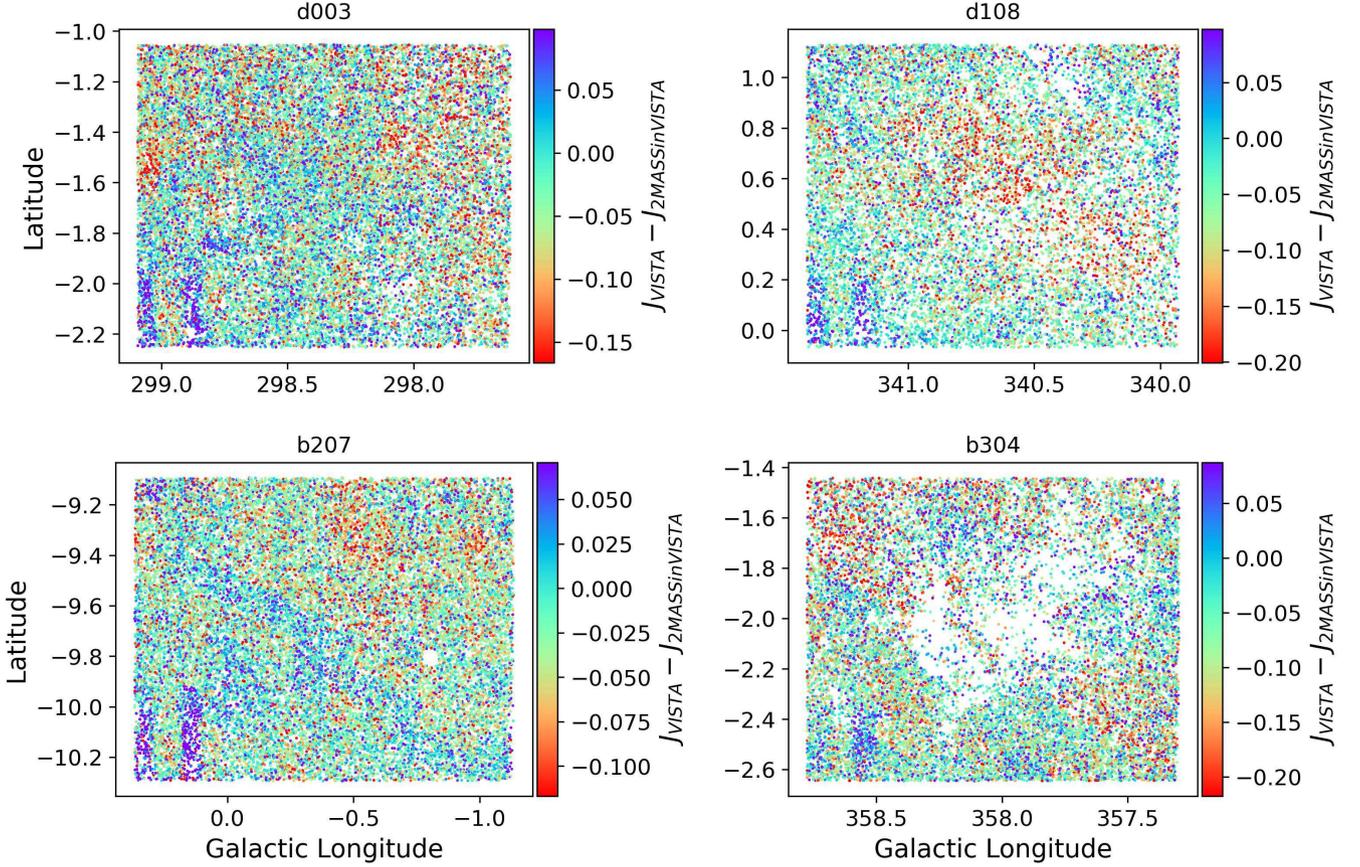}
    \caption{Photometric difference between VVV DaoPHOT $J$-band magnitude and converted 2MASS magnitude in four tiles.}
    \label{fig:zmag-tiles}
\end{figure*}

\begin{figure*}
    \centering
    \includegraphics[width=1.0\linewidth]{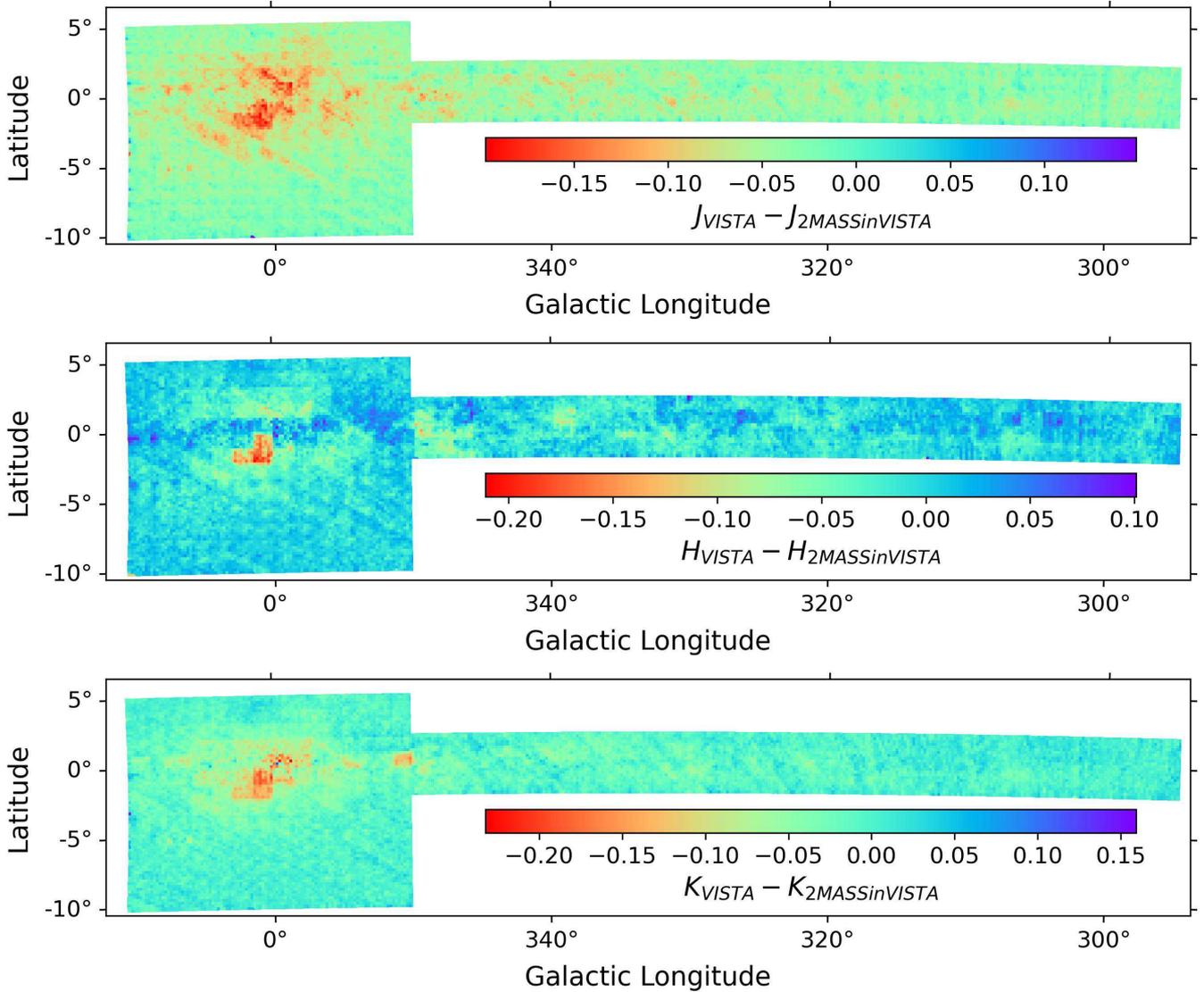}
    \caption{The $J$, $H$, and $K$ \mz{band} smoothed maps of photometric difference between the VVV and converted 2MASS magnitudes over the whole VVV survey area.}
    \label{fig:zmag-wholecoverage}
\end{figure*}

We re-calibrated the VVV DaoPHOT catalog \citep{vvvdaophotmypub} using the method suggested by \citet{hajdu2020}. First, the 2MASS photometry was converted to the VISTA system with the transformations suggested by \citet{vistacali2018} and \citet{hajdu2020}.
Here we only consider the high reliable 2MASS sources with the photometric quality flags\footnote{\url{https://old.ipac.caltech.edu/2mass/releases/allsky/doc/sec1_6b.html\#phqual}} of "AAA". 
Second, the converted 2MASS catalog was cross-matched with the DaoPHOT catalog with a tolerance of 0.15\arcsec~that was suggested by \citet{hajdu2020}. Such a small tolerance helps to filter out the mismatch between 2MASS and VVV because several VVV sources can be blended into the same 2MASS source in the crowded fields if using a large tolerances. \citet{vistacali2018} allowed a relatively large cross-match radius of 1\arcsec~between 2MASS and VVV catalogs to calculate the zero magnitude of VVV aperture photometric catalog, which resulted in the mismatches and thus bias of zero points in the crowded fields \citep{hajdu2020}.  Figure~\ref{fig:zmag-tiles} shows the photometric difference between DaoPHOT $J$ magnitudes and converted 2MASS magnitudes in four tiles that are located in the disk and bulge areas, covering different Galactic latitudes and thus represent regions with different crowding conditions. 
We can see obvious patterns that show the systematic large offsets in the bottom-left corner of each tile. We believed that this bias was due to the variable quantum efficiency (QE) of detector 16\footnote{\url{http://casu.ast.cam.ac.uk/surveys-projects/vista/technical/known-issues\#section-2}}, which was quite small in $K_s$ band but very noticeable in the bluer bands such as $J$ band. Then we smoothed the photometric difference between the VVV and converted 2MASS magnitudes of all cross-matched sources with a gaussian beam of 0.2\degr~that was roughly the size of a detector. Figure~\ref{fig:zmag-wholecoverage} shows the $J$, $H$, and $K_s$ smoothed maps over the whole VVV survey area. Obviously, there is large bias in the crowed fields, especially the region close to the Galactic center. 

\begin{figure*}
    \centering
    \includegraphics[width=1.0\linewidth]{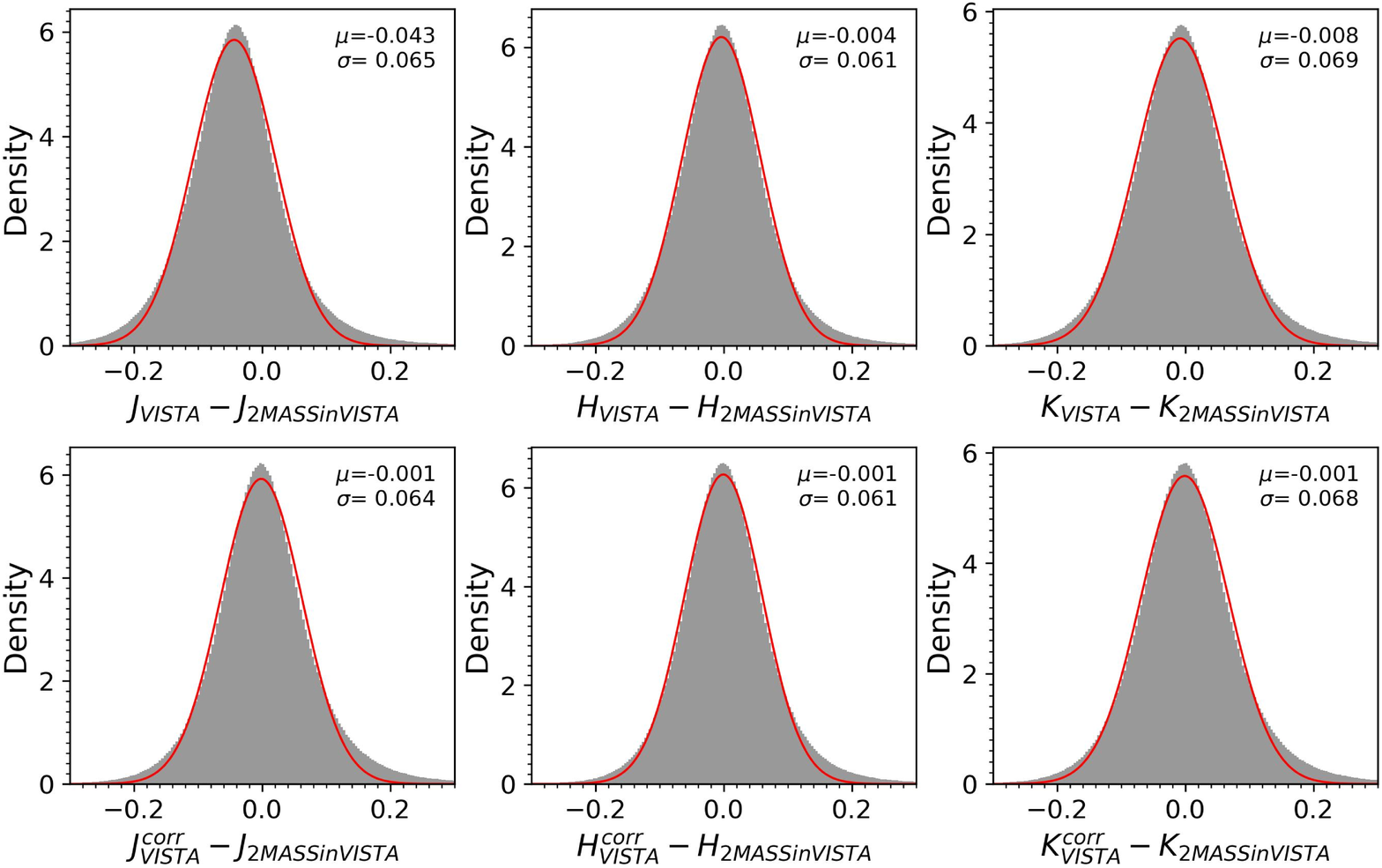}
    \caption{Histograms of photometric difference between VVV DaoPHOT $J$- (left), $H$- (middle), $K$- (right) magnitudes and converted 2MASS magnitudes in the whole VVV survey area before (top) and after (bottom) re-calibration. The red solid lines show the gaussian fittings to the histograms and the fitting parameters, $\mu$ and $\sigma$ (gaussian mean and sigma), are also marked in the corresponding panels.}
    \label{fig:zmag-correction}
\end{figure*}

Finally we applied the smoothed maps shown in Fig~\ref{fig:zmag-wholecoverage} as the photometric correction factor to the DaoPHOT catalog, i.e., the corrected magnitudes of sources in the DaoPHOT catalog were obtained by subtracting the correction factor 
from the uncorrected photometry. Figure~\ref{fig:zmag-correction} shows the distributions of photometric difference in $J$-, $H$-, and $K_s$-band before and after the re-calibration. We found that these distributions can be approximately described with the gaussian functions. 
Apparently, there is a systematic photometric offset between converted 2MASS photometry and raw DaoPHOT catalog released by \citet{vvvdaophotmypub}, especially in the $J$-band. Our re-calibration process decreases the systematic offsets down to only 0.001\,mag level. We also note that the photometric accuracy of the DaoPHOT catalog is $\sim$60-70\,mmag for bright sources compared with 2MASS. 

\subsection{StarHorse catalogs}\label{sect:starhorse-catalog}
\mz{In this paper, we also use the StarHorse catalogs \citep{starhorse2019,starhorse2020} as ancillary data. In the StarHorse catalogs, there are several hundreds of million sources with the Bayesian stellar parameters, distances, and extinctions. We use these stars as the reference sources in our extinction mapping process (see Sect. \ref{sect:method-singlestar}).}

\mz{The StarHorse catalogs were constructed with the StarHorse code \citep{starhorse2016,starhorse2018}. By comparing a number of observed quantities, e.g., multi-band photometry, spectroscopically determined parameters like effective temperatures, or parallaxes, to the stellar evolutionary models \citep{parsec2012,basti2018}, the StarHose code can be}  used to estimate the stellar parameters, distances and extinctions of stars with the Bayesian method. 

\citet[][StarHorse2019 hereafter]{starhorse2019} applied the StarHorse code to the precise parallaxes and multi-band photometry released by the Gaia DR2 \citep{gaia2018}, Pan-STARRS1 \citep{pan-starrs2016}, 2MASS, and AllWISE \citep{wise2010,gaiamatched2019} surveys. They finally obtained the distances and extinctions for $\sim$265 million sources \mz{in the whole sky} with a median accuracy of $\sim$5-16\% in distance and $\sim$0.2\,mag in visual extinction.

\citet[][StarHorse2020 hereafter]{starhorse2020} also obtained the distances and extinctions for more than 5 million sources by applying the StarHorse code to the  spectroscopic surveys of APOGEE DR16 \citep{apogee2020}, GALAH DR2 \citep{galah2018}, LAMOST DR5 \citep{lamost2019}, RAVE DR6 \citep{rave2020a,rave2020b}, and GES DR3 \citep{ges2012}, together with the photometric catalogs of Gaia DR2, Pan-STARRS1, APASS \citep{apass2014}, 2MASS and AllWISE. The typical uncertainties of distances and visual extinctions are $\sim$3-5\% and 0.05-0.16\,mag, individually.

We combined the StarHorse2019 and StarHorse2020 catalogs to produce a new StarHorse catalog. Specifically, the sources in StarHorse2019 that have counterparts in StarHorse2020 were replaced with the entries of StarHorse2020 due to the relatively small uncertainties of distances and extinctions of StarHorse2020. The StarHorse2020 sources that have no StarHorse2019 counterparts were just added as new entries to the new StarHorse catalog.

\subsection{\jk{Final input catalog for the extinction mapping}}
\label{sect:input-catalog}

We generated a matched catalog by combining the re-calibrated VVV DaoPHOT catalog (see Sect~\ref{sect:vvv-daophot-catalog}) and the combined StarHorse catalog (see Sect~\ref{sect:starhorse-catalog}) as the input of our extinction mapping procedure that was described in the subsequent section (Sect~\ref{sect:method}). We used a matching radius of 0.5\arcsec~to match the StarHorse catalog to the VVV DaoPHOT catalog. We also required the photometric uncertainties of $<$0.35\,mag in all $J$, $H$, and $K_s$ bands and excluded the possible spurious detections, i.e., sources that have any of nine spurious detection flags equaling 1 (see Sect. 3.2.4 of \citealt{vvvdaophotmypub} for details). Finally we obtained $\sim$820 million NIR sources in the VVV survey area, of which $\sim$3\% have StarHorse extinction estimates.

\section{METHOD}\label{sect:method}
Our extinction mapping technique, XPNICER, is based on the PNICER \citep{pnicer2017} and X percentile \citep{dobashi2008} methods. In brief, the extinctions towards individual stars are estimated by comparing their observed colors to intrinsic colors that are inferred statistically \mz{from the} StarHorse sources in the same area \mz{(elaborated below in Sect. \ref{sect:method-singlestar})}. Then we group stars into discrete sightlines and select possible background sources along different lines of sights with the X percentile method \mz{(elaborated below in Sect. \ref{sect:background-and-mapping})}. The selected background stars are used to map the spatial distribution of the integrated dust extinction. 

\subsection{Single-star extinction estimation}\label{sect:method-singlestar}

In our input catalog (see Sect.~\ref{sect:input-catalog}), about 23 million sources have StarHorse extinction estimates. We checked the quality flags for these sources in the original StarHorse2019 and StarHorse2020 catalogs. The StarHorse2019 catalog used "SH\_OUTFLAG" \citep[see Sect.~3.4.4 of][]{starhorse2019} to mark the sources with unreliable extinction values or large $A_V$ uncertainties while the StarHorse2020 catalog labeled the sources with bad extinction estimates using the flag of "SH\_OUTPUTFLAGS" \citep[see Table A.2 of][]{starhorse2020}. We note that the StarHorse extinction and distance of each source were offered with 5\%, 16\%, 50\%, 84\%, and 95\% quantiles ([av$_{05}$, av$_{16}$, av$_{50}$, av$_{84}$, av$_{95}$] and [dist$_{05}$, dist$_{16}$, dist$_{50}$, dist$_{84}$, dist$_{95}$]). We also excluded the sources with av$_{05}=$~av$_{16}$ because the StarHorse code had difficulties for them \citep{leike2020}. After removing the unreliable extinction estimates mentioned above, there are about 17.6 million sources remaining.

We then used the quantiles to calculate the means and standard deviations of extinction ($\mu_{A_V}$, $\sigma_{A_V}$) and distance ($\mu_{\textrm{dist}}$, $\sigma_{\textrm{dist}}$) assuming a Gaussian error for them:
\begingroup
\allowdisplaybreaks
\begin{align}
    \mu_{A_V}&=\frac{1}{2}(\textrm{av}_{\textrm{16}}+\textrm{av}_{\textrm{84}})\\
    \sigma_{A_V}&=\frac{1}{2}(\textrm{av}_{\textrm{84}}-\textrm{av}_{\textrm{16}})\\
    \mu_{\textrm{dist}}&=\frac{1}{2}(\textrm{dist}_{\textrm{16}}+\textrm{dist}_{\textrm{84}})\\
    \sigma_{\textrm{dist}}&=\frac{1}{2}(\textrm{dist}_{\textrm{84}}-\textrm{dist}_{\textrm{16}}).
\end{align}
\endgroup
We also obtained the dereddened NIR colors of $[J-H]_0$ and $[H-K_s]_0$ using $\mu_{A_V}$ based on the extinction law suggested by \citet{wangchen2019}. Furthermore, we applied the following criteria to the selected 17.6 million StarHorse sources:
\begin{align}
    &-1<[J-H]_0<1.5\\
    &-1<[H-K_s]_0<1.5\\
    &\mu_{\textrm{dist}}>\sigma_{\textrm{dist}} \label{cria10}\\ 
    &\mu_{\textrm{dist}}<20\,\textrm{kpc} \label{cria11}.
\end{align}
Finally we obtained about 17.47 million DaoPHOT sources which have high-quality (hq) StarHorse extinction estimates.

For the DaoPHOT sources without hq StarHorse extinction estimates, we used PNICER method to estimate their extinction values. \citet{pnicer2017} presented an unsupervised machine learning technique, PNICER, which determined the probability  distribution of extinction through fitting the features in color space of sources in the reference field. PNICER decreases the variance of intrinsic color measurements and thus can offer more accurate extinction estimation than NICER \citep{nicer2001}. \citet{meingast2018} also successfully applied PNICER to Orion A based on the VIenna Survey In OrioN \citep[VISION,][]{vision2016} data.
A necessary input of PNICER is the intrinsic colors of sources, which are usually inferred with an extinction-free region close to the target field. However, it is quite difficult to identify such extinction-free reference fields in the Galactic plane. Therefore, we directly adopted the DaoPHOT sources with hq StarHorse estimates as the reference stars. In specific, the extinctions towards DaoPHOT souces without hq StarHorse estimates ($A_{V,\textrm{PNICER}}^{\textrm{control:Starhorse}}$) in one tile were obtained by comparing their observed colors to the de-reddened colors of DaoPHOT sources with hq StarHorse extinction estimates in the same tile using the PNICER python package\footnote{\url{https://github.com/smeingast/PNICER}} based on the extinction law suggested by \citet{wangchen2019}. The uncertainties of $A_{V,\textrm{PNICER}}^{\textrm{control:Starhorse}}$ were mainly due to the photometric uncertainties and the variations of intrinsic colors.

The above PNICER process relied on the assumption that the probability density distributions of intrinsic colors of observed DaoPHOT sources can be well-represented by the de-reddened color distributions of reference stars. However, the DaoPHOT sources with hq StarHorse extinction estimates as reference stars were selected from all DaoPHOT sources with a bias to the bright and nearby stellar objects, which, obviously, introduced additional uncertainties in the PNICER extinction determination process. Subsequently, we tried to quantify these additional uncertainties using the Besan{\c c}on model of stellar population synthesis of the Galaxy \citep{besancon2003}. Here we have ignored the contamination of galaxies in the DaoPHOT catalogs because the high star, dust, and gas density of the Galactic plane obscure most of extragalactic objects \citep{extragalactic2020}. Actually, \citet{galaxies2012} identified 204 galaxy candidates in tile "d003" by visual inspection on the VVV images, resulting in a surface density of $\sim$124.7 galaxies per deg$^2$. \citet{galaxies2018} also searched extragalactic sources in VVV tiles "d010" and "d115" by their photometric procedures and finally detected 345 and 185 extragalactic candidates in 752,233 and 310,283 photometric sources. \mzrevi{\citet{baravalle2021} then applied \citet{galaxies2018}'s method to the disk parts of $\sim$220 deg$^2$ covered
by the VVV survey and finally identified 5563 galaxies.} \citet{galaxies2012} estimated that $\sim$15,000 galaxies can be detected among about a billion point sources in the whole VVV survey area based on the semi-analytic galaxy formation model \citep{bower2006} and the Galactic interstellar extinction model \citep{amores2005}, which suggested that only 1 (0.02\%) out of $\sim$5000 objects was a galaxy. Thus the contamination of galaxies in the DaoPHOT catalogs is negligible.

\begin{figure*}
    \centering
    \includegraphics[width=1.0\linewidth]{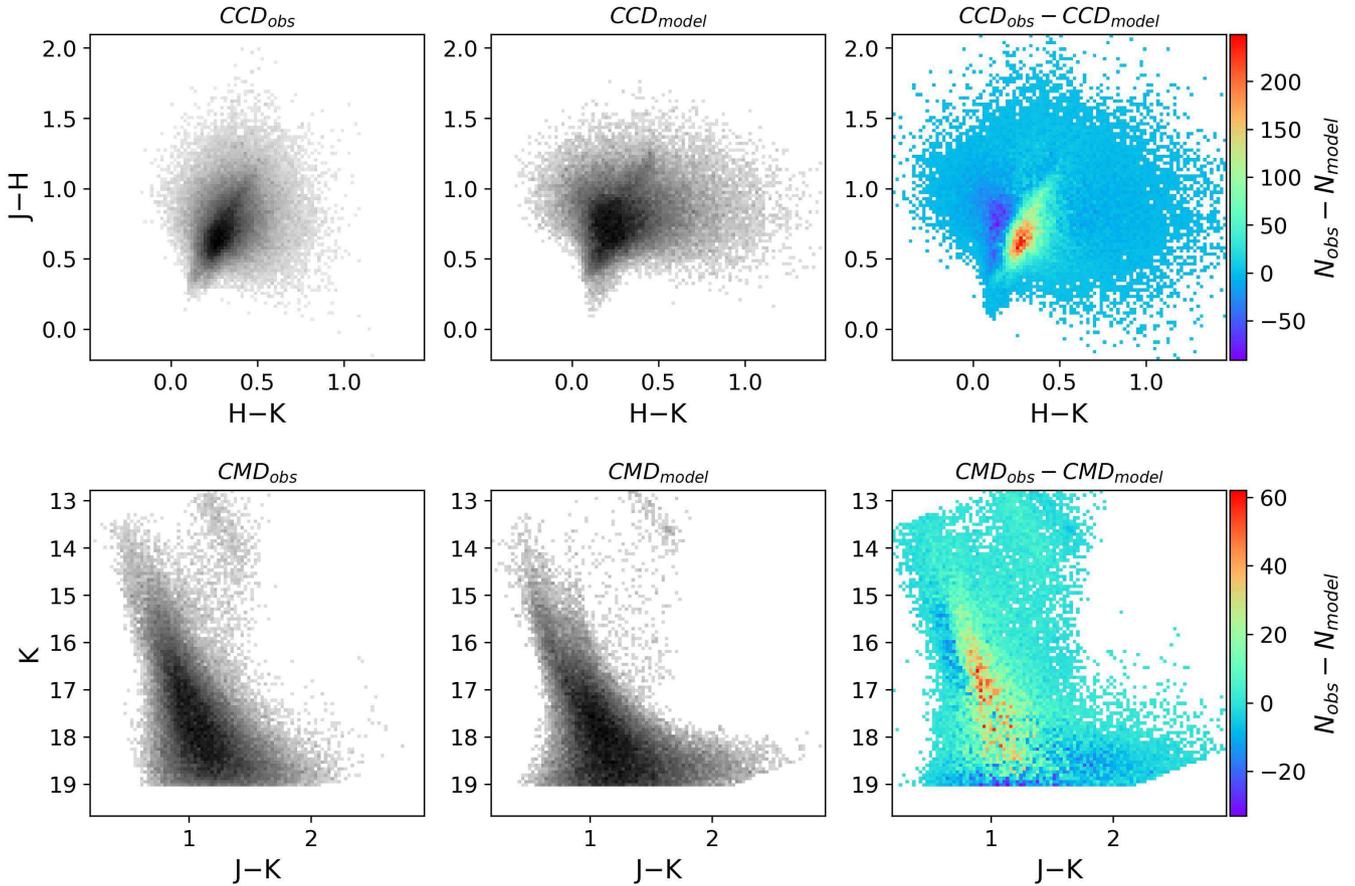}
    \caption{$H-K$ vs. $J-H$ CCDs (upper) and $J-K$ vs. $K$ CMDs (bottom) of the 15\arcmin$\times$15\arcmin~central region of tile "d003". Left panels: CCD and CMD constructed with observed DaoPHOT sources; middle panels: the synthetic CCD and CMD built with the Besan{\c c}on Galactic model \citep{besancon2003} assuming a diffuse extinction distribution of 0.6\,mag/kpc in the visual band; right panels: differential CCD and CMD obtained by subtracting the number density map built with Besan{\c c}on Galactic model from that constructed with the observed DaoPHOT sources.}
    \label{fig:ccd-cmd-d003}
\end{figure*}

\begin{figure*}
    \centering
    \includegraphics[width=1.0\linewidth]{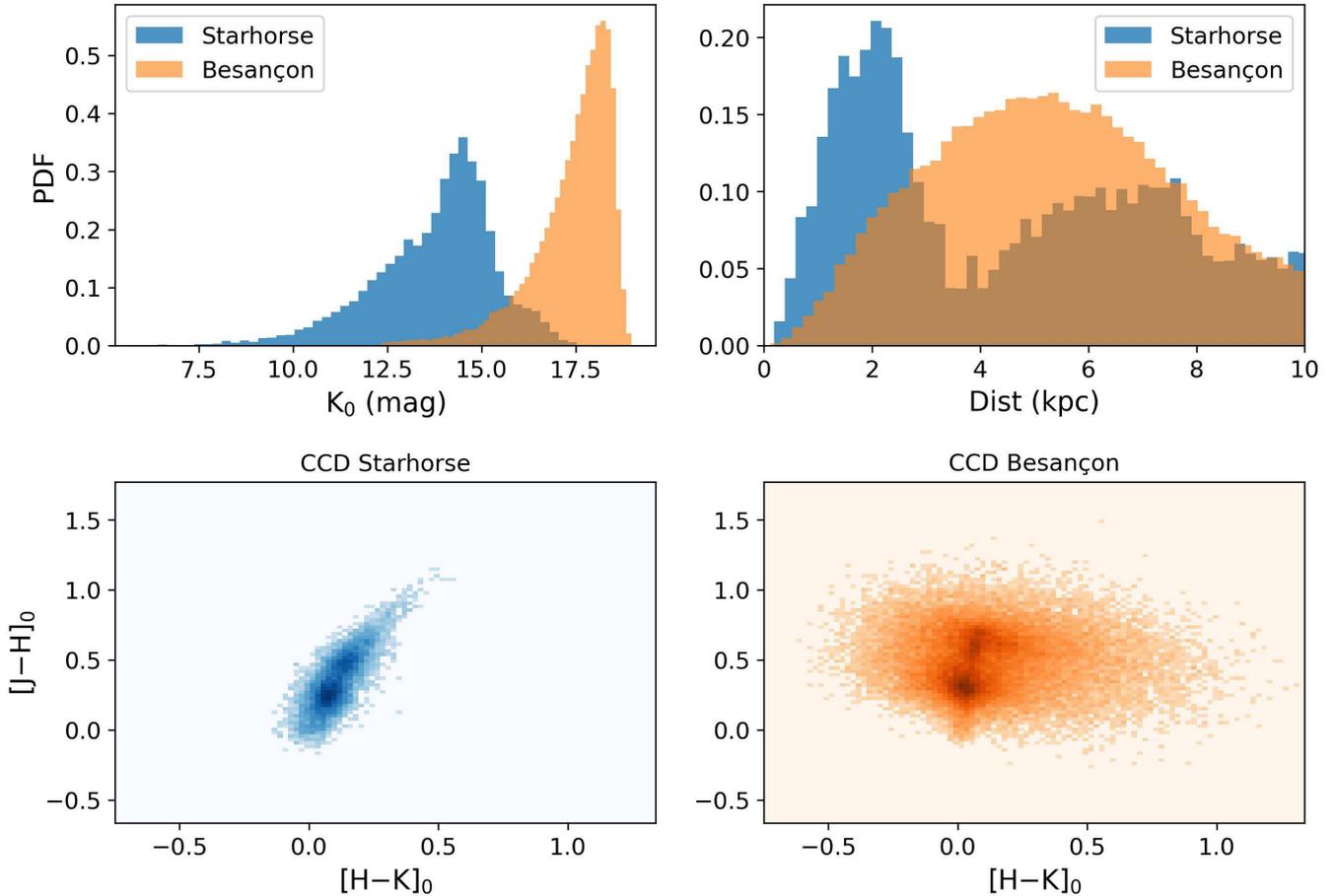}
    \caption{Top panels: PDFs of de-reddened $K$-band magnitudes (top left) and distances (top right) for the DaoPHOT sources with hq Starhorse estimates and the pseudo-stars from the best-fitting Besan{\c c}on Galactic model; bottom panels: intrinsic $[H-K]_0$ vs. $[J-H]_0$ CCDs built with DaoPHOT sources with hq Starhorse estimates (bottom left) and pseudo-stars from the best-fitting Besan{\c c}on Galactic model (bottom right); in the 15\arcmin$\times$15\arcmin~central region of tile "d003".}
    \label{fig:model-to-obs}
\end{figure*}

\begin{figure}
    \centering
    \includegraphics[width=1.0\linewidth]{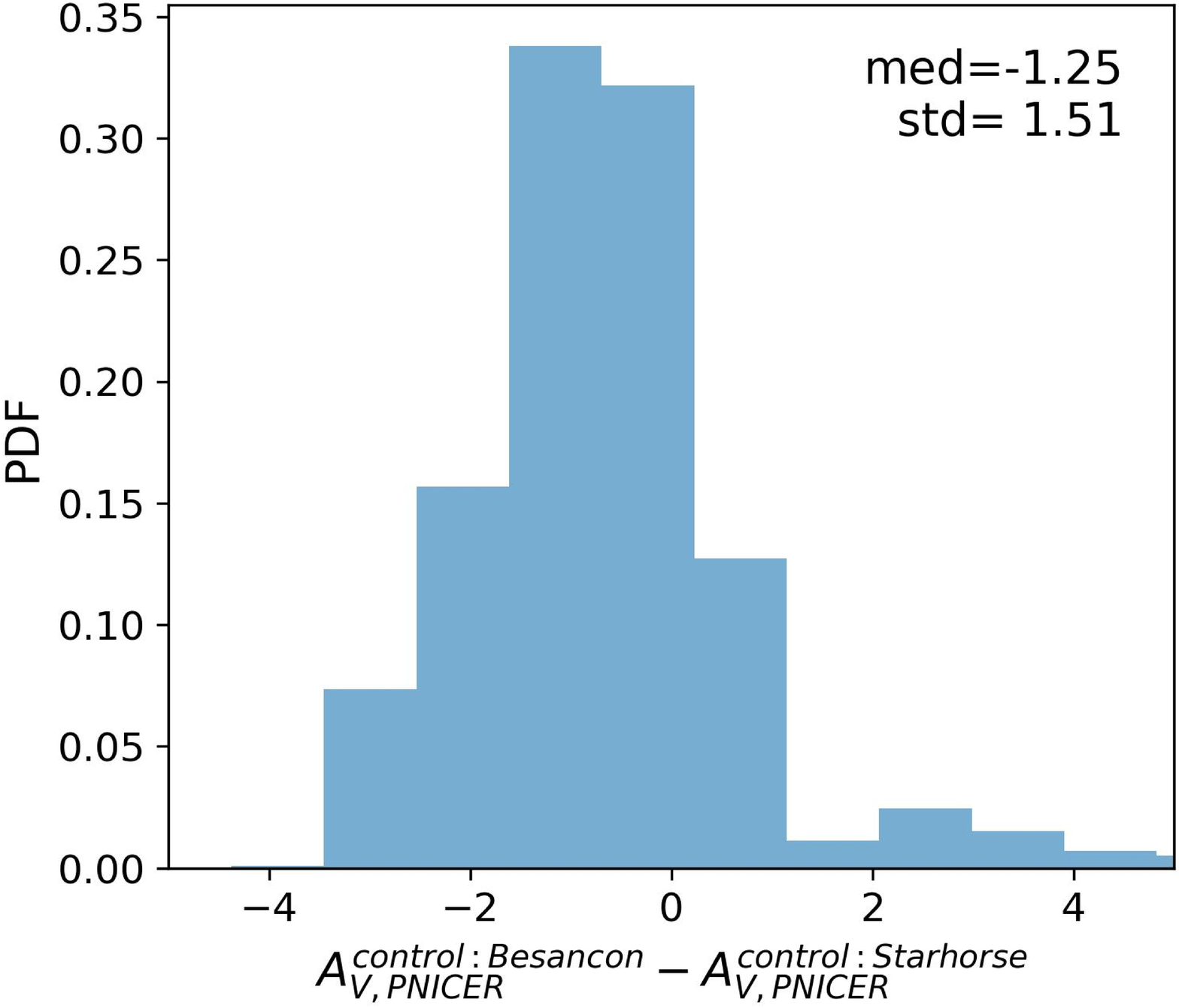}
    \caption{Difference between visual extinctions calculated using the DaoPHOT sources with hq Starhorse extinction estimates as reference stars ($A_{V,\textrm{PNICER}}^{\textrm{control:Starhorse}}$) and that calculated using the pseudo-stars from the Besan{\c c}on model as reference stars ($A_{V,\textrm{PNICER}}^{\textrm{control:Besan{\c c}on}}$) in the 15\arcmin$\times$15\arcmin~central region of tile "d003".}
    \label{fig:ext-model-to-obs}
\end{figure}

We selected a 15\arcmin$\times$15\arcmin~central region in each tile and extracted the pseudo-stars from the Besan{\c c}on Galactic model \citep{besancon2003}. Due to the lack of information about the extinction distribution profile along the line of sight, we assumed a simple linear relation of $A_V=aD$, where $A_V$ is the total extinction along one sightline till the distance of $D$\,kpc. We constructed 60 synthetic color-color diagrams (CCDs) and color-magnitude diagrams (CMDs) with the pseudo-stars based on the different $a$ values from 0.1 to 6 with the step of 0.1\,mag per kpc. The best pseudo-star model of observed DaoPHOT sources can be obtained by minimizing the differential CCDs and CMDs that were built by subtracting the synthetic number density map from the observed number density map in the color space. Figure~\ref{fig:ccd-cmd-d003} shows the CCDs and CMDs constructed with observed DaoPHOT sources and best pseudo-stars from the Besan{\c c}on Galactic model in the 15\arcmin$\times$15\arcmin~central region of tile "d003". We used these best pseudo-star models to describe the intrinsic color distribution of the corresponding DaoPHOT sources. Figure~\ref{fig:model-to-obs} shows the distributions of de-reddened magnitudes, distances, and intrinsic colors for the DaoPHOT sources with hq Starhorse estimates and for the pseudo-stars from the best-fitting Besan{\c c}on Galactic model in the 15\arcmin$\times$15\arcmin~central region of tile "d003". Obviously, the DaoPHOT sources with hq Starhorse extinction estimates were significantly biased to the bright and close stars as a selected sub-sample from all DaoPHOT sources. We also calculated extinctions towards DaoPHOT sources ($A_{V,\textrm{PNICER}}^{\textrm{control:Besan{\c c}on}}$) in the 15\arcmin$\times$15\arcmin~central region of each tile with PNICER method using the pseudo-stars predicted by the best-fitting Besan{\c c}on model as the reference stars. Figure~\ref{fig:ext-model-to-obs} shows the histogram of difference between $A_{V,\textrm{PNICER}}^{\textrm{control:Starhorse}}$ and $A_{V,\textrm{PNICER}}^{\textrm{control:Besan{\c c}on}}$ in the 15\arcmin$\times$15\arcmin~central region of tile "d003". We adopted the standard deviation ($\sigma_{\Delta A_V}$) of $\Delta A_V = A_{V,\textrm{PNICER}}^{\textrm{control:Besan{\c c}on}}-A_{V,\textrm{PNICER}}^{\textrm{control:Starhorse}}$ as the additional uncertainty introduced by the bias of reference stars in each tile, e.g., $\sigma_{\Delta A_V}=$~1.51\,mag in tile "d003" which has been marked in Fig.~\ref{fig:ext-model-to-obs}. Therefore, $\sigma_{\Delta A_V}$ in the whole VVV survey area was a function of tile positions. We finally smoothed the $\sigma_{\Delta A_V}$ map using a Gaussian kernel with the full width at half maximum (FWHM) of 1.5\degr~to avoid the edge effect. The total error ($\sigma_{\textrm{total}}$) of $A_V$ of the DaoPHOT sources without hq Starhorse extinction estimates was the combination of error given by PNICER method ($\sigma_{\textrm{PNICER}}$) and that due to the bias of reference stars ($\sigma_{\Delta A_V}$), i.e., $\sigma_{\textrm{total}}=\sqrt{\sigma_{\textrm{PNICER}}^2+\sigma_{\Delta A_V}^2}$. During the above process, we have ignored the systematic difference of $\Delta A_V$ that can be seen in Fig.~\ref{fig:ext-model-to-obs} because it is sensitive to the assumed dust distribution along the line of sight and thus difficult to quantify without the knowledge of 3D dust distribution. Because the systematic difference affected the zero point of our extinction map, we revisited it in Sect.~\ref{sect:zmag}.
 
\subsection{Background source selection and extinction mapping}\label{sect:background-and-mapping}

\mz{To map the integrated Galactic extinction, we must isolate the background sources that are as far as possible, i.e., are on the background of as much dust as possible.}
In principle, \jk{the optimal way would be to use}
the stars at the far-side of the Galaxy, which is of course, impossible without the distance information of stars. However, a reasonable approach is to select the most extincted stars as the background stars\jk{, based on the fact that on average the total extinction increases with the distance from the Sun}. We note that this approach ignores any structure within the resolution element.

\begin{figure}
    \centering
    \includegraphics[width=1.0\linewidth]{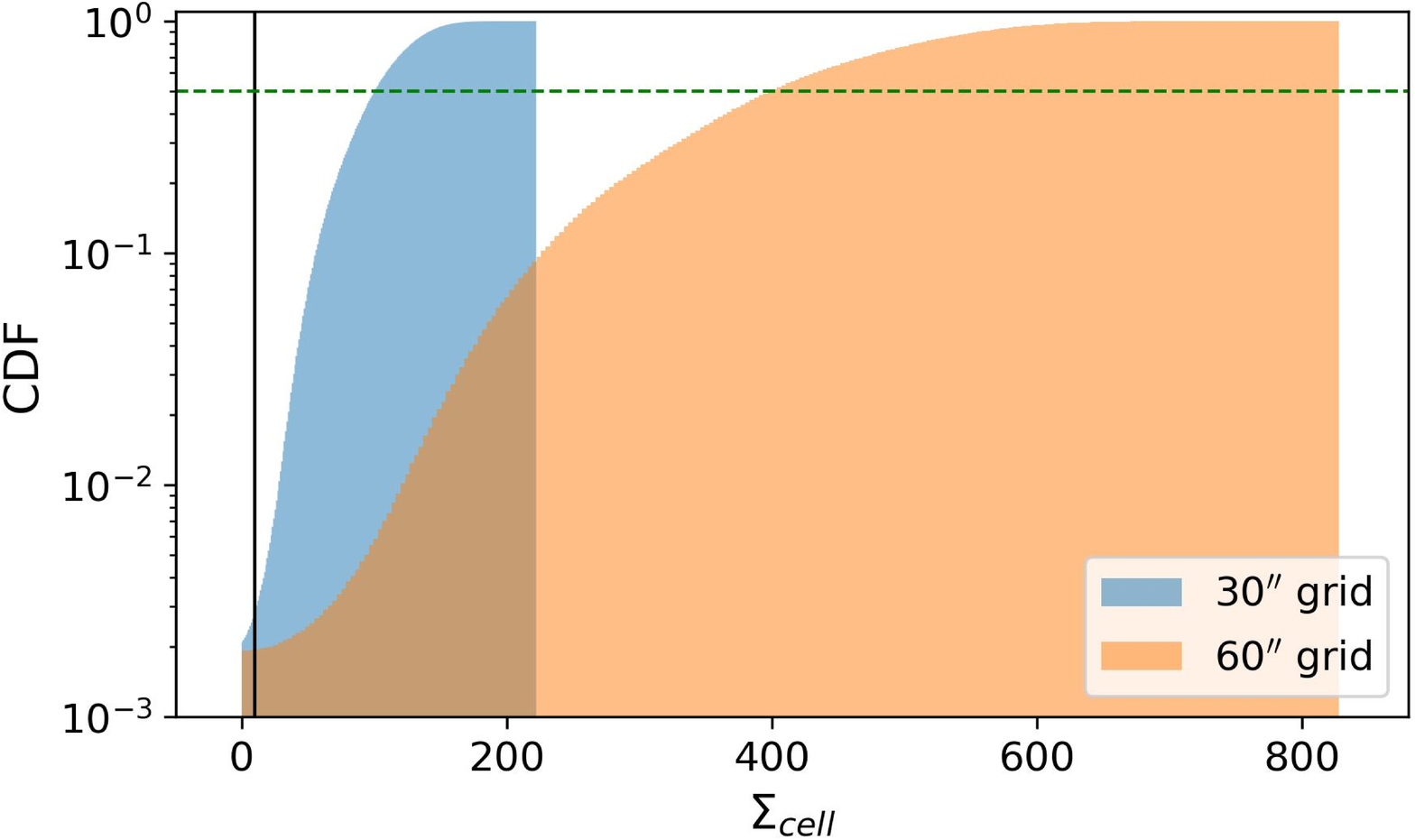}
    \caption{The cumulative density distributions of DaoPHOT source number density ($\Sigma_{\textrm{cell}}$) at the 30\arcsec~and 60\arcsec~grids of the whole VVV survey area. The black solid line marks $\Sigma_{\textrm{cell}}=$~10 while the green dashed line labels 50\% percentile value.}
    \label{fig:hist_densmap}
\end{figure}

We used the "X percentile method" to derive the extinction maps. The key of this method is to utilize the color of the X percentile reddest star to measure the color excess of a sightline cell. The full description of the X percentile method can be found in \citet{dobashi2008}, we only summarize our implementation here. First, we gridded the whole VVV survey area into discrete square sightline cells. The size of grid cell was selected to be 30\arcsec~and 60\arcsec~based on our experience. Figure~\ref{fig:hist_densmap} shows the cumulative density function (CDF) of DaoPHOT source number density ($\Sigma_{\textrm{cell}}$) at 30\arcsec~and 60\arcsec~grids. The 50\% percentiles of $\Sigma_{\textrm{cell}}$ were 100 and 401 counts at 30\arcsec~and 60\arcsec~grids, respectively. About 2-3\textperthousand~of grid cells have $\Sigma_{\textrm{cell}}\leq$~10. The 30\arcsec~grid can offer higher spatial resolution while the 60\arcsec~grid can give lower $A_V$ noise level for the final extinction maps. Actually, the selection of these two grids (30\arcsec~and 60\arcsec) can make sure that there is at least one DaoPHOT source in $>$99.97\% beams of the final extinction maps. 

Second, we located the DaoPHOT sources in each grid cell. Then we sorted the stars according to their $A_V$ values \mz{in the ascending order} and denoted the \textit{q}-th percentile of $A_V$ as $A_V(q)$. Considering $X_0$- and $X_1$-th percentiles (0\%$\leq X_0 < X_1 \leq$100\%) of $A_V$, i.e., $A_V(X_0)$ and $A_V(X_1)$, we can select DaoPHOT sources in each grid cell with $A_V(X_0)\leq A_V\leq A_V(X_1)$ as the background sources. \mz{The classical NICER method \citep{nicer2001} usually uses the spatial averages (or medians) of extinctions towards all detected stars, which is equivalent to setting $X_0 =$~0\% and $X_1 =$~100\%}. In general, higher $X_0$ and $X_1$ values are more helpful to select the true background sources, especially for the clouds that are heavily contaminated by the foreground stars. Assuming a cloud located at a distance with the uniform density distribution along a line of sight, \citet{dobashi2008} modelled the cloud color excess as a function of $X_0$, i.e., $E(X_0)$, and found that $E(X_0)$ is close to the true cloud color excess when $X_0\geq $70\%. \citet{dobashi2008} and \citet{dobashi2009} obtained the extinction maps for the Large Magellanic Cloud (LMC) and Small Magellanic Cloud (SMC) using the X percentile method with different $X_0$ based on 2MASS source catalog. By comparing to the CO maps of LMC and SMC \citep{mizuno2001lmc,mizuno2001smc}, they also found that $X_0=$~80\% gives a reasonable moderate noise level and should trace the cloud extinction best. Therefore, we decided to adopt the suggestions from \citet{dobashi2008} and \citet{dobashi2009} and use $X_0=$~80\% to select the background sources. However, for some distant and/or very dense dust clouds, $X_0=$~80\% was still not high enough to select the true background sources. Thus we also used the configuration of $X_0=$~90\%, but only for 60\arcsec~grid. 

\begin{figure}
    \centering
    \includegraphics[width=1.0\linewidth]{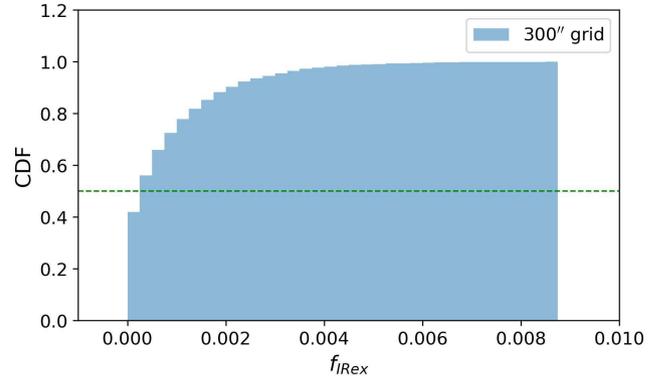}
    \caption{The cumulative density distributions of the infrared excess source fraction per grid cell($f_{\textrm{IRex}}$) at the 300\arcsec~grid of the regions with GMFs \citep{mypub2019}. The green dashed line labels 50\% percentile value.}
    \label{fig:hist-ysofraction}
\end{figure}

We also set $X_1=$~95\% to remove the sources with infrared excess. The intrinsically red sources such as young stellar objects (YSOs) and asymptotic giant branch stars (AGBs) also contaminated our selected background sources because their infrared excess results in a serious overestimation of their true color excess and extinctions. \citet{marton2016} presented an all-sky catalog of sources with infrared excess as the YSO candidates using the support vector machine algorithm based on 2MASS and AllWISE catalogs \citep{wise2010}. The whole catalog includes 133, 980 Class I/II candidates and 608, 606 Class III candidates, of which 363, 564 YSO candidates are located in the VVV survey area. However, considering the limited sensitivity of 2MASS and WISE, we should expect more infrared excess sources in our VVV DaoPHOT source catalogs. \citet{vision32019} revisited the YSO candidates in the Orion A molecular clouds based on the NIR VISION data and mid- and far-infrared survey data such as Spitzer \citep{megeath2012}, WISE, and Herschel \citep{hops2016}. They compared the YSO candidates selected with Spitzer/VISTA data and that identified with WISE/VISTA data in L1641 and found that the WISE selection method recovers about 59\% of the YSOs identified with Spitzer/VISTA data. Therefore, there could be about 0.6 million  contaminants with infrared excess of our DaoPHOT sources, which gives a low contamination fraction of $\sim$1\textperthousand. The Galactic AGB stars could have an universal spatial distribution \citep{jackson2002,rob2008}, but the YSOs usually cluster in the dense gas regions \citep{lada2010}, which suggests that the above contamination fraction increases towards the higher extinction regions. \citet{mypub2019} searched through several tens of million VVV/Spitzer/Herschel sources and identified 36,~394 sources with the infrared excess in 57 giant molecular filaments (GMFs) that are roughly defined as the filamentary structures with $A_V>$~3\,mag. We located their YSO sample in a 5\arcmin~grid of the VVV survey area and calculated the infrared excess source fraction per grid cell, i.e., $f_{\textrm{IRex}}$. Figure~\ref{fig:hist-ysofraction} shows the cumulative density distribution of $f_{\textrm{IRex}}$, which has a median value of  $<$1\textperthousand, but can be close to 1\% in some regions with active star formation. Therefore, $X_1=$~95\% can efficiently exclude the sources with infrared excess in each grid cell. We note that the $X_1$ value of 95\% is also the same as that suggested by \citet{dobashi2008}, \citet{dobashi2009}, and \citet{dobashi2011}. Using the configuration of $X_0=$~80\% (90\%) and $X_1=$~95\% we obtained about 120 (41) million DaoPHOT background sources in the VVV survey area.

Finally, we smoothed the $A_V$ values of selected background sources with a Gaussian kernel to obtain the extinction maps. The FWHMs of Gaussian kernels were 30\arcsec~and 60\arcsec~for the background sources selected from 30\arcsec~and 60\arcsec~grids, \mz{respectively}. To decrease the calculation time, we also set a radius of 2$\times$FWHM to truncate the smoothing. The pixel size of the extinction map was defined as the half of the FWHM of Gaussian kernel. This smoothing process actually assumed that the dust density distribution was smooth on small scales, i.e., if a dust cloud appears in one sightline it is likely to appear in neighboring sightlines.

The uncertainties of the extinction maps were estimated with a Monte Carlo method as suggested by \citet{dobashi2008}. Assuming a Gaussian error for $A_V$ values of DaoPHOT sources, we can generate a random $A_V$ measurement for each DaoPHOT source and produce a simulated DaoPHOT catalog. Then we located the simulated DaoPHOT sources into a grid and selected the background sources in each grid cell with the XPNICER technique described above. The extinction maps can be obtained by smoothing the $A_V$ measurements of the simulated background sources. By repeating the simulation 100 times we can obtain 100 extinction maps. The final extinction map and the associated uncertainty map were adopted as the mean and standard deviation of these 100 simulated extinction maps.

\section{RESULTS}
\label{sect:results}

\subsection{Dust extinction maps}
\label{sect:extmap}

\begin{figure*}
    \centering
    \includegraphics[width=1.0\linewidth]{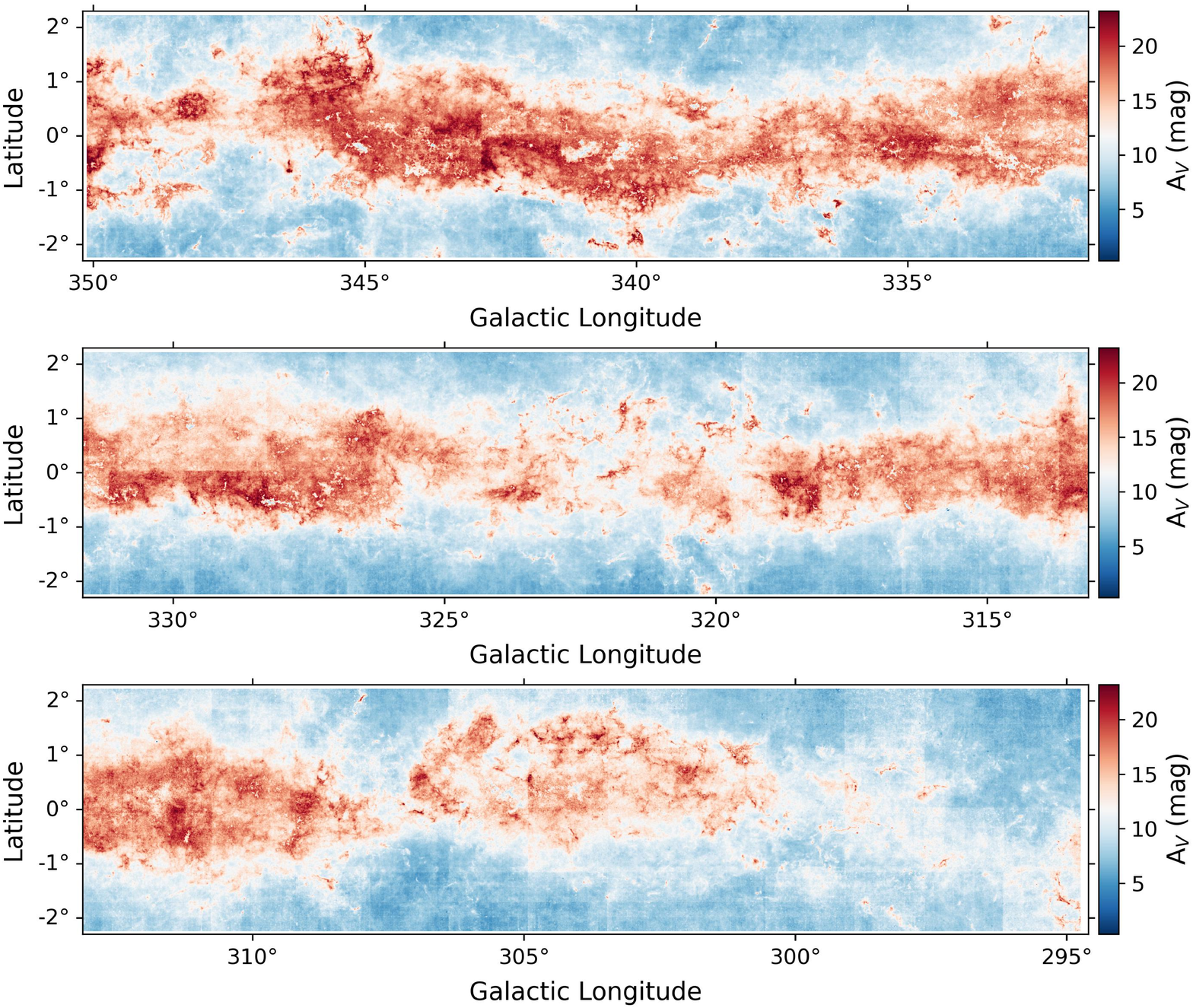}
    \caption{The XPNICER extinction maps obtained using $X_0=$~80\% and $X_1=$~95\% with the spatial resolution of 30\arcsec~for the Galactic disk area.}
    \label{fig:extmap_disk}
\end{figure*}

\begin{figure*}
    \centering
    \includegraphics[width=1.0\linewidth]{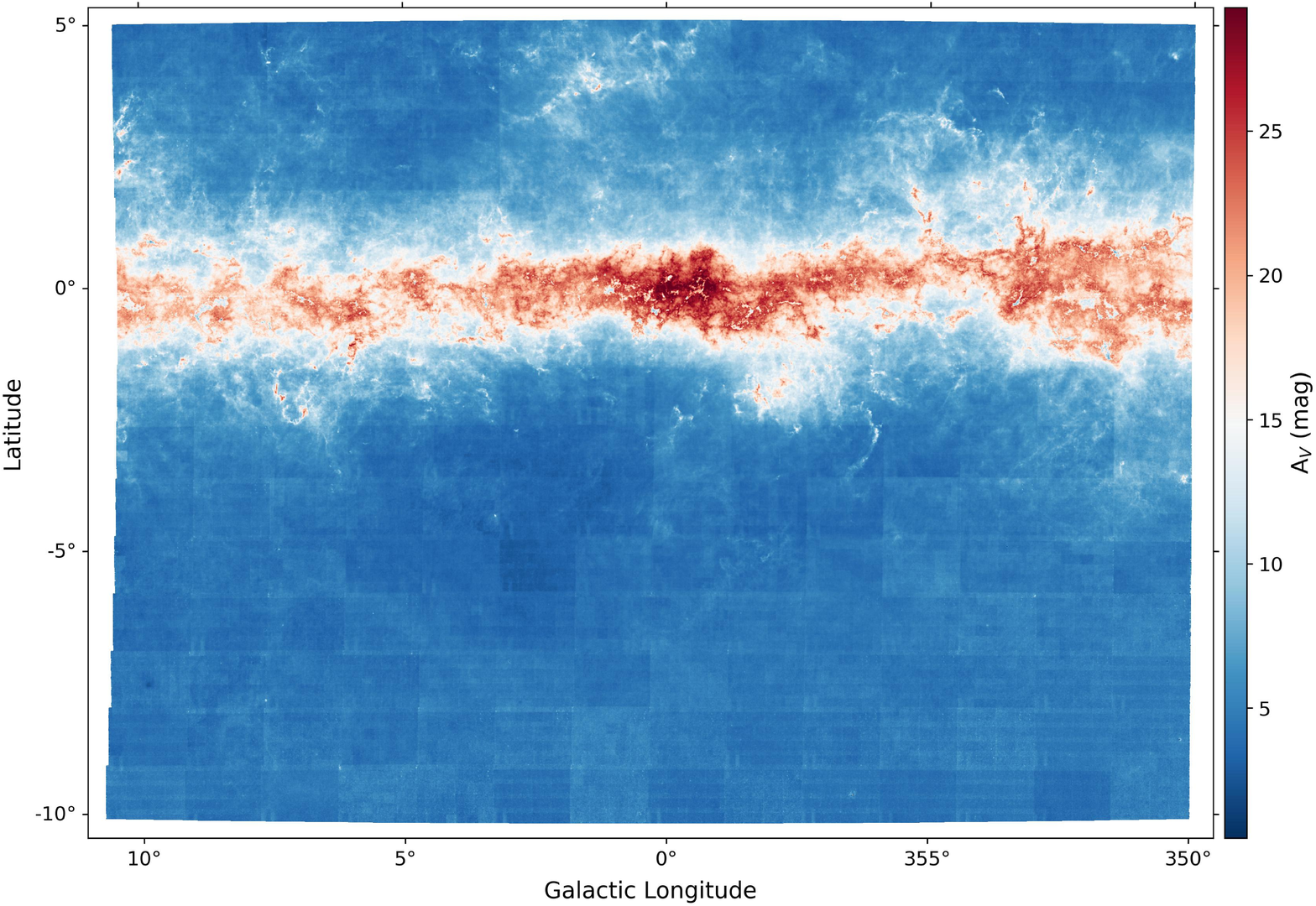}
    \caption{The XPNICER extinction map obtained using $X_0=$~80\% and $X_1=$~95\% with the spatial resolution of 30\arcsec~for the Galactic bulge area.}
    \label{fig:extmap_bullge}
\end{figure*}

\begin{figure*}
    \centering
    \includegraphics[width=1.0\linewidth]{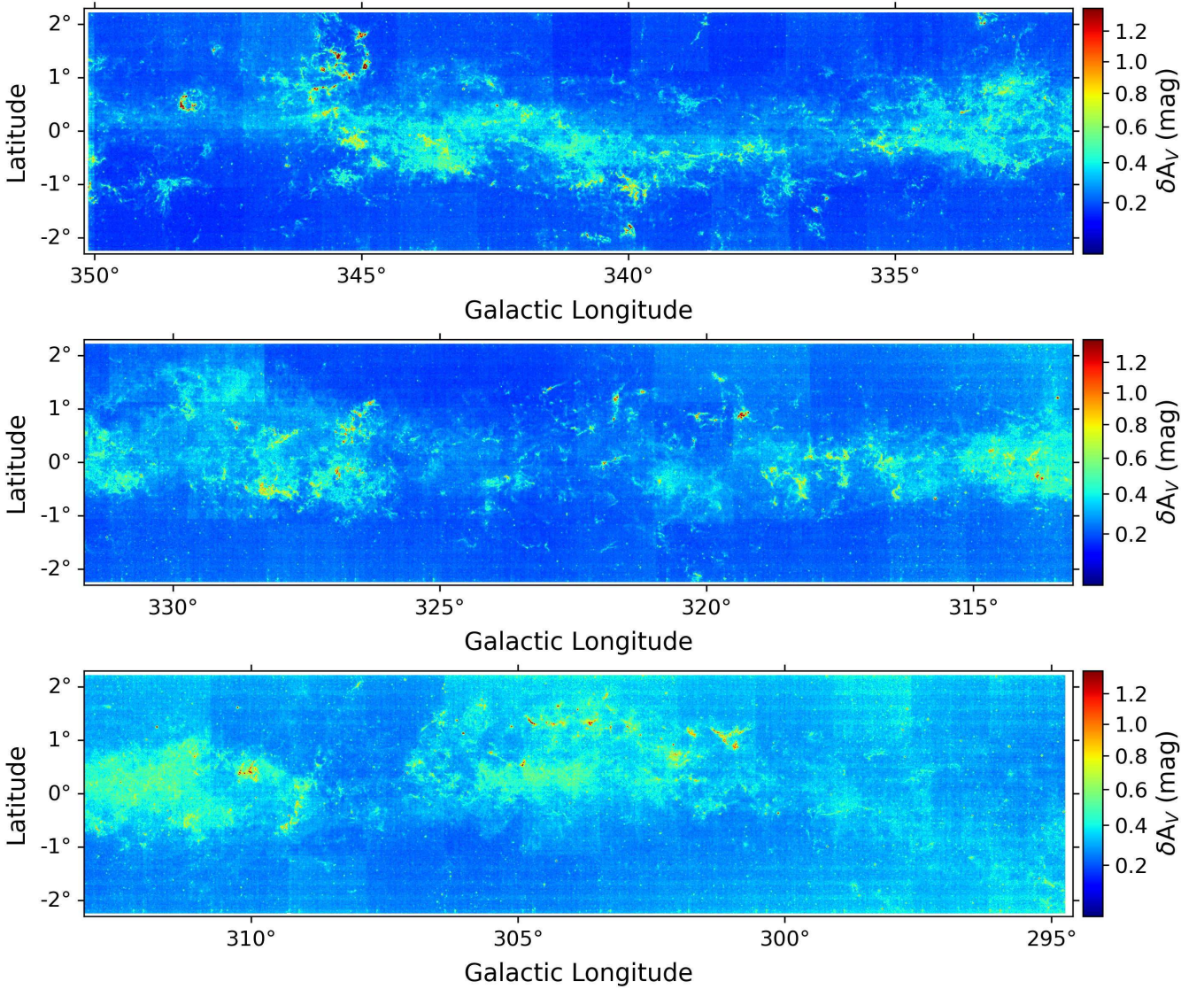}
    \caption{The uncertainty maps of the extinction maps shown in Fig.~\ref{fig:extmap_disk}.}
    \label{fig:errmap_disk}
\end{figure*}

\begin{figure*}
    \centering
    \includegraphics[width=1.0\linewidth]{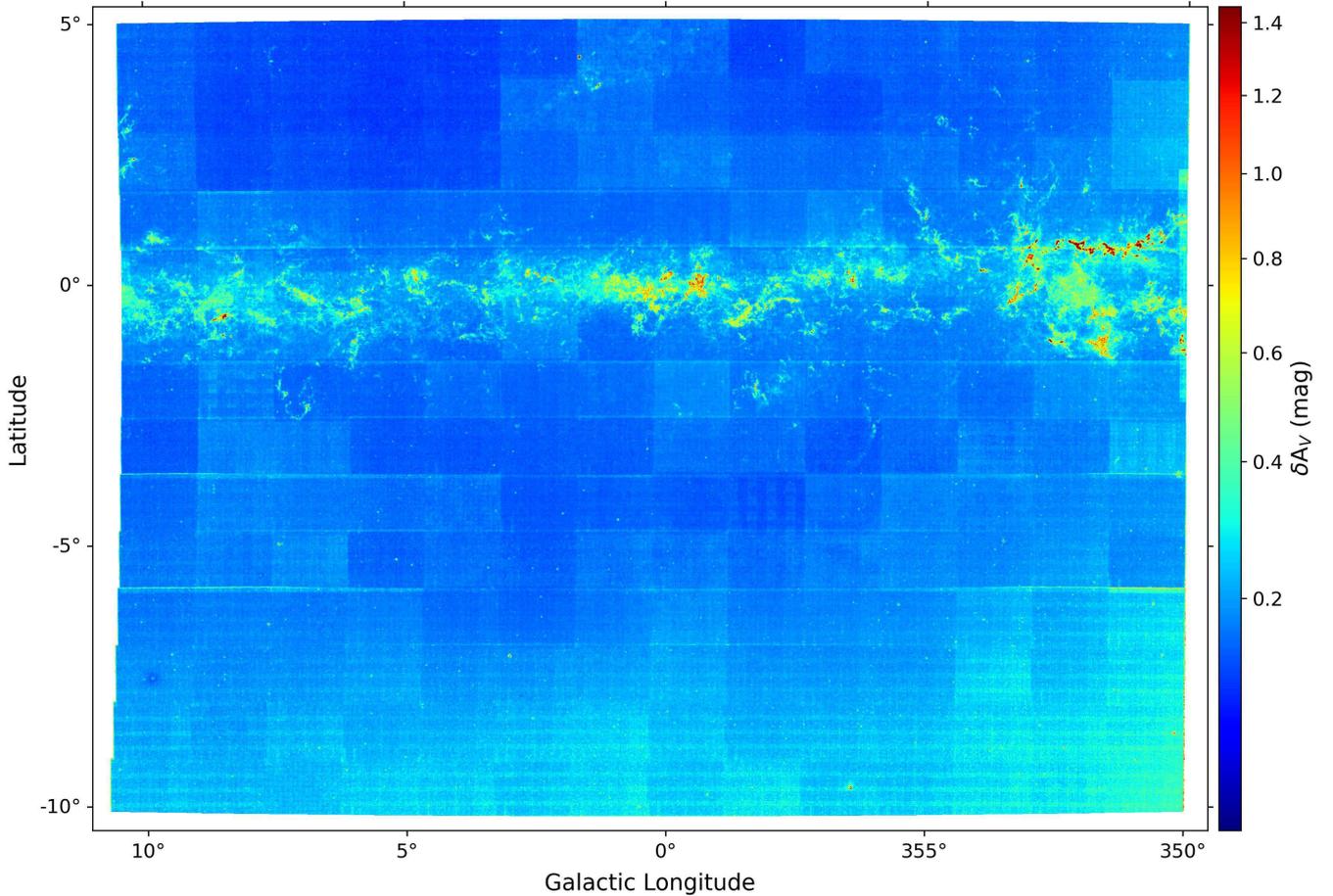}
    \caption{The associated uncertainty map of the extincon map shown in Fig.~\ref{fig:extmap_bullge}.}
    \label{fig:errmap_bulge}
\end{figure*}

Figures~\ref{fig:extmap_disk} and \ref{fig:extmap_bullge} show our XPNICER extinction maps of the Galactic disk and bulge areas covered by the VVV survey in units of $A_V$. Figures~\ref{fig:errmap_disk} and \ref{fig:errmap_bulge} show the corresponding uncertainty. The maps were obtained using the configuration of $X_0=$~80\% and $X_1=$~95\% with the spatial resolution of 30\arcsec. In the following, we refer to the extinction and uncertainty maps as $A_V^{30}(X_0=80)$ and $\delta A_V^{30}(X_0=80)$, respectively. 

The basic statistics of $A_V^{30}(X_0=80)$ and $\delta A_V^{30}(X_0=80)$ are shown in Table.~\ref{tab:tab1}. The maps can trace dust extinction up to $A_V\sim$~30\,mag with a typical uncertainty of $\sim$0.2\,mag. 
Figures~\ref{fig:extmap_disk} and \ref{fig:extmap_bullge} reveal a wealth of complex structures at different scales. 

By construction, the maps describe the extinction integrated to some \mz{limitting} distance, $d_{\textrm{limit}}$, along the line of sight. Obviously, $d_{\textrm{limit}}$ is a function of Galactic longitude and latitude and related to the detection limit of the VVV survey and the 3D dust distribution. It is impossible to fully quantify $d_{\textrm{limit}}$ without \mz{an actual 3D model} of the Galactic dust. \mz{Therefore,} here we only gave the rough estimation of \mz{the range within which $d_{\textrm{limit}}$ is likely to be}. \mz{To make the estimates, we} selected two 1\degr$\times$1\degr~low extinction regions towards [$l=$~0\degr, $b=-$9\degr] and [$l=$~330\degr, $b=-$1.5\degr]. Assuming a constant dust distribution profile of $A_V\sim$~0.7\,mag\,kpc$^{-1}$ \citep{besancon2003} in these two regions, we estimated the $d_{\textrm{limit}}$ that corresponds to the detection limit, $K=$~18.7\,mag, of the VVV DaoPHOT catalog \citep{vvvdaophotmypub} to be 12-16\,kpc. The estimate was done using the TRILEGAL\footnote{\url{http://stev.oapd.inaf.it/cgi-bin/trilegal}} model, \mz{which is a stellar population code that can simulate the the stellar photometry of any Galactic field} \citep{girardi2005}. Thus, the VVV survey can see the stars as far as $\sim$12-16\,kpc, if only considering a simple model for the diffuse dust distribution in the Galaxy. \mz{This} can be treated as the upper limit of $d_{\textrm{limit}}$. 

\begin{figure}
    \centering
    \includegraphics[width=1.0\linewidth]{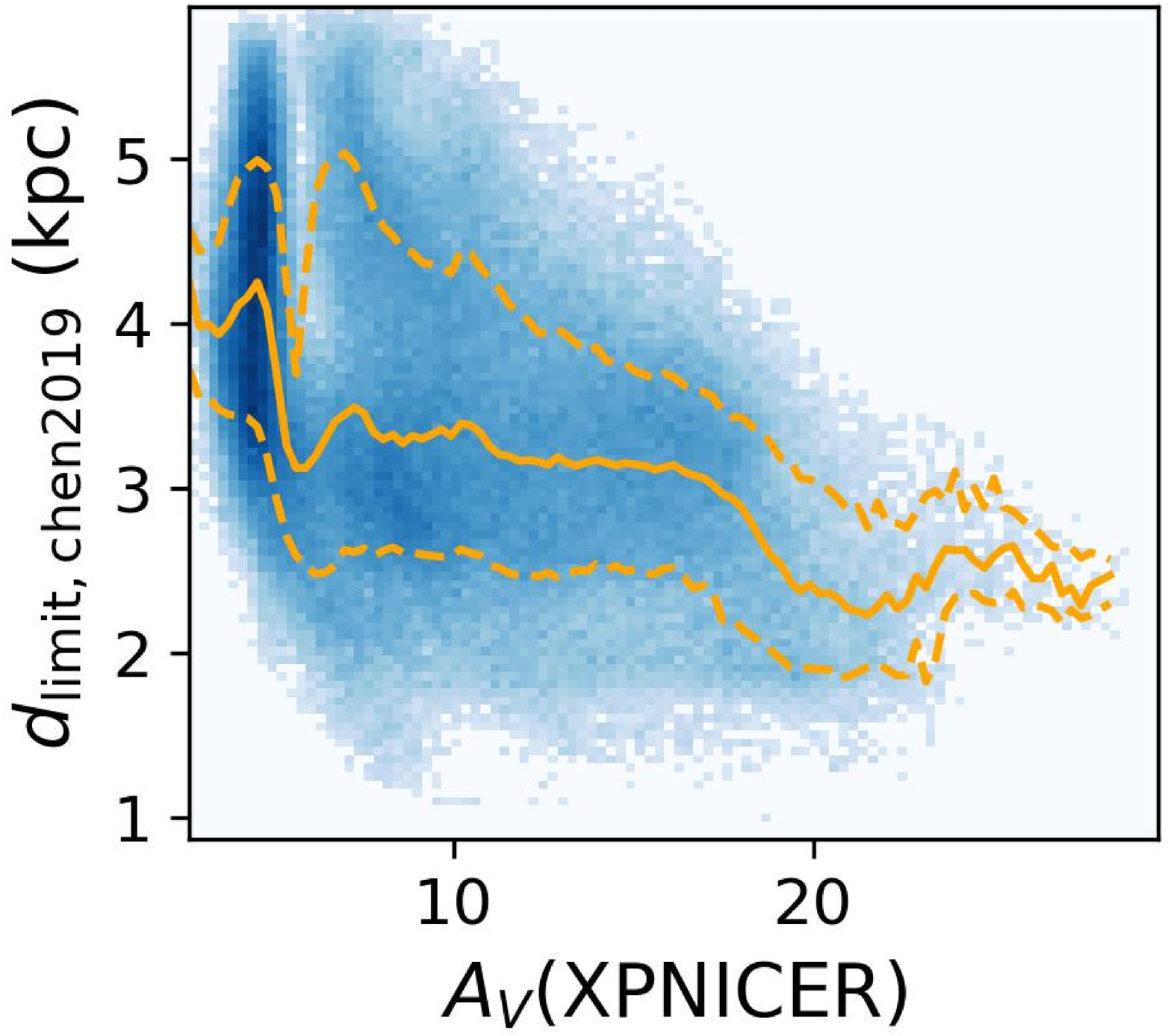}
    \caption{Pixel-to-pixel comparison of our XPNICER extinction map at 6\arcmin~resolution and \citet{chen2019}'s reliable depth map ($d_{\mathrm{limit,chen2019}}$). The orange solid line represent medians in the individual bins while the orange dashed lines mark 16\% and 84\% percentiles in the individual bins.}
    \label{fig:ext2chendmax}
\end{figure}

\mz{We also use an indirect way to estimate the lower limit of $d_{\textrm{limit}}$ using a published 3D extinction map}. \citet{chen2019} presented a 3D dust extinction map of the Galactic plane (0\degr$<l<$360\degr, $|b|<$10\degr) based on the Gaia DR2, 2MASS, and WISE data. They obtained the color excesses of over 56 million stars and mapped the Galactic dust distribution up to $\sim$4-6\,kpc at 6\arcmin~resolution. \citet{chen2019} defined the reliable depth of each pixel as the maximum distance, $d_{\textrm{limit,chen2019}}$, of all stars in that pixel. \mz{Fig. \ref{fig:ext2chendmax} shows the pixel-to-pixel comparison of our XPNICER map at 6\arcmin~resolution to the $d_{\textrm{limit,chen2019}}$ map}. We found the value of $d_{\textrm{limit,chen2019}}$ in the range of 0.8-6\,kpc with a median value of $\sim$3.5\,kpc. Compared with Gaia and 2MASS data, the VVV survey detected much more faint and distant stars. Thus $d_{\textrm{limit,chen2019}}$ can be treated as the \mz{conservative} lower limit of $d_{\textrm{limit}}$, which results in 0.8-6\,kpc$<d_{\textrm{limit}}<$12-16\,kpc depending on the line of sight. \jk{One can get the exact estimate of $d_{\textrm{limit}}$ for individual lines of sight from \citet{chen2019}.} 


\mz{There are several types of defects and artifacts in our XPNICER extinction maps, including defects by bright stars, absorption-like features in high extinction regions, artifacts caused by detector defects, and tile patterns in the maps. The detailed description and examples of the defects can be found in the Appendix~\ref{sect:defects}.}

\begin{table}
    \centering
    \caption{Statistics of the visual extinction maps and their associated uncertainty maps of the Galactic plane covered by the VVV survey}
    \begin{tabular}{lcccc}
    \hline\hline
       Maps  & min & max & mean & median \\
        & (mag) & (mag) & (mag) & (mag) \\
        \hline
        disk $A_V^{30}(X_0=80)$ & -6.66 & 28.89& 11.55& 10.84\\
        bulge $A_V^{30}(X_0=80)$ & -4.64 & 34.12 & 7.25 & 5.08\\
        disk $\delta A_V^{30}(X_0=80)$ & 1.55$\times$10$^{-4}$&4.72 & 0.27& 0.26\\
        bulge $\delta A_V^{30}(X_0=80)$ & 2.83$\times$10$^{-5}$&4.52 &0.19 & 0.18\\
        \hline
    \end{tabular}
    \label{tab:tab1}
\end{table}

\subsection{Uncertainties of the derived maps}
We have presented the uncertainty maps (see Fig.~\ref{fig:errmap_disk} and \ref{fig:errmap_bulge}) in Sect.~\ref{sect:extmap}. The sources of uncertainties are also discussed in Sect.~\ref{sect:method-singlestar}, including 1) observed photometric uncertainties, 2) variations of intrinsic colors, and 3) bias of reference stars, which, however, do not include the systematic uncertainties. In this section we discussed two main sources of the systematic uncertainties.

\subsubsection{Uncertainties of extinction map zero point}\label{sect:zmag}
We used DaoPHOT sources with hq StarHorse extinction estimates to infer the intrinsic colors of all DaoPHOT sources. Because the StarHorse sources were biased to bright and nearby stars, there was systematic difference between the de-reddened colors of StarHorse sources and the true intrinsic colors of DaoPHOT sources, which also resulted in the systematic error of zero point of our extinction maps. Due to the lack of 3D image of Galactic dust we did not investigate the systematic difference with the Besan{\c c}on Galactic model in Sect.~\ref{sect:method-singlestar}. Here we give a rough estimation of the systematic error of zero point of our extinction maps by comparing the PNICER extinction measurements of background DaoPHOT sources with the corresponding StarHorse2020 estimates.

\begin{figure*}
    \centering
    \includegraphics[width=1.0\linewidth]{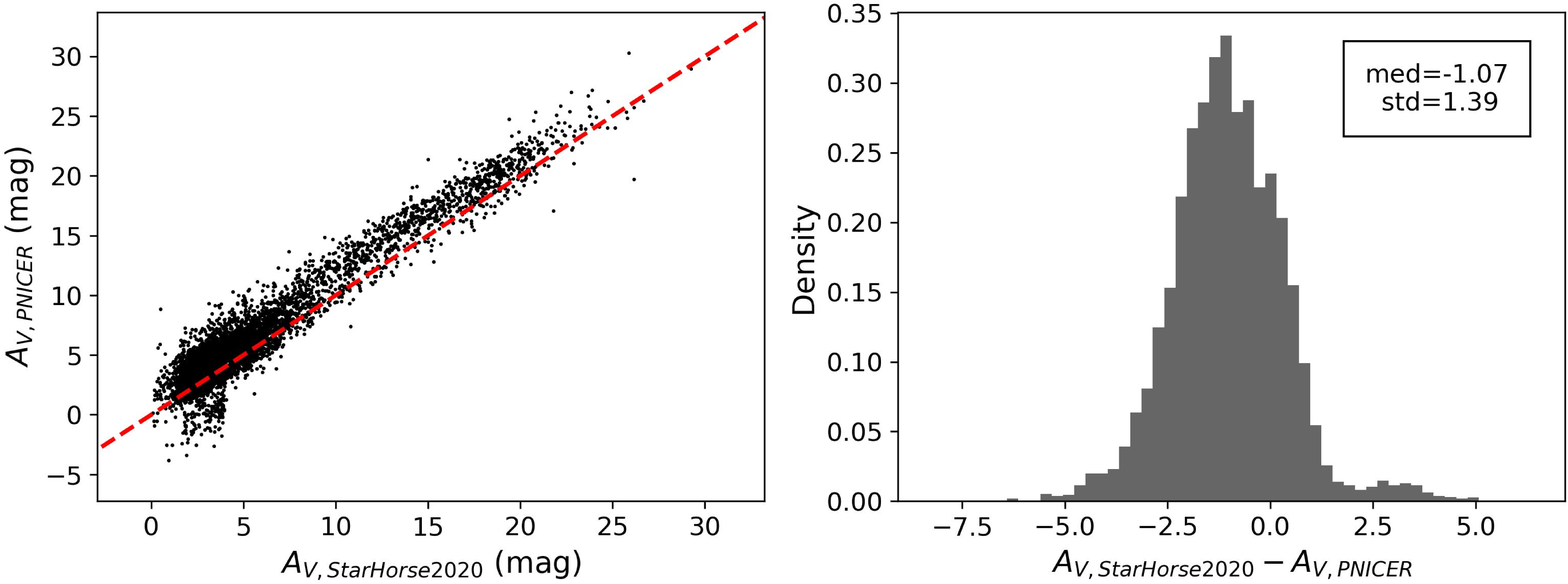}
    \caption{Left: comparison of DaoPHOT source extinction estimates obtained with PNICER method and from StarHorse2020 catalog \citep{starhorse2020}. The red dashed line shows the one-to-one relation. Right: histogram of extinction difference between PNICER measurements and StarHorse2020 estimates. The median value and standard deviation are also marked on the panel.}
    \label{fig:pnicer2sh20}
\end{figure*}

There are 5688 sources with hq StarHorse2020 extinction estimates out of 120 million DaoPHOT background sources selected with $X_0=$~80\% and $X_1=$~95\% from 30\arcsec~grid. Figure~\ref{fig:pnicer2sh20}  shows the comparison of extinction values obtained by PNICER method ($A_{V,\textrm{PNICER}}$) and from StarHorse2020 catalog ($A_{V,\textrm{StarHorse2020}}$). Considering that $A_{V,\textrm{StarHorse2020}}$ was estimated based on the spectroscopical observations and more reliable than $A_{V,\textrm{PNICER}}$, our PNICER method overestimated the source extinction by about 1\,mag. Therefore, the zero point of our $A_V$ extinction maps is about 1$\pm$1.4\,mag, which means that our extinction maps could systematically overestimate the integrated extinction.

\subsubsection{Uncertainties due to extinction laws}

Our work currently assumed an universal NIR extinction law suggested by \citet{wangchen2019} for the entire VVV survey area. The extinction laws in the NIR were commonly described with a power-law relation as $A_{\lambda}\propto \lambda^{-\alpha}$, where $\lambda$ is the wavelength and $\alpha$ is a dimensionless parameter that defines a specific extinction law. \citet{wangchen2019} combined the data from different large-scale spectroscopic, astrometric, and photometric surveys to derive a relative extinction curve from optical to mid-infared bands, which gave $\alpha =$~2.07$\pm$0.03 at NIR. 

We note that other works usually suggested different values of $\alpha$, from 1.61 \citep{rieke1985}, 1.99 \citep{nishiyama2006}, 2.11 \citep{fritz2011}, up to 2.47 \citep{alonso2017}. Apparently, the variation of NIR extinction law can also cause uncertainties of the extinction estimation. A direct idea is to use the standard deviation of $\alpha$ values from different works to estimate the variation of the NIR extinction law. However, because the different $\alpha$ values were obtained based on different datasets and methods, we can not use them to quantify the variation of NIR extinction law without the cross-calibration between them \citep{wangjiang2014,extlaw2020}. In fact, whether the NIR extinction law is universal is still under debate.

\citet{wangjiang2014} determined accurately the intrinsic colors of $\sim$6000 giants based on the APOGEE spectroscopic survey that covered the area of 0\degr$<l<$~220\degr. They revisited the NIR extinction law and obtained an $\alpha$ value of about 1.95$\pm$0.02. \citet{wangjiang2014} also found that the NIR extinction law showed no dependence on the color excess of $E(J-K_s)$ in the range of 0.3$<E(J-K_s)<$~4$-$5, which indicated an universal NIR extinction law from diffuse to dense (up to $A_V\sim$~20-30\,mag) regions. However, \citet{wangjiang2014} could underestimate the uncertainty of $\alpha$ due to their small sample size, especially for the dense regions. \citet{extlaw2020} also analyzed the NIR extinction with the all-sky 2MASS stellar objects with high-quality photometry. They tried to select a clean red giant sample with the help of Gaia DR2 data and finally reached a NIR extinction of $\alpha \sim$~2.27 with the uncertainty of a few hundredths of a magnitude. \citet{extlaw2020} found no significant spatial variation of NIR extinction law towards different sightlines. Therefore, above large-scale investigations suggested an universal NIR extinction law in most regions with typical environments in the Galactic plane although they gave different $\alpha$ values. On the other hand, it is possible that there are different NIR extinction laws in some specific environments such as the molecular clouds associated with star formation. For example, \citet{racca2002} obtained a NIR extinction law with $\alpha=$~2.52 in the Coalsack globule 2 that is the highest-density region in the Coalsack molecular cloud. \citet{wang2013} obtained the different $\alpha$ values for the diffuse ($\alpha=$~1.73) and dense ($\alpha>$~2) environments in the Coalsack nebula. \citet{meingast2018} investigated the NIR extinction curve in Orion A and found a significant evidence that the extinction law varied $\sim$3\% across the cloud. They suggested that this variation could be due to the influence of massive stars on their ambient medium.

Therefore, our assumption of an universal NIR extinction law should be valid in most VVV survey area. Based on the value of $\alpha=$~2.07$\pm$0.03 that we adopted, the uncertainties of visual extinctions due to the error of $\alpha$ were within 3\%. From diffuse to dense environments, the variation of NIR extinction law can be from $<$2\% \citep{wangjiang2014} to up to $\sim$30\% \citep{wang2013}, which was still not well quantified. However, the dense regions in our extinction maps tend to suffer the problem due to the contamination of foreground sources (see Sect~\ref{sect:absorption-like_features}), which should be the main source of uncertainties of extinctions. Of course, adopting other NIR extinction laws rather than the one with $\alpha=$~2.07 results in the systematic under- or over-estimation of the extinction. The systematic error due to different $\alpha$ values from different works can be up to 20-30\% \citep{wangchen2019}.

\section{DISCUSSION}\label{sect:discussion}

\subsection{Comparisons with previous dust based maps}\label{sect:comparison}

\mz{In this section, we compare our XPNICER extinction maps with \mzrevi{ten} existing dust-based maps, including the extinction maps derived by \citet{dobashi2011} and \citet{juvela16}, the widely-used SFD map \citep{sfd1998}, the Planck dust map \citep{planck-dust-2014}, the Herschel PPMAP \citep{ppmap2017}, \mzrevi{the reddening map of the Galactic bulge presented by \citet{beam2012,beam2018}, the Rayleigh-Jeans Color Excess (RJCE) extinction maps obtained by \citet{rjce2012} and \citet{rjce2019}, and the reddening map of the Galactic bulge derived by \citet{surot2020}.}}  

\mz{For brevity, we refer to our XPNICER extinction map as $A_V$(XPNICER) and to the other dust-based maps by \citet{dobashi2011}, \citet{juvela16}, \citet{sfd1998}, \citet{planck-dust-2014}, \citet{ppmap2017}, \mzrevi{\citet{beam2012,beam2018}, \citet{rjce2012}, \citet{rjce2019}, \citet{schultheis2014}, and \citet{surot2020} as $A_V$(Dobashi), $A_V$(M2b), $A_V$(SFD), $\tau_{\mathrm{353}}$(Planck), $\tau_{\mathrm{353}}$(PPMAP), $A_V$(BEAM), $A_V$(RJCE12), $A_V$(RJCE19), $A_V$(Schultheis), and $A_V$(Surot), respectively.} Table~\ref{tab:tab2} summarizes the basic information of the \mzrevi{ten} maps. \mztwo{We note that the maps have been derived with different techniques and datasets \mzrevi{and thus have different units}. To better understand the differences between them and our XPNICER map, we summarize the different mapping processes in Appendix~\ref{ap:summaryext} to the degree relevant to perform meaningful comparisons. The detailed descriptions can be found in the respective papers.} \mzrevi{We also converted the units of all previous dust extinction maps to the visual extinction ($A_V$). The details of the conversion process can be found in the following individual comparison sections (Sect.~\ref{sect:comparison2dobashi}-\ref{sect:compare2surot}).} }

\jk{We first present visual comparisons of the $A_V$(XPNICER) and \mzrevi{ten} other maps in Sect.~\ref{sect:visualcomp}. Then, the sections~\ref{sect:comparison2dobashi}-\ref{sect:compare2surot} describe the quantitative pixel-to-pixel comparisons, both for the whole $A_V$(XPNICER) coverage area and a few selected sub-regions.}

\begin{table*}
    \centering
    \caption{Basic information of the \mzrevi{ten} dust-based maps compared with our data.}
    \label{tab:tab2}
    \begin{tabular}{llccl}
    \hline\hline
       Maps  & Type & Resolution & Units & References \\
        \hline
        $A_V$(Dobashi) & dust extinction & 1\arcmin-12\arcmin& $A_V$& \citet{dobashi2011}\\
        $A_V$(M2b) & dust extinction & 3\arcmin & $A_J$ & \citet{juvela16}\\
        $A_V$(SFD) & dust emission & 6\arcmin& $E(B-V)$ &\citet{sfd1998}\\
        $\tau_{\mathrm{353}}$(Planck) & dust emission&5\arcmin &optical depth & \citet{planck-dust-2014}\\
        $\tau_{\mathrm{353}}$(PPMAP)& dust emission & 12\arcsec & optical depth & \citet{ppmap2017}\\
        $A_V$(BEAM) & dust extinction & 1\arcmin-6\arcmin&$E(J-K_s)$&\citet{beam2012,beam2018}\\
        $A_V$(RJCE12) & dust extinction & 2\arcmin &$A_{K_s}$&\citet{rjce2012}\\
        $A_V$(RJCE19) & dust extinction & 1\arcmin&$A_{K_s}$&\citet{rjce2019}\\
        $A_V$(Schultheis) & dust extinction & 6\arcmin & $E(J-K_s)$ & \citet{schultheis2014}\\
        $A_V$(Surot) & dust extinction & $\sim$10\arcsec-2\arcmin& $E(J-K_s)$ & \citet{surot2020}\\
        \hline
    \end{tabular}
\end{table*}

\subsubsection{Visual comparison}\label{sect:visualcomp}

\begin{figure*}
    \centering
    \includegraphics[width=1.0\linewidth]{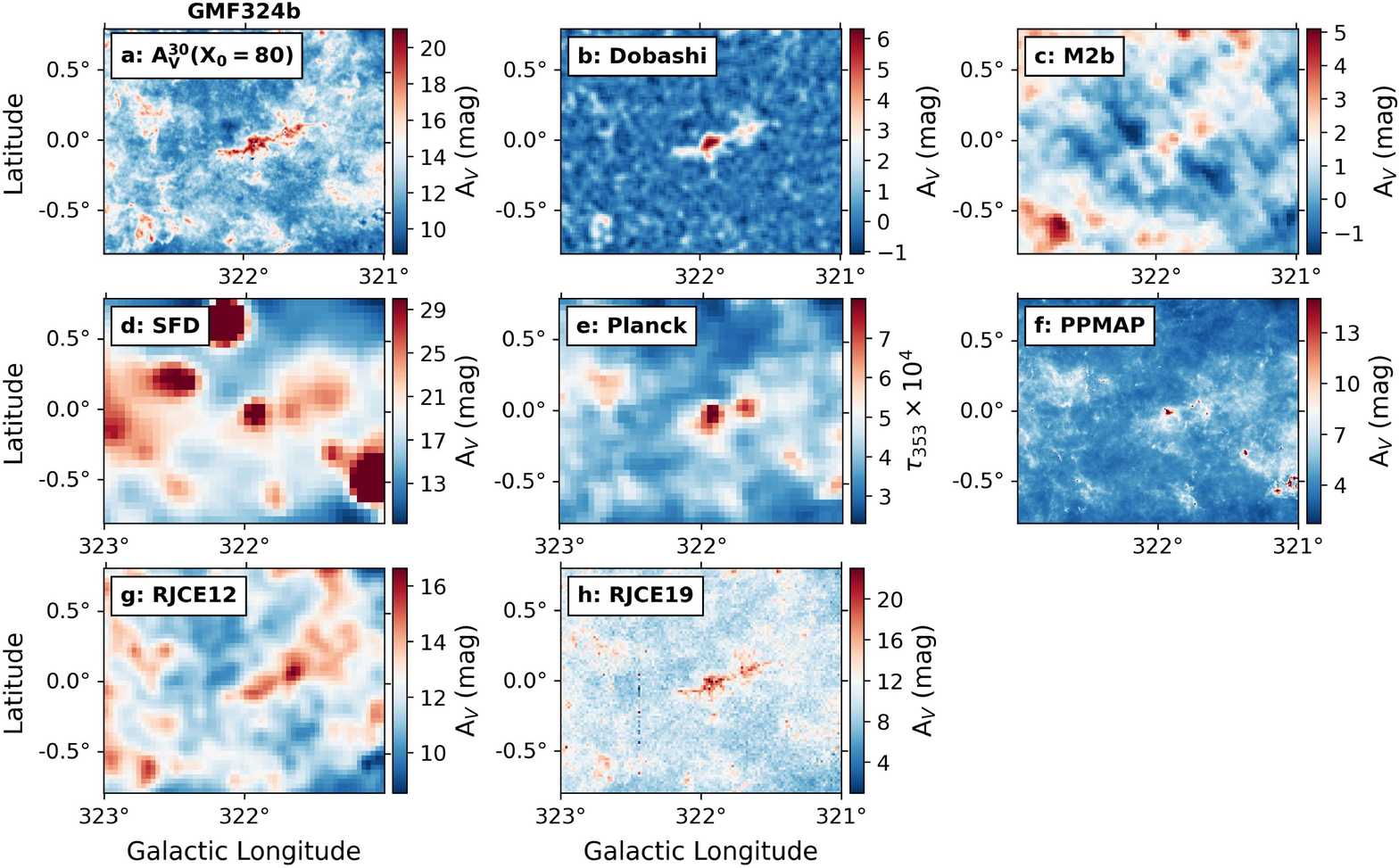}
    \caption{Zoom-in view of a region with the giant filament GMF 324b of (a): our $A_V^{30}(X_0=80)$ map; (b): extinction map by \citet{dobashi2011}; (c): extinction map by \citet{juvela16}; (d): SFD map \citep{sfd1998}; (e): Planck dust map \citep{planck-dust-2014} in units of dust optical depth at 353\,GHz; (f): Herschel column density map obtained with PPMAP technique \citep{ppmap2017}\mzrevi{; (g): extinction map constructed with the RJCE method \citep{rjce2011} by \citet{rjce2012}; and (h): extinction map obtained with the RJCE method by \citet{rjce2019}.} The PPMAP has been converted to the $A_V$ units with the relation of $N(H_2)=0.94\times10^{21}A_V$ \citep{bohlin1978}.}
    \label{fig:extmap_zoomin}
\end{figure*}

\mz{Figure~\ref{fig:extmap_zoomin} shows the zoom-in view of \mzrevi{a region with the giant filament GMF 324.5-321.4b \citep[GMF 324b,][]{abreu2016}} in $A_V$(XPNICER), $A_V$(Dobashi), $A_V$(M2b), $A_V$(SFD), $\tau_{\mathrm{353}}$(Planck), $\tau_{\mathrm{353}}$(PPMAP), \mzrevi{$A_V$(RJCE12), and $A_V$(RJCE19),} respectively. We also selected six other regions for visual comparisons, shown in Appendix~\ref{app:vcom}. \mztwo{These selected regions are located at different Galactic longitudes and latitudes, representing different conditions in the Galactic plane.}} 

\mz{Compared to $A_V$(Dobashi), $A_V$(M2b), $A_V$(SFD),  $\tau_{\mathrm{353}}$(Planck), \mzrevi{$A_V$(BEAM), $A_V$(RJCE12), $A_V$(RJCE19), and $A_V$(Schultheis)}, our $A_V$(XPNICER) has higher spatial resolution and it can trace finer structures. \mzrevi{We note that $A_V$(Surot) reaches the significant higher resolution ($<$10\arcsec) than our $A_V$(XPNICER) map in the area close to the Galactic midplane ($b=$~0\degr), but $A_V$(Surot) only covers the Galactic bulge area.}}
\mz{When comparing $A_V$(XPNICER) with $\tau_{\mathrm{353}}$(PPMAP), we can see two main differences. First, $\tau_{\mathrm{353}}$(PPMAP) shows more small-scale structures due to its higher resolution (12\arcsec). Second, the very dense regions in $\tau_{\mathrm{353}}$(PPMAP) commonly correspond to absorption-like features in our $A_V$(XPNICER) that are artifacts due to the contamination of foreground sources (see Appendix~\ref{sect:absorption-like_features}). This is due to the limitations of our XPNICER method that runs out of background stars at high column densities. The dust emission based $\tau_{\mathrm{353}}$(PPMAP) can better trace the total column density, \mzrevi{reaching regions with higher densities}.} \jk{Overall, \mzrevi{considering the spatial resolution and coverage}, the comparison of the dust extinction based maps indicates that our $A_V$(XPNICER) map provides \mzrevi{a high-fidelity} extinction map of the Galactic plane.}

\mztwo{Another obvious difference is the systematic offset between maps. The $A_V$ values of $A_V$(XPNICER) are usually \mzrevi{significantly} higher than that of $A_V$(Dobashi), $A_V$(M2b), and $\tau_{\mathrm{353}}$(PPMAP), but lower than that of $A_V$(SFD). \mzrevi{When comparing with $A_V$(BEAM), $A_V$(RJCE12), $A_V$(RJCE19), $A_V$(Schultheis), and $A_V$(Surot), $A_V$(XPNICER) usually has slightly higher $A_V$ values.} The offsets result from the different dust tracers and mapping techniques. In sections~\ref{sect:comparison2dobashi}-\ref{sect:compare2surot} we will discuss these offsets quantitatively.}



\subsubsection{Detailed comparisons with NIR extinction maps by \citet{dobashi2011} and \citet{juvela16}}\label{sect:comparison2dobashi}


\jk{Both $A_V$(Dobashi) and $A_V$(M2b) significantly underestimate, by construction, the extinction from diffuse dust (see Appendix~\ref{ap:summaryext}). In contrast, our XPNICER map measures the integrated extinction until some limiting depth (see Section \ref{sect:extmap})}. \mztwo{The offsets between $A_V$(XPNICER), $A_V$(Dobashi), and $A_V$(M2b), shown in Fig.~\ref{fig:extmap_zoomin}, should mainly reflect the extinction from diffuse dust.} 

\jk{To better understand the difference between our XPNICER map and $A_V$(Dobashi) and $A_V$(M2b), we need to understand the effect of the mapping techniques on the diffuse dust component. To study this, we make a simple model for the diffuse dust component and subtract it from the XPNICER map}. Of course, the precise modelling needs the full knowledge of 3D dust distribution in the Galactic plane, which is beyond the scope of this paper. Here we simply assumed that the diffuse dust distribution causes a smooth large-scale background in $A_V$(XPNICER) and used two different methods (see Appendix~\ref{ap:backgroundest}) to estimate this background, i.e., $A_V$(BG1) and $A_V$(BG2).

\begin{figure*}
    \centering
    \includegraphics[width=1.0\linewidth]{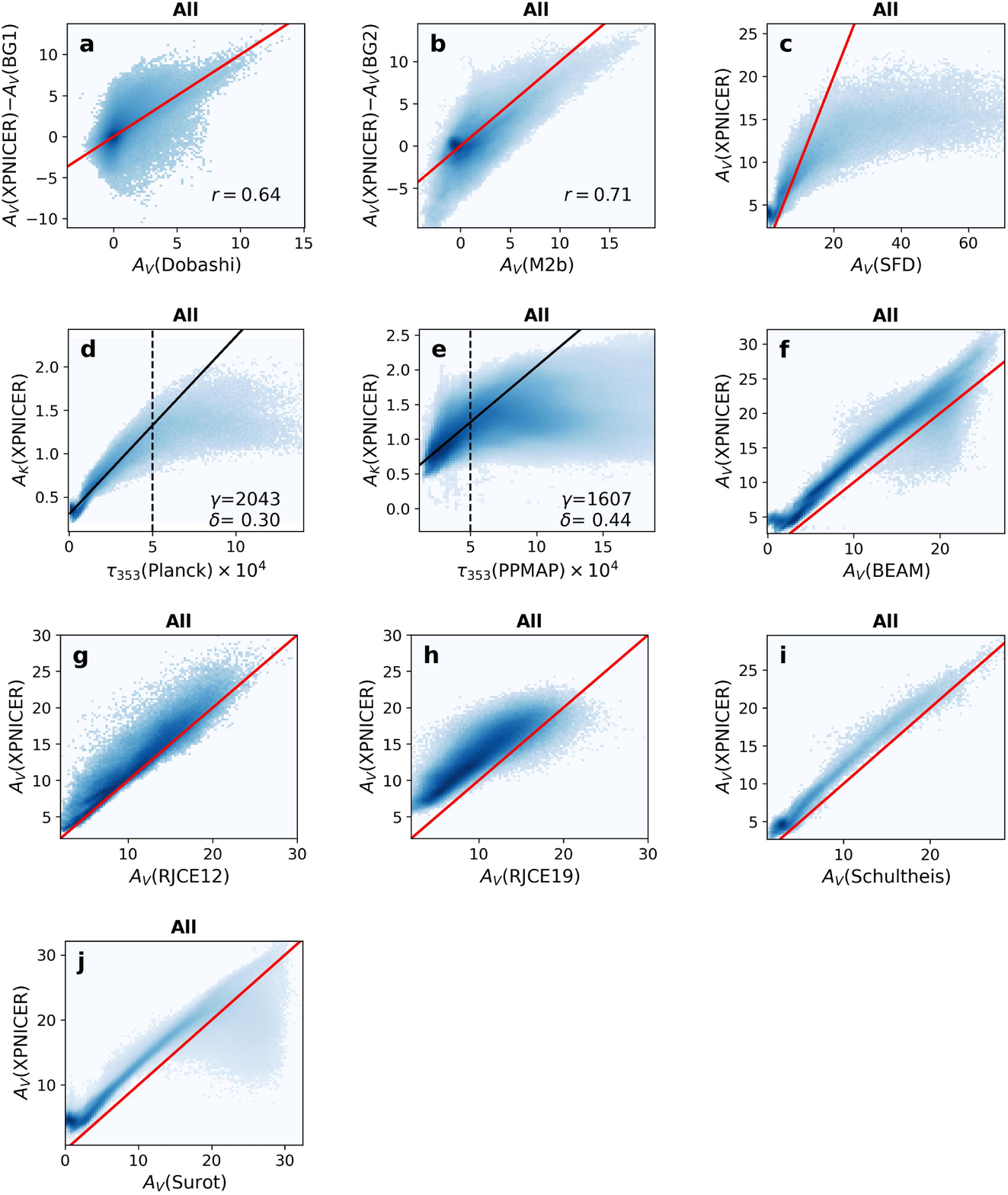}
    \caption{\mzrevi{Pixel-to-pixel comparisons of our XPNICER extinction map with (a): extinction map by \citet{dobashi2011}; (b): extinction map by \citet{juvela16}; (c): SFD dust map \citep{sfd1998}; (d): Planck dust map \citep{planck-dust-2014}; (e): Herschel PPMAP \citep{ppmap2017}; (f): BEAM extinction map by \citet{beam2012,beam2018}; (g): RJCE extinction map by \citep{rjce2012}; (h): RJCE extinction map by \citet{rjce2019}; (i): 2D extinction map integrated from the 3D dust map by \citet{schultheis2014}; and (j): extinction map by \citet{surot2020}. The comparison is for the whole map. The red lines mark the one-to-one relation. The black solid lines show the linear fits with the slope of $\gamma$ and the intercept of $\delta$. In the panels d and e, we limited the fit to $\tau_{353}<$~$\tau_{\mathrm{cut}}\times$10$^{-4}$ that were marked with the vertical black dashed lines. The Pearson correlation coefficient ($r$) or the fitting parameters are also labeled on the corresponding panels. }}
    \label{fig:compare2pre}
\end{figure*}

\begin{figure*}
    \centering
    \includegraphics[width=1.0\linewidth]{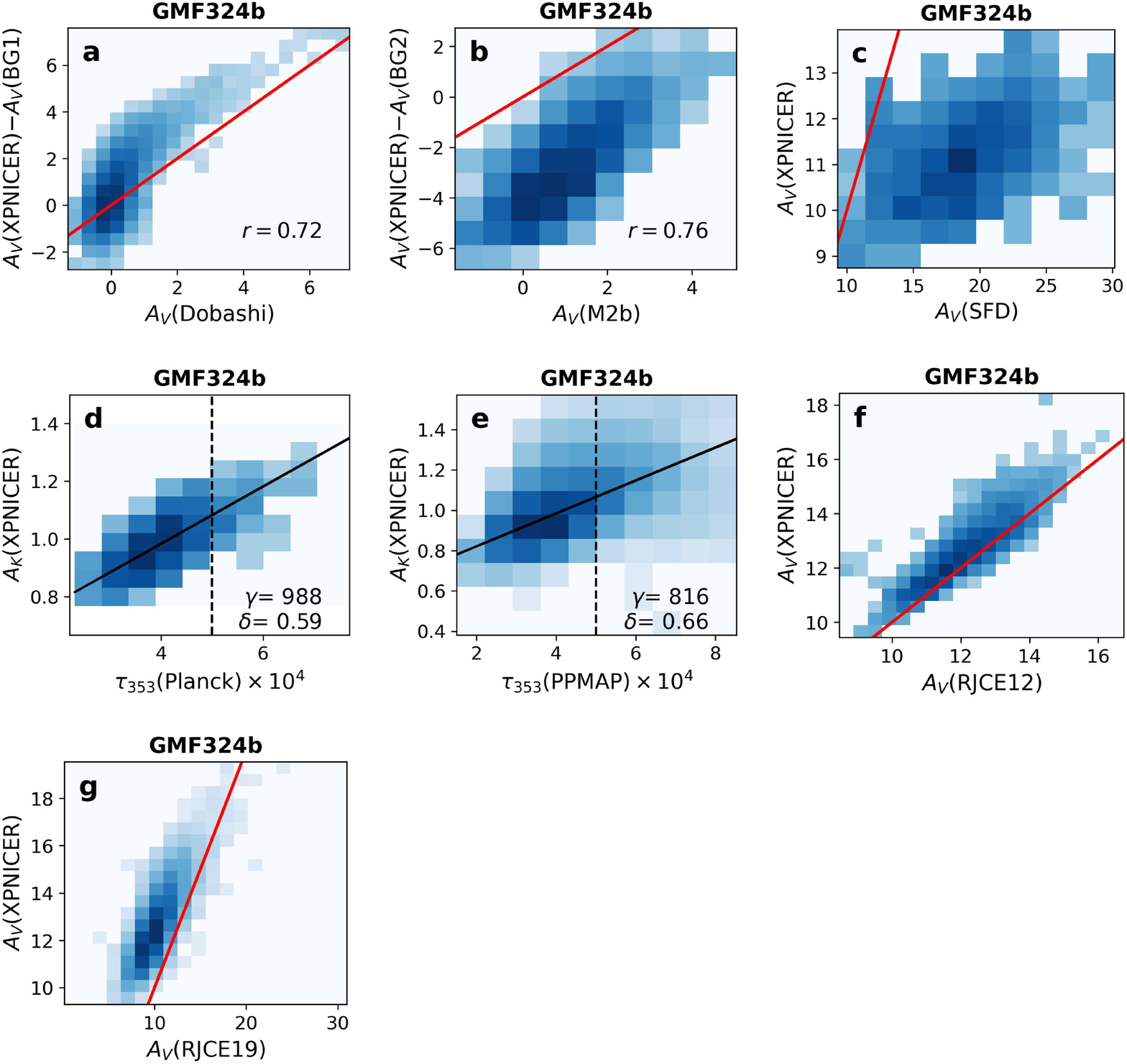}
    \caption{\mzrevi{Pixel-to-pixel comparisons of our XPNICER extinction map with (a): extinction map by \citet{dobashi2011}; (b): extinction map by \citet{juvela16}; (c): SFD dust map \citep{sfd1998}; (d): Planck dust map \citep{planck-dust-2014}; (e): Herschel PPMAP \citep{ppmap2017}; (f): RJCE extinction map by \citep{rjce2012}; and (g): RJCE extinction map by \citet{rjce2019}. The comparison is for GMF 324b. The red lines mark the one-to-one relation. The black solid lines show the linear fits with the slope of $\gamma$ and the intercept of $\delta$. In the panels d and e, we limited the fit to $\tau_{353}<$~$\tau_{\mathrm{cut}}\times$10$^{-4}$ that were marked with the vertical black dashed lines. The Pearson correlation coefficient ($r$) or the fitting parameters are also labeled on the corresponding panels.}}
    \label{fig:compare2pre-gmf324b}
\end{figure*}

\begin{figure*}
    \centering
    \includegraphics[width=1.0\linewidth]{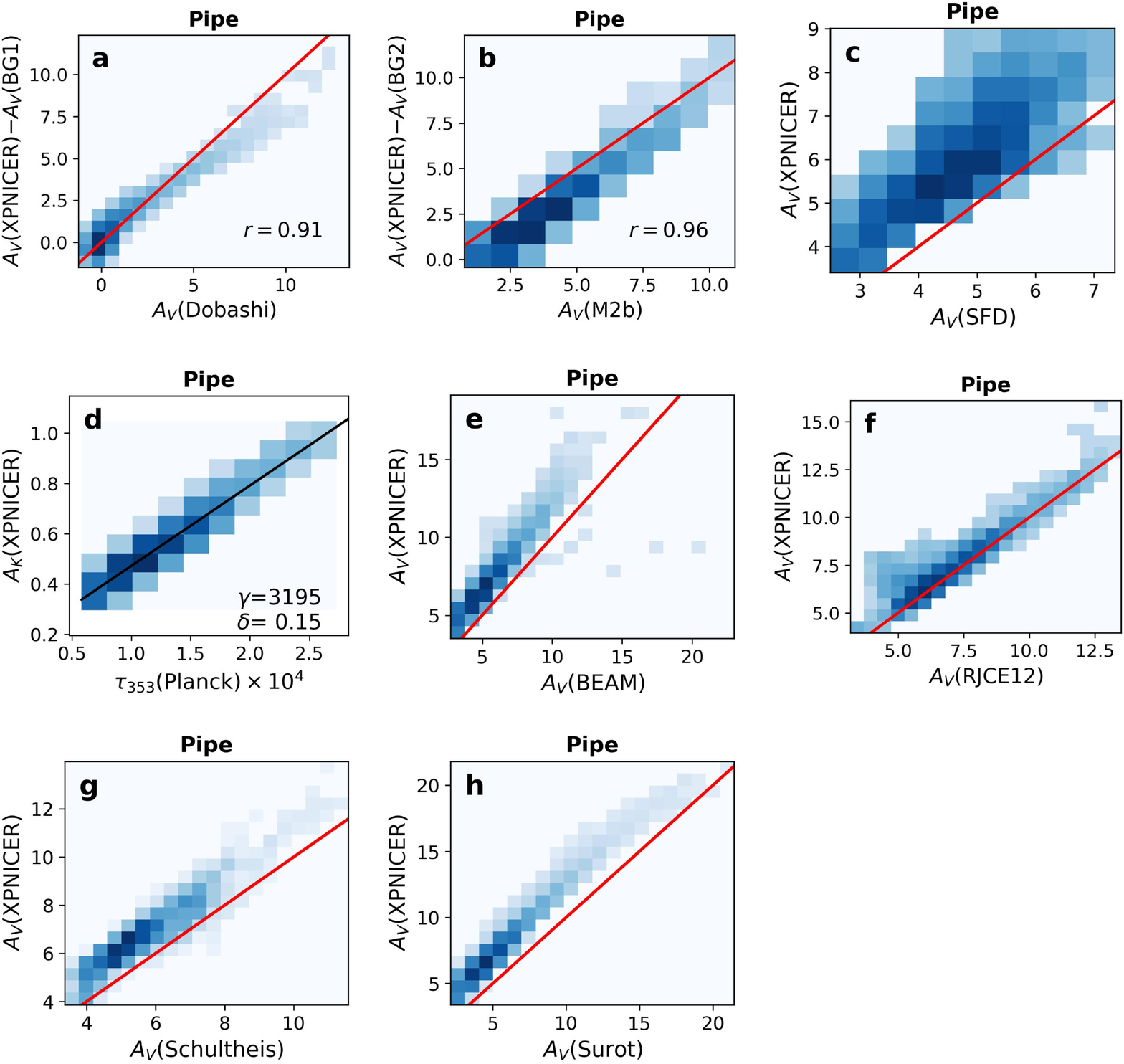}
    \caption{\mzrevi{Pixel-to-pixel comparisons of our XPNICER extinction map with (a): extinction map by \citet{dobashi2011}; (b): extinction map by \citet{juvela16}; (c): SFD dust map \citep{sfd1998}; (d): Planck dust map \citep{planck-dust-2014}; (e): BEAM extinction map by \citet{beam2012,beam2018}; (f): RJCE extinction map by \citep{rjce2012}; (g): 2D extinction map integrated from the 3D dust map by \citet{schultheis2014}; and (h): extinction map by \citet{surot2020}. The comparison is for the Pipe molecular cloud. The red lines mark the one-to-one relation. The black solid lines show the linear fits with the slope of $\gamma$ and the intercept of $\delta$. The Pearson correlation coefficient ($r$) or the fitting parameters are also labeled on the corresponding panels.}}
    \label{fig:compare2pre-pipe}
\end{figure*}

Figure~\ref{fig:compare2pre}a shows the comparisons of $A_V$(XPNICER)$-A_V$(BG1) with $A_V$(Dobashi) for the whole map. We calculated the Pearson correlation coefficient ($r$) between them and obtained a value of 0.64, which indicated a weak correlation. The median and \mzrevi{standard} deviation of $A_V$(XPNICER)$-A_V$(BG1)$-A_V$(Dobashi) are 0.05 and 0.96 magnitude, \mz{respectively}. \mztwo{\mzrevi{Figures~\ref{fig:compare2pre-gmf324b}a, and \ref{fig:compare2pre-pipe}a} show the similar comparisons for two sub-regions, GMF 324b and Pipe (see Fig.~\ref{fig:reg_pipe}). Their $r$ values also indicate correlations. However, there are systematic offsets between $A_V$(XPNICER)$-A_V$(BG1) and $A_V$(Dobashi) in sub-regions (also see \mzrevi{Figs.~\ref{fig:ap1com}-\ref{fig:ap5com}}).} 

Figure~\ref{fig:compare2pre}b shows the relation between $A_V$(XPNICER)$-A_V$(BG2) and $A_V$(M2b) for the whole $A_V$(XPNICER) map. We can see a correlation between them ($r=$~0.71). The median and standard deviation of $A_V$(XPNICER)$-A_V$(BG2)$-A_V$(M2b) are -0.58 and 1.63 magnitude, respectively. \mztwo{Similar correlations can be also found in the sub-regions and the systematic offsets are also significant (see \mzrevi{Figs.~\ref{fig:compare2pre-gmf324b}b, \ref{fig:compare2pre-pipe}b, and Figs.~\ref{fig:ap1com}-\ref{fig:ap5com})}.}  

\mztwo{Overall, our XPNICER map is roughly in agreement with the extinction maps presented by \citet{dobashi2011} and \citet{juvela16}, if a smooth background is subtracted from our XPNICER map. However, there are systematic offsets at different scales that could be due to the oversimplified diffuse dust model we used.}

\subsubsection{Detailed comparison with SFD map}
\label{sect:compare2sfd}

The original SFD map has been calibrated to the color excess units (see Appendix~\ref{ap:summaryext}). We retrieved the original SFD map with the python interface of DUSTMAPS \citep{dustmaps} and converted it to $A_V$ units using $A_V=$~2.742$E(B-V)$. 
We note that the extinction law used to calibrate SFD map \citep{schlafly2011} was close to the law suggested by \citet{fitzpatrick1999} and different from the one that we used \citep{wangchen2019}. To decrease the systematic difference due to extinction laws, we scaled $A_V$(XPNICER) by a factor of $\frac{2.742}{3.1}$.

Figures~\ref{fig:compare2pre}\mzrevi{c, \ref{fig:compare2pre-gmf324b}c, and \ref{fig:compare2pre-pipe}c} show the pixel-to-pixel comparisons between $A_V$(XPNICER) and $A_V$(SFD) \mztwo{for the whole map and sub-regions of GMF 324b and Pipe, respectively (also see \mzrevi{Figs.~\ref{fig:ap1com}-\ref{fig:ap5com}} for the comparisons in other sub-regions)}. We can immediately get two main results: \mz{first}, there is an approximate one-to-one relation between $A_V$(XPNICER) and $A_V$(SFD) in the range of \mztwo{$A_V\lesssim$~10$-$20\,mag (depending on different sub-regions)}, but with a systematic difference of $A_V$(XPNICER)$-$$A_V$(SFD)$\sim$1\,mag. The systematic difference could result from the adopted different extinction laws, uncertainty of zero point of our extinction map (see Sect.~\ref{sect:zmag}), and \mztwo{the uncertainties of the SFD map (see Appendix~\ref{ap:summaryext})}; \mz{second}, $A_V$(SFD) is significantly higher than $A_V$(XPNICER) when \mztwo{$A_V\gtrsim$~10$-$20\,mag}. In Sect.~\ref{sect:extmap} we discussed that $A_V$(XPNICER) traces the total dust extinction integrated to the distance of $d_{\textrm{limit}}$ and that $d_{\textrm{limit}}$ could be only a few kpc towards the regions with dense molecular clouds (also see Appendix~\ref{sect:absorption-like_features}). However, it was commonly believed that $A_V$(SFD) trace the total dust column density. Therefore, the discrepancy between $A_V$(XPNICER) and $A_V$(SFD) at \mztwo{$A_V\gtrsim$~10$-$20\,mag} should reflect the limitation of \mz{our XPNICER} extinction mapping technique. 

Overall, our XPNICER extinction map is roughly consistent with the SFD map in the range of \mztwo{$A_V\lesssim$~10$-$20\,mag}, but there is significant discrepancy when \mztwo{$A_V\gtrsim$~10$-$20\,mag}. \mzrevi{The discrepancy could result from, on the one hand, the large uncertainties of SFD map at low Galactic latitude \citep[see Appendix~\ref{ap:summaryext}, also see][]{sfd1998,beam2012}. On the other hand, }considering that our XPNICER map can not trace the total dust extinction along some lines of sights towards the dense molecular clouds, the discrepancy \mzrevi{also} reflects the limitation of our XPNICER extinction mapping technique.


\subsubsection{Detailed comparison with Planck dust map}\label{sect:compare2planck}

We downloaded the $\tau_{353}$(Planck) with the Python interface of DUSTMAPS \citep{dustmaps}. 
For convience, we also converted $A_V$(XPNICER) back to the $A_K$ units, i.e., $A_K$(XPNICER), with the reddening law suggested by \citet{wangchen2019}.
Figure~\ref{fig:compare2pre}d shows the pixel-to-pixel comparison between $A_K$(XPNICER) and $\tau_{353}$(Planck) for the whole $A_K$(XPNICER) map. There is roughly a linear relation in the range of $\tau_{353}\mathrm{(Planck)}\lesssim$~5$\times$10$^{-4}$, but significant discrepancy when $\tau_{353}\mathrm{(Planck)}\gtrsim$~5$\times$10$^{-4}$. \mztwo{When we focus on the sub-regions (\mzrevi{Figs.~\ref{fig:compare2pre-gmf324b}d, \ref{fig:compare2pre-pipe}d and Figs.~\ref{fig:ap1com}-\ref{fig:ap5com}}), the approximating linear relations can retain in the range of $\tau_{353}\mathrm{(Planck)}\lesssim$~(4$-$7)$\times$10$^{-4}$}. 
As mentioned in Sect.~\ref{sect:compare2sfd}, the discrepancy between $A_K$(XPNICER) and $\tau_{353}$(Planck) in the dense regions could result from the bias of $A_K$(XPNICER) due to the limitation of our extinction mapping technique. 

We also fitted $A_K$(XPNICER) versus $\tau_{353}$(Planck) with a linear relation:
\begin{align}\label{eq:tau2ak}
    A_K\mathrm{(XPNICER)} = \gamma \tau_{353}\mathrm{(Planck)} + \delta.
\end{align}
The fitting was limited in the range of $\tau_{353}$(Planck)~$<$~$\tau_{\mathrm{cut}}\times$10$^{-4}$, \mztwo{where $\tau_{\mathrm{cut}}$ has different values of 4$-$7 in different sub-regions. The returned $\gamma$ has values of ~1000$-$3000 while $\delta$ has values of ~0.1$-$0.6.} 

The intercept 
\mztwo{$\delta\sim$~0.1$-$0.6} indicates that $A_K$(XPNICER) could systematically overestimate the $A_K$ values by about 0.1$-$0.6 in the low extinction regions, corresponding to a \mztwo{$A_V$ value of $\sim$1$-$7\,mag}. \mztwo{This systematic offset can be explained by the zero point uncertainty of our extinction map (see Sect.~\ref{sect:zmag}) and the uncertainty of the Planck dust maps.} However, the slope of the linear fit, $\gamma$, was proportional to the ratio of dust opacity and extinction coefficient as suggested by \citet{lombardi2014}:
\begin{align}\label{eq:gammafit}
    \gamma \approx 1.0857 \frac{C_{2.2}}{\kappa_{850}} = 1.0857\frac{C_{2.2}}{\kappa_{250}}\left( \frac{850\,\mu m}{250\,\mu m} \right)^{\beta},
\end{align}
where $C_{2.2}$ was the extinction coefficient at 2.2\,$\mu$m, $\kappa_{850}$ and $\kappa_{250}$ were the dust opacity at 850 and 250\,$\mu$m, respectively. 
$C_{2.2}$ and $\kappa_{850}$ were dependant on the dust properties such as composition and grain size distribution \citep{ossenkopf1994}.
Thus different values of $\gamma$ could be related to the variation of dust properties, which has been observed in many different regions. For example, \citet{kramer2003} measured variation of opacity ratios towards different prestellar cores, corresponding $\gamma$ value from $\sim$\mztwo{1000} to 5700. \citet{lombardi2014} obtained $\gamma=$~2640 in Orion A and $\gamma=$~3460 in Orion B while \citet{meingast2018} obtained a slightly larger $\gamma$ value of 3042 in Orion A. \citet{zari2016} also obtained $\gamma=$~3931 in Perseus. \mztwo{The $\gamma$ values of $\sim$1000$-$3000 obtained by us are consistent with the previous studies.}

\jk{A linear fit for the entire VVV survey area yielded an average value of $\gamma=$~2043.The median $\beta$ value for the area is $\sim$1.68, }based on the Planck $\beta_{\mathrm{obs}}$ map (see Appendix~\ref{ap:summaryext}). If adopting $\frac{C_{2.2}}{\kappa_{250}}\approx$~254.7 from \citet{mathis1990}, we can calculate a $\gamma$ value of $\sim$2160 with Eq.~\ref{eq:gammafit}. We note that our fitting value of $\gamma=$~2043 is very close to this value of 2160, which indicates that our XPNICER map is globally consistent with the Planck dust model \citep{planck-dust-2014}.


\subsubsection{Detailed comparison with Herschel PPMAP}\label{sect:compare2ppmap}

We downloaded the Herschel PPMAPs\footnote{\url{https://www.astro.cf.ac.uk/research/ViaLactea/}}, but here only used the integrated column density maps. 
We also note that \citet{ppmap2017} assumed $\kappa(\lambda)=0.1\,\mathrm{cm}^{2}\,\mathrm{g}^{-1}(\frac{\lambda}{ 300\,\mu m})^{-\beta}$ and $\beta =$~2 to convert the dust optical depth to the column density. To decrease the uncertainty introduced by the former conversion, we transformed the Herschel PPMAPs from column densiy units back to the optical depth and obtained $\tau_{353}$(PPMAP).


\mztwo{To compare $A_K$(XPNICER) to $\tau_{353}$(PPMAP), we firstly calibrate $\tau_{353}$(PPMAP) with $\tau_{353}$(Planck) that was obtained in Sect.~\ref{sect:compare2planck} using a simple method described in Appendix~\ref{ap:calipp}. \mzrevi{Figures~\ref{fig:compare2pre}e and \ref{fig:compare2pre-gmf324b}e} show the pixel-to-pixel comparison for the whole map and the region with GMF 324b (see \mzrevi{Figs.~\ref{fig:ap1com}-\ref{fig:ap5com}} for the comparison in five other sub-regions). The behavior of $A_K$(XPNICER) versus $\tau_{353}$(PPMAP) is similar to that of $A_K$(XPNICER) versus $\tau_{353}$(Planck), but with larger scatter. We also fitted the data with $\tau_{353}$(PPMAP)~$<$~$\tau_{\mathrm{cut}}\times 10^{-4}$ with a linear function as describe in Eq.~\ref{eq:tau2ak} and obtained similar $\gamma$ and $\delta$. Thus our XPNICER map is also roughly consistent with the Herschel PPMAPs in the low extinction area. However, due to the oversimplified calibration method (see Appendix~\ref{ap:calipp}), we did not discuss the pixel-to-pixel comparison between $A_K$(XPNICER) and $\tau_{353}$(PPMAP) in detail.}


\subsubsection{\mzrevi{Detailed comparison with reddening map by \citet{beam2012,beam2018}}}\label{sect:compare2beam}
\mzrevi{We downloaded the reddening map presented by \citet{beam2012,beam2018} through the Bulge Extinction and Metallicity (BEAM) Calculator\footnote{\url{https://www.oagonzalez.net/beam-calculator}}. The $E(J-K_s)$ map obtained by \citet{beam2012} and \citet{beam2018} were constructed based on the aperture photometric catalog \citep{vvvdr1} and the DoPHOT catalog \citep{vvvdophot2018}, respectively. As mentioned in Sect.~\ref{sect:vvv-daophot-catalog}, these catalogs have bias in the photometric zero-points. Thus we firstly corrected the $E(J-K_s)$ using our correction maps shown in Fig.~\ref{fig:zmag-wholecoverage} and then converted the color excess units to $A_V$ with the extinction law suggested by \citet{wangchen2019}.}

\mzrevi{Figures~\ref{fig:compare2pre}f and \ref{fig:compare2pre-pipe}e show the pixel-to-pixel comparisons between $A_V$(XPNICER) and $A_V$(BEAM) for the whole map and sub-region of Pipe, respectively (also see Fig.~\ref{fig:ap5com} for the comparison in another sub-region). We can see an approximate one-to-one relation between $A_V$(XPNICER) and $A_V$(BEAM), but with a systematic difference. The median value of $A_V$(XPNICER)$-A_V$(BEAM) is about 2.8 mag for the whole map while about 1.6-2.5 mag for the sub-regions. This $\sim$1-3 mag systematic offset could result from the uncertainty of zero point of $A_V$(XPNICER) and the uncertainties of $A_V$(BEAM). We also note that $A_V$(XPNICER)  values are significantly lower than $A_V$(BEAM) in some high extinction regions. This discrepancy could result from different stellar populations that were used to construct the maps. Our XPNICER map is produced with all stars and underestimate the $A_V$ values of many high density regions due to the contamination of foreground sources (e.g., main sequence (MS) stars, see Appendix.~\ref{sect:absorption-like_features}). However, $A_V$(BEAM) was constructed with the red clump (RC) giants, which can naturally eliminate the contamination of foreground MS stars and thus trace the high extinction regions more reasonably in the bulge area.}

\mzrevi{Overall, considering that $A_V$(XPNICER) overestimates the $A_V$ values of $\sim$1$\pm$1.4 mag (see Sect.~\ref{sect:zmag}), our XPNICER map is consistent with $A_V$(BEAM) in the majority of the Galactic bulge area.}

\subsubsection{\mzrevi{Detailed comparisons with RJCE maps by \citet{rjce2012} and \citet{rjce2019}}}\label{sect:compare2rjce}

\mzrevi{We downloaded the $A_{K_s}$(RJCE12)\footnote{\url{ http://www.astro.virginia.edu/rjce/}} and $A_{K_s}$(RJCE19)\footnote{\url{http://www.astro.uda.cl/msoto/extinction/ab.php}} that were presented by \citet{rjce2012} and \citet{rjce2019}. The RJCE method is based on the extinction laws suggested by \citet{ind2005} and \citet{rjce2009}. To match the reddening law that we used \citep{wangchen2019}, the RJCE maps were scaled by a factor of $\frac{0.743}{0.918}$ and then converted to the visual extinction units.}

\mzrevi{Figures~\ref{fig:compare2pre}g, \ref{fig:compare2pre-gmf324b}f, and \ref{fig:compare2pre-pipe}f show the pixel-to-pixel comparisons between $A_V$(XPNICER) and $A_V$(RJCE12) for the whole map and two sub-regions with GMF 324b and Pipe, respectively (also see Figs.~\ref{fig:ap1com}-\ref{fig:ap5com}  for the comparisons in other sub-regions). There is a good correlation between $A_V$(XPNICER) and $A_V$(RJCE12), but with a systematic offset of $A_V$(XPNICER)$-A_V$(RJCE12)$\sim$0.3$-$2 mag at different scales. This systematic difference can be explained with the uncertainty of zero-point of our XPNICER map (see Sect.~\ref{sect:zmag}). Therefore, our XPNICER map agrees well with the RJCE map by \citet{rjce2012}.}

\mzrevi{Figures~\ref{fig:compare2pre}h shows the comparison between $A_V$(XPNICER) and $A_V$(RJCE19) for the whole map. We can also see a correlation. However, the median value of $A_V$(XPNICER)$-A_V$(RJCE19) is about 3.4 mag for the whole map, which is significantly higher than that of $A_V$(XPNICER)$-A_V$(RJCE12). For the sub-regions (see Fig.~\ref{fig:compare2pre-gmf324b}g and Figs.~\ref{fig:ap1com}-\ref{fig:ap4com}), the medians of $A_V$(XPNICER)$-A_V$(RJCE19) are in the range of $\sim$2$-$4 mag which is also $\sim$1$-$2.5 mag higher than that of $A_V$(XPNICER)$-A_V$(RJCE12) individually. Actually, \citet{rjce2019} also compared $A_{K_s}$(RJCE19) with $A_{K_s}$(RJCE12) and found a systematic offset of $A_{K_s}$(RJCE12)$-A_{K_s}$(RJCE19)$\sim$0.2 mag, corresponding to $A_V\sim$2 mag in a VVV tile (b328). They suggested that the discrepancies could arise from the selection effects
based on the techniques and data employed, e.g., the different stellar samples used to constructed the maps. Therefore, considering the uncertainties of zero-point of $A_V$(XPNICER) and $A_V$(RJCE19), there should be a good agreement between our XPNICER map and the RJCE map by \citet{rjce2019}.}

\subsubsection{\mzrevi{Detailed comparison with 3D reddening map by \citet{schultheis2014}}}

\mzrevi{We accessed the 3D $E(J-K_s)$ reddening map through the VizieR server\footnote{\url{https://vizier.cds.unistra.fr/viz-bin/VizieR}} and then integrated it up to a distance of 10 kpc to obtain a projected $E(J-K_s)$ map. Because this 3D map is constructed with the VVV aperture photometric catalog \citep{vvvdr1}, we corrected the 2D $E(J-K_s)$ map for the bias in the photometric zero-points as described in Sect.~\ref{sect:compare2beam}. Finally we obtained the 2D extinction map of $A_V$(Schultheis) using the reddening law suggested by \citet{wangchen2019}.}

\mzrevi{Figure~\ref{fig:compare2pre}i, \ref{fig:compare2pre-pipe}g show the pixel-to-pixel comparisons between $A_V$(XPNICER) and $A_V$(Schultheis) for the whole map and the sub-region with the Pipe, respectively (also see Fig.~\ref{fig:ap5com} for the comparison in another sub-region). The relations between $A_V$(XPNICER) and $A_V$(Schultheis) are close to the one-to-one relation, but with a systematic offset of $\sim$1$-$2 mag. Considering that $A_V$(XPNICER) overestimates the extinction values by $\sim$1 mag, there should be a good agreement between $A_V$(XPNICER) and $A_V$(Schultheis).}

\subsubsection{\mzrevi{Detailed comparison with reddening map by \citet{surot2020}}}\label{sect:compare2surot}

\mzrevi{We downloaded the $E(J-K_s)$ reddening map presented by \citet{surot2020} through the VizieR server. We did not try to correct for the photometric bias as described in Sect.~\ref{sect:compare2beam} because \citet{surot2020} has applied the internal and absolute calibrations to their color excess maps (see Appendix.~\ref{sect:summarysurot}). We just converted $E(J-K_s)$ to the visual extinction units using the extinction law suggested by \citet{wangchen2019}.}

\mzrevi{Figures~\ref{fig:compare2pre}j and \ref{fig:compare2pre-pipe}h show the pixel-to-pixel comparisons between $A_V$(XPNICER) and $A_V$(Surot) for the whole map and the sub-region with the Pipe, respectively (also see Fig.~\ref{fig:ap5com} for the comparison in another sub-region). The behavior of $A_V$(XPNICER) versus $A_V$(Surot) is similar to that of $A_V$(XPNICER) versus $A_V$(BEAM) because $A_V$(Surot) was calibrated with $A_V$(BEAM) \citep{surot2020}. The medians of $A_V$(XPNICER)$-A_V$(Surot) in the whole map and sub-regions are also in the range of $\sim$1$-$3 mag, similar to that of $A_V$(XPNICER)$-A_V$(BEAM). In some high extinction regions with $A_V\sim$15$-$30 mag, the values of $A_V$(XPNICER) are significantly lower than that of $A_V$(Surot). As mentioned in Sect.~\ref{sect:compare2beam}, these discrepancies could result from the uncertainties of zero-points of $A_V$(XPNICER) and $A_V$(Surot) and the different stellar population used to map the extinction. }

\subsection{On the advantages and caveats of the XPNICER maps}

We performed comparisons between our XPNICER maps and \mzrevi{ten} previous dust based maps in Sect.~\ref{sect:comparison}. Here we summarize the main advantages and caveats of our XPNICER map and give some suggestions for its use.

%


One advantage of the XPNICER map over the previous extinction maps is that it traces the total dust extinction integrated to a distance of $d_{\mathrm{limit}}$. In low extinction regions, the value of $d_{\mathrm{limit}}$ can be up to $\sim$12-16\,kpc (see Sect.~\ref{sect:extmap}), which \mz{approaches} the "total Galactic dust extinction". Previous \mzrevi{large-scale} NIR extinction maps such as that by \citet{dobashi2011} and \citet{juvela16} 
do not measure the total dust extinction along the line of sight and usually significantly underestimate the extinction from diffuse dust component.   

Another advantage of our XPNICER map is its high spatial resolution (30\arcsec) and large coverage. Previous dust-based maps with large coverage such as that presented by \citet{sfd1998,dobashi2011,rjce2012,planck-dust-2014,juvela16} usually have spatial resolutions of $>$1\arcmin. 
\mzrevi{The reddening map presented by \citet{surot2020} can archieve $\sim$10\arcsec~resolution in the area close to the Galactic midplane, but their map only covers the Galactic bulge.}
The Herschel PPMAP data presented by \citet{ppmap2017} have the resolution of 12\arcsec~and a coverage of $\sim$720 deg$^2$, but it only covered the central Galactic midplane with $|b|\lesssim$~1\degr~\mz{($|b|\lesssim$~2.2\degr~for our XPNICER map)}. The \mz{higher Galactic latitudes are crucial for a variety of studies, e.g., of the stellar populations in the disk, gas content of the Milky Way, nearby molecular clouds, and studies of extragalactic sources. Our map can be highly beneficial in these regions}. 

\mz{Finally, for studies of Galactic extinction, our XPNICER map measures actual extinction. The emission-based maps, i.e., SFD, Planck, and PPMAP maps measure emission that needs to be converted to column density and further to extinction. Thus, our XPNICER map provides a highly complementary, direct way to measure Galactic extinction. The extinction and emission based methods can be significantly different in specific regions, such as} supernova remnants or star-forming regions, because dust extinction and emission are sensitive to dust properties such as composition, grain size distribution, temperature distribution, and interstellar radiation field \citep{draine2003}. \mz{As a result}, compared with many previous dust-based maps, \mztwo{our XPNICER map offers a better estimation for the "total Galactic dust extinction"}. 

The main disadvantage of our XPNICER map is that it can not trace the dense and/or distant dust features because the sensitivity of the VVV survey is not good enough to detect sufficient number of stars behind the dense clouds. A very dense nearby cloud could result in an almost total lack of sources and thus a "hole" in our XPNICER map while a distant dense cloud could result in a absorption-like feature in our XPNICER map as described in Appendix~\ref{sect:absorption-like_features}. Identifying or masking such artifacts is a nontrivial task, as it needs the knowledge of the dense dust structure distribution. Comparisons with other dust tracers can be used to study reliability of different lines of sight. For example, the comparisons with previous dust based maps in Sect.~\ref{sect:comparison} suggested linear correlations between our XPNICER map and several dust emission based maps in the range of \mztwo{$A_V\lesssim$~10$-$20\,mag or the optical depth $\tau_{\mathrm{353}}<$~(4$-$7)$\times$10$^{-4}$. The pixels in our XPNICER map with values that correspond to $A_V\gtrsim$~10$-$20\,mag or $\tau_{\mathrm{353}}>$~(4$-$7)$\times$10$^{-4}$ have relatively large uncertainties and should be treated with caution.}


The main objective of this work is to provide a data product in which the dust extinction towards any line of sight of the inner Galactic plane can be easily and conveniently obtained. We believed our XPNICER map to be useful in many studies, at least as the first step towards more detailed analysis of the Galactic dust distribution or stellar populations. For example, our XPNICER map can be used to identify the dust structures at cloud scale as a searching map or investigate the diffuse dust distribution at large scales. Many studies with the aim to investigate the stellar properties can also use our map as a quick check to estimate the initial values of the reddenings of stars. By comparing with other gas tracer maps \citep{gass2009,thrumms2015,sedigism2017}, our XPNICER map can be also used to analyze the relation between gas and dust, and even constrain the properties of "CO-dark" gas \citep{darkgas2010}.

\section{SUMMARY AND CONCLUSIONS}\label{sect:summary}
We have presented a 2D dust extinction map with the spatial resolution of 30\arcsec~based on the previously released DaoPHOT photometric catalog \citep{vvvdaophotmypub}, covering the whole VVV survey area of $\sim$562 deg$^2$ in the Galactic plane. The map was derived with the XPNICER technique that \mz{was developed} based on the previous PNICER  \citep{pnicer2017} and Xpercentile \citep{dobashi2008} methods. Our main results and conclusions are as follows:

\begin{itemize}
    \item[1.]{
    To decrease the bias in the photometric zero-points, we re-calibrated the \citet{vvvdaophotmypub}'s DaoPHOT catalogs with the method suggested by \citet{hajdu2020}. The new calibrated DaoPHOT catalogs have no significant systematic photometric offset compared with the 2MASS photometry and the photometric accuracy of new DaoPHOT catalog is about 60-70\,mmag.}

    \item[2.]{We derived a novel dust extinction map that covers the VVV survey area. To do this, we developed the XPNICER method (see Sect.~\ref{sect:method}). The map has a spatial resolution of 30\arcsec~and it can trace the dust extinction up to $A_V\sim$~30\,mag\jk{, although, it is generally less reliable above $A_V\sim$~10-20\,mag. The map describes the total extinction up to the distance of $d_{\mathrm{limit}}$. This limiting distance varies strongly, but is typically higher than 3 kpc and can reach up to 12-16 kpc, depending on the line of sight.}
    }
    
    \item[3.]{The typical uncertainty of our XPNICER $A_V$ map is about 0.2\,mag. The sources of uncertainties include the observed photometric uncertainties, the variations of intrinsic colors, and the bias of reference stars. In addition to these, the zero point uncertainty of our XPNICER map is about 1$\pm$1.4\,mag, and the systematic uncertainty due to different extinction laws can be up to 20-30\%. \jk{In addition to the uncertainties, we describe the typical artifacts in the map that can make it locally unreliable.} 
    }
    
    \item[4.] Compared to most previous dust based maps, our XPNICER map has higher spatial resolution and it can better trace the total dust extinction, including the extinction from diffuse dust that is usually underestimated by the previous extinction maps. Thus, it provides useful data for a variety of studies ranging from individual objects to the stellar population of the Milky Way. It represents \mzrevi{a high} fidelity extinction-based map and can be highly complementary as an independent measure of dust column densities.
    
\end{itemize}

\jk{Even if our extinction map represents a clear step onwards, it }still cannot trace well the dense dust structures, especially at the Galactic latitude of $|b|\lesssim$~1\degr. In the future, one solution is to perform deeper optical or infrared imaging surveys, which can be expected from the ground-based or space projects such as LSST and JWST. Another way forward is to combine dust extinction and emission data as suggested by \citet{lombardi2014}. We are working on the combination of the extinction map presented in this work and the Herschel PPMAP mentioned in Sect.~\ref{sect:compare2ppmap}. This work is part of the PROMISE\footnote{\url{http://promise.jounikainulainen.com}} program that derives high-dynamical-range column density data for molecular clouds by combing mid-infrared extinction, far-infrared dust emission, and near-infrared extinction data.

\section*{Acknowledgements}

This work was supported by the National Natural Science Foundation of China (grants No. 12073079). This project has received funding from the European Union’s Horizon 2020 research and innovation programme under grant agreement No. 639459 (PROMISE). We acknowledge the science research grants from the China Manned Space Project with No. CMS-CSST-2021-B06. This research made use of Astropy,\footnote{\url{http://www.astropy.org}} a community-developed core Python package for Astronomy \citep{astropy:2013,astropy:2018}. \mzrevi{This research has made use of the VizieR catalogue access tool, CDS,
 Strasbourg, France (DOI : 10.26093/cds/vizier). The original description 
 of the VizieR service was published in 2000, A\&AS 143, 23}

\section*{Data Availability}
We released the extinction maps and their associated uncertainty maps derived with XPNICER technique based on the VVV photometric catalogs in 
\url{http://paperdata.china-vo.org/miaomiaozhang/VVVextmap/VVVextmap\_mef.fits}. It is a multi-extension FITS file and self-explained in its header.
 



\bibliographystyle{mnras}
\bibliography{myref} 



\appendix

\section{Defects and artifacts in our XPNICER extinction maps}
\label{sect:defects}

\subsection{Pixels with negative extinction values}\label{sect:negpixs}

\begin{figure*}
    \centering
    \includegraphics[width=1.0\linewidth]{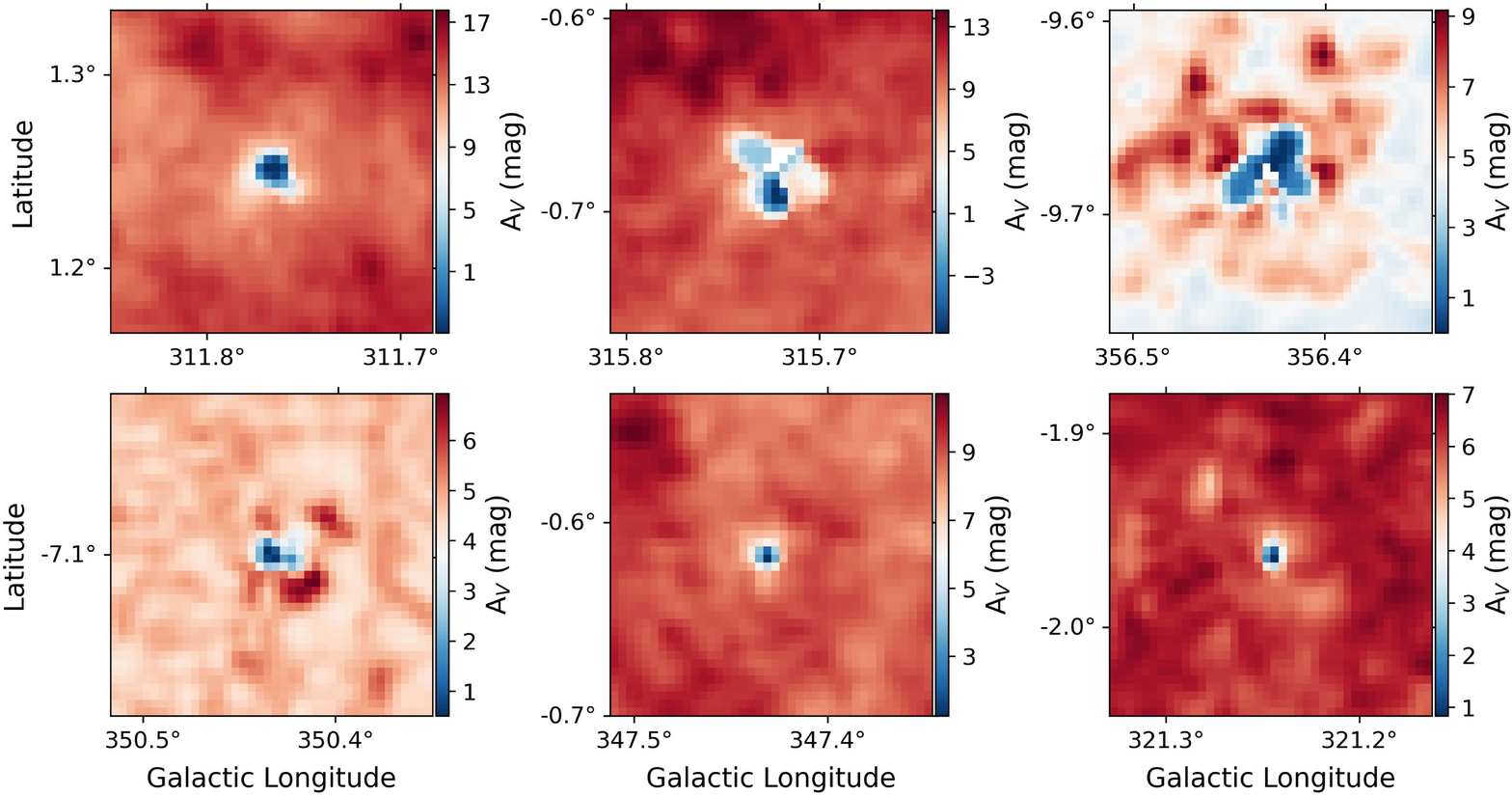}
    \caption{The "hole" defects due to the bright stars.}
    \label{fig:holes}
\end{figure*}

We note that there are 66 pixels with negative values in the $A_V^{30}(X_0=80)$ map, about half of which are located at the edge of maps and thus treated as the unreliable measurements. Other half of negative pixels are concentrated in several "holes" as shown in Fig.~\ref{fig:holes} (top panels). We found that these negative "holes" are in the vicinity of very bright sources that are saturated in the VVV survey images and thus are artifacts due to the lack of reliable photometry around the saturated sources. 

\subsection{Defects by bright stars}
\label{sect:defectsofbrightstars}

The bright sources usually cause the artifacts in the detectors like saturation, non-linearity, and diffraction patterns. Thus there are often many spurious detections and very few reliable photometric sources around them. Such a lack of stars in the vicinity of the bright sources results in the "holes" in our extinction maps. Figure~\ref{fig:holes} shows some examples in the $A_V^{30}(X_0=80)$ map. The "holes" usually have point-like morphology and their sizes are related to the brightness of the corresponding bright stars. Several very bright stars ($K_s<$~2\,mag) result in the negative "holes" (also see Sect.~\ref{sect:negpixs}). 

\subsection{Absorption-like features in high extinction regions}\label{sect:absorption-like_features}

There are also many irregular "holes" in the center of high extinction regions, especially the area at low Galactic latitudes (see Fig.~\ref{fig:extmap_disk} and~\ref{fig:extmap_bullge}), which look like the absorption-like features on the high extinction background. These absorption-like features should arise from an underestimation of the true integrated extinction due to contamination of foreground sources and can be improved by using higher $X_0$ percentile values.

\begin{figure*}
    \centering
    \includegraphics[width=1.0\linewidth]{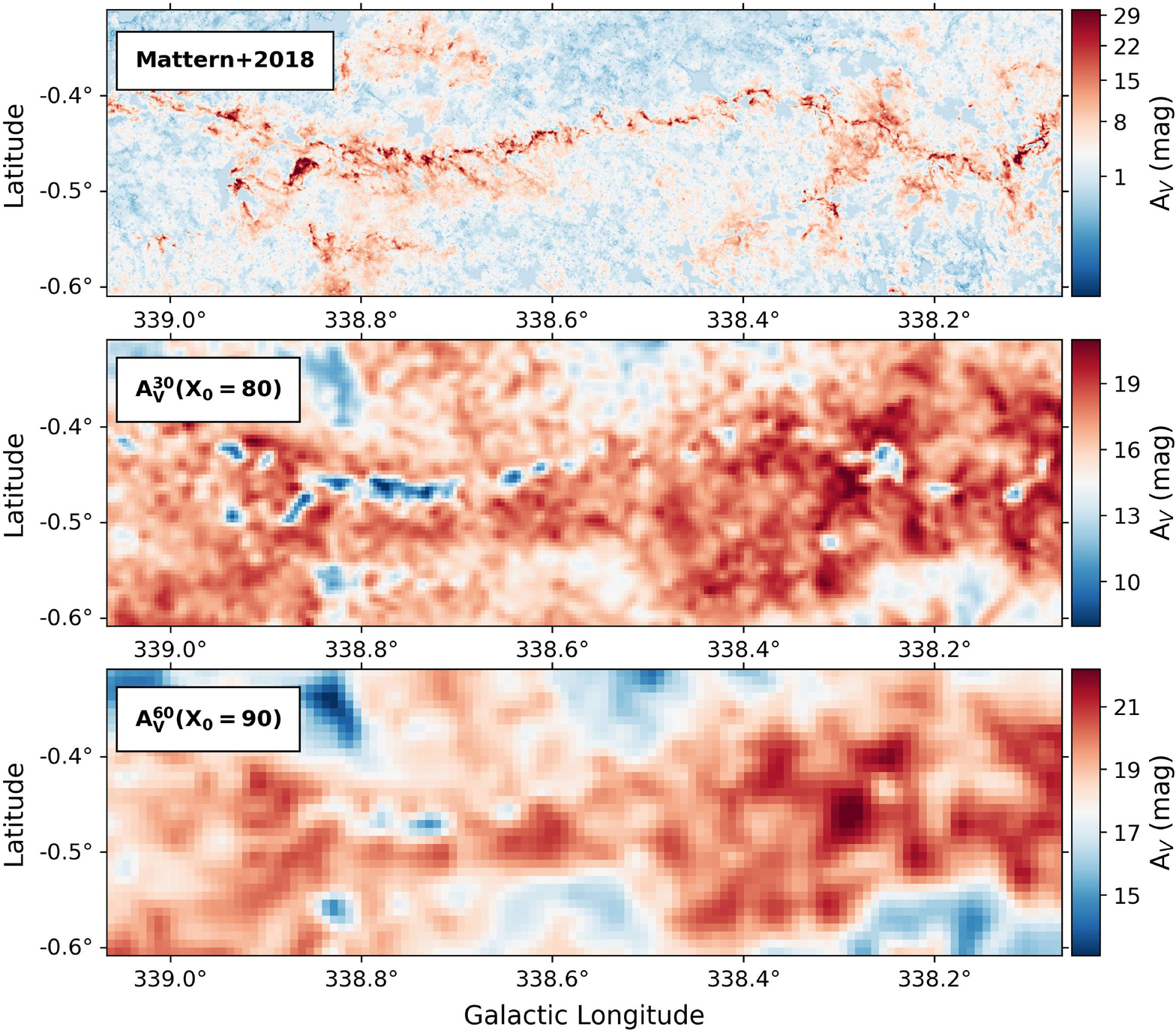}
    \caption{The high-resolution extinction map obtained by \citet{mattern2018} (top panel); our $A_V^{30}(X_0=80)$ map (middle panel); and $A_V^{60}(X_0=90)$ map (bottom panel); for the region with Nessie \citep{nessie2010}.}
    \label{fig:nessie}
\end{figure*}

Figure~\ref{fig:nessie} shows a zoom-in region with the Nessie that was identified as a giant filament with the length of $\sim$80\,pc and width of only $\sim$0.5\,pc \citep{nessie2010,ragan2014}. \citet{mattern2018} presented a high resolution ($\sim$2\arcsec) extinction map of Nessie by combining the NIR and mid-infrared (MIR) extinction maps with the method suggested by \citet{kainulainen11alves}, \citet{but12}, and \citet{kt2013}. Their combined extinction map has a large dynamical range and can trace the dust extinction up to $A_V\sim$~100\,mag. Figure~\ref{fig:nessie} (top panel) shows their map and we can see the high-extinction skinny filamentary structures. Figure~\ref{fig:nessie} (middle panel) shows our extinction map, $A_V^{30}(X_0=80)$. Obviously, the absorption-like features in $A_V^{30}(X_0=80)$ exactly match the high-extinction filamentary structures in \citet{mattern2018}'s map, which means that $X_0=80$ is not high enough to select the background sources that are located behind the dense part of Nessie. We also obtained the extinction map using $X_0=90$, i.e.,  $A_V^{60}(X_0=90)$, which has a lower spatial resolution of 60\arcsec~and is shown in Fig.~\ref{fig:nessie} (bottom panel). Apparently, some absorption-like features in $A_V^{30}(X_0=80)$ have disappeared in $A_V^{60}(X_0=90)$, but there are still absorption-like features remaining in $A_V^{60}(X_0=90)$. Therefore, $X_0=90$ is not high enough yet to exclude the contamination of foreground sources for the very dense part of Nessie. 

In general, the absorption-like features in high extinction regions of our map are artifacts due to the contamination of foreground sources. They usually correspond to the very dense and/or distant molecular clouds. Increasing $X_0$ percentile value is helpful to remove some of absorption-like features. However, if a cloud has a total extinction sufficient to darken the background stars beyond the detection limit of the VVV survey, its density distribution can not be mapped by our extinction mapping technique. 

\subsection{Artifacts caused by detector defects}
\label{sect:artifactsbydetector}

\begin{figure}
    \centering
    \includegraphics[width=1.0\linewidth]{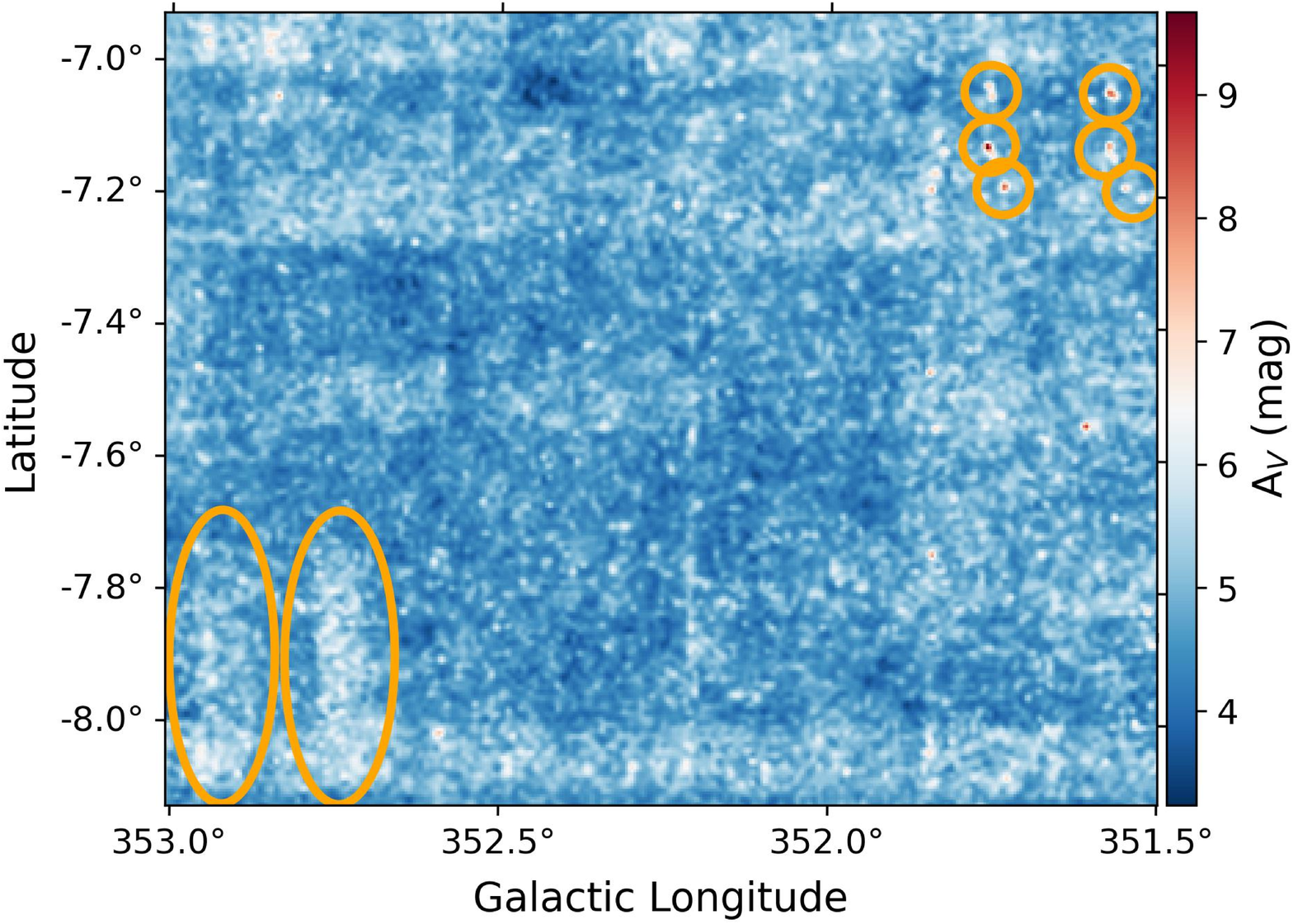}
    \caption{The $A_V^{30}(X_0=80)$ map for the tile region "b230". The orange ellipses mark the artifacts caused by the defects of detector VIRCAM.}
    \label{fig:defects-detector}
\end{figure}

The defects of VIRCAM also result in some artifacts in our extinction maps. Figure~\ref{fig:defects-detector} shows the $A_V^{30}(X_0=80)$ map for a bulge tile "b230". The six spots marked with orange circles in the top-right corner of Fig.~\ref{fig:defects-detector} are due to the bad pixels of VIRCAM detector \#1\footnote{\url{http://casu.ast.cam.ac.uk/surveys-projects/vista/technical/known-issues\#section-2}}. There are also two stripes labeled with orange ellipses in the bottom-left corner of Fig.~\ref{fig:defects-detector}. They arise from the variable QE of VIRCAM detector \#16. Although we tried to remove these QE variant through re-calibrating the DaoPHOT catalog (see Sect~\ref{sect:vvv-daophot-catalog}), the residuals still exist in our extinction maps. We note that these artifacts appear in many tile regions of our extinction maps. However, their unchanging patterns can be used to distinguish them from the true dust features.

\subsection{Tile patterns in the maps}
\label{sect:tilepatterns}

Figure~\ref{fig:extmap_disk}-\ref{fig:errmap_bulge} show obvious tile patterns in the $A_V^{30}(X_0=80)$ and $\delta A_V^{30}(X_0=80)$ maps, especially the bulge area at high Galactic latitude. The tile patterns should arise from the varying sensitivity of the VVV survey. \citet{vvvdaophotmypub} performed the DaoPHOT photometry on the stacked VVV survey images that were obtained through combining the pawprints observed at different time. The different characteristics of VIRCAM detectors and atmospheric conditions at different observing time result in different exposure time at different pixels of stacked images. The sensitivity variation causes the patterns at the VVV tile size scale as shown in Fig~\ref{fig:extmap_disk}-\ref{fig:errmap_bulge} and also the patterns at the VIRCAM detector size scale which can be seen in Fig.~\ref{fig:defects-detector}. We note that these patterns can not be removed even if we only use DaoPHOT sources with photometric uncertainties of $<$0.1\,mag to produce the extinction maps.

\section{Short descriptions of previous ten dust-based maps}\label{ap:summaryext}
\subsection{\citet{dobashi2011}}
\citet{dobashi2011} obtained an all-sky extinction map with the "X percentile" method based on the 2MASS point source catalog (PSC). 
\mz{To estimate the intrinsic colors of the background sources, \citet{dobashi2011}} assumed that bright and faint stars were nearby and distant stars, respectively, and estimated the intrinsic colors of faint stars with the intrinsic color of nearby stars and a color correction item. 
The intrinsic color of nearby stars was resembled with the colors of stars in the polar area ($|b|>$~80\degr) while the color correction item was estimated using stars in the high latitude regions without the molecular clouds.
However, \citet{dobashi2013} found that the assumption of a constant intrinsic color of nearby stars was not valid at large scales and they used the Besan{\c c}on Galactic model to calibrate the intrinsic colors of background sources. Because the Besan{\c c}on Galactic model can not precisely simulate the 2MASS PSC without the full information of 3D dust distribution, \citet{dobashi2013} found that the obtained color excess maps significantly underestimated the extinction from large-scale diffuse dust. 

\subsection{\citet{juvela16}}
\citet{juvela16} also presented an all-sky extinction map using the NICER technique \citep{nicer2001} based on the 2MASS PSC. They offered three extinction maps based on the different methods that were used to estimate the intrinsic colors of background stars. The M1 map used the average colors of all 2MASS stars at Galactic latitude of $|b|>$~60\degr. The M2a map employed the average colors of all simulated stars from the Besan{\c c}on model within a distance of 8\,kpc as the reference colors. For the M2b map, the intrinsic colors of the simulated Besan{\c c}on stars were firstly reddened by a simple diffuse dust model of $f(z) \propto 1.0\,\mathrm{mag}\,\mathrm{kpc}^{-1} \times e^{-\frac{|z|}{150\,\mathrm{pc}}}$, where $z$ was the Galactic height. Then the reddened colors were used as the reference colors, which also meant that M2b map did not include the extinction from diffuse dust component. M1 map was not appropriate to trace the dust extinction in the Galactic plane because it ignored the variation of the intrinsic colors of background stars at low Galactic latitude. Compared with M2a, M2b considered the 3D dust distribution and thus the simulated Besan{\c c}on stars used for M2b were more analogous to the 2MASS PSC. 
Therefore, in this paper we only consider the M2b map.

\subsection{\citet{sfd1998}}
The SFD dust map was \mz{derived from} the 100\,$\mu$m thermal emission maps from IRAS and DIRBE/COBE surveys. \citet{sfd1998} used the flux ratio map of DIRBE 100\,$\mu$m to 240\,$\mu$m to constrain the dust temperature and 
transformed the IRAS 100\,$\mu$m map to a normalized dust column density map which was then calibrated to the color excess $E(B-V)$ units using about 400 galaxies with Mg$_2$ indices and photometric color measurements \citep{faber1989}. \citet{schlafly2011} investigated the reddenings towards individual stars based on the SDSS photometric and spectroscopic survey data \citep{sdss2011}, which were used to recalibrate the SFD map. They provided a set of conversion coefficients from the SFD $E(B-V)$ units to extinction in different passbands, e.g., $A_V=$~2.742$E(B-V)$ assuming $R_V=$~3.1.

\mztwo{\citet{sfd1998} mentioned that most contaminating unresolved sources such as stars, planetary nebulae, and extragalactic sources were not removed from the SFD dust map at low Galactic latitudes ($|b|<$~5\degr), which usually results in anomalous bright blobs in the map. Moreover, the temperature distribution of the Galaxy is not well resolved \citep[see Appendix C of][]{sfd1998}. Thus the SFD map at low Galactic latitudes has large uncertainties.}

\subsection{\citet{planck-dust-2014}}
\citet{planck-dust-2014} fitted the emission from Planck 353, 545, 857 GHz, and IRAS 100\,$\mu$m survey data with an all-sky dust model. Their model described the SED of emission 
with a modified blackbody assuming thermal equilibrium in the optically thin limit:
\begin{align*}
    I_{\nu}=\tau_{\nu_0}B_{\nu}(T_{\mathrm{obs}})\left( \frac{\nu}{\nu_0} \right)^{\beta_{\mathrm{obs}}},
\end{align*}
where $I_{\nu}$ is the specific intensity at frequency $\nu$, $B_{\nu}(T_{\mathrm{obs}})$ is the Planck function at temperature $T_{\mathrm{obs}}$, $\tau_{\nu_0}$ is the optical depth at a reference frequency $\nu_0$, and $\beta_{\mathrm{obs}}$ defines a commonly-used power law relation of $\tau_{\nu}=\tau_{\nu_0}(\nu/\nu_0)^{\beta_{\mathrm{obs}}}$. \citet{planck-dust-2014} selected a reference frequency $\nu_0=$~353\,GHz ($\sim$850\,$\mu$m) and presented the all-sky maps of $\beta_{\mathrm{obs}}$ at 30\arcmin, $T_{\mathrm{obs}}$ and $\tau_{353}$ at 5\arcmin~resolution. 

\subsection{\citet{ppmap2017}}
\citet{marsh2015} used the point process mapping (PPMAP) technique to decompose an astrophysical structure into a set of small building-block components that were parameterized by the variables of angular positions and dust temperature. 
\citet{ppmap2017} applied this technique to all Herschel infrared Galactic Plane (Hi-GAL) survey \citep{herschel-higal} images and constructed a set of separate column density maps in twelve dust temperature intervals at a spatial resolution of 12\arcsec. 

\subsection{\citet{beam2012,beam2018}}
\mzrevi{\citet{beam11} developed a method of obtaining the reddening maps based on the VVV photometric catalogs, which derives the mean color excess $E(J-K_s)$ maps by comparing the observed $J-K_s$ colors of the RC giants to the color of RC stars in Baade's window. The RC giants can be selected on the CMDs with the appropriate criteria and Baade's window is a low extinction region at $l\simeq$1.14\degr, $b\simeq-$4.18\degr~with the known reddening value of $E(B-V)=$~0.55 \citep{zoc2008}.  \citet{beam2012} applied this method to the VVV aperture photometric catalog \citep{vvvdr1} and obtained a $E(J-K_s)$ reddening map for the bulge area, i.e., $-$10.0\degr$\leq l\leq$10.4\degr, $-$10.3\degr$\leq b \leq$~5.1\degr. The reddening map has variant resolutions: 6\arcmin~for $-$10\degr$\lesssim b \lesssim$$-$7\degr; 4\arcmin~for $-$7\degr$\lesssim b \lesssim -$3.5\degr; and 2\arcmin~for $-$3.5\degr$\lesssim b \lesssim$~5\degr. \citet{beam2018} also applied \citet{beam11}'s method to the VVV DoPHOT PSF photometric catalog \citep{vvvdophot2018} and derived a reddening map with the spatial resolution of 1\arcmin~for the central Galactic bulge area, i.e., $-$10\degr$<l<$~10\degr, $-$1.5\degr$<b<$~1.5\degr.}

\subsection{\citet{rjce2012}}\label{ap:rjceext}
\mzrevi{\citet{rjce2011} developed the RJCE method to obtain the foreground reddening of any normal star by combing the NIR photometry such as 2MASS and the mid-infrared (MIR) photometry such as {\it Spitzer}/IRAC. The MIR filters sample the Rayleigh-Jeans tail of a star's SED. Because the slope of the tail is nearly independent of stellar temperature, the variation of intrinsic colors in MIR or NIR-MIR is significantly smaller than that in NIR. Taking advantage of these NIR-MIR color properties, the RJCE method can produce reliable estimates of the foreground reddening towards each individual star through adopting a optimal single NIR-MIR color, i.e., $H-[4.5]$, where $[4.5]$ represents {\it Spitzer} 4.5\,\micron~band. Using the extinction law suggested by \citet{rjce2009} and \citet{ind2005}, the extinction of an individual star can be obtained with:}
\begin{align}
    A_{K_s} = 0.918\times(H-[4.5]-0.08) \label{eq:rjce}
\end{align}
\mzrevi{\citet{rjce2012} applied the RJCE method to the 2MASS and {\it Spitzer} GLIMPSE data \citep{glimpse2003,glimpse2009,majewski2007} and obtained a 2\arcmin~resolution extinction map in the Galactic plane, covering the area of 256\degr$<l<$65\degr~and $|b|\leqslant$1\degr$-$1.5\degr (extending $|b|\leqslant$4\degr~in the bulge). Specifically, \citet{rjce2012} split the stellar sample into three sub-samples: MS dwarfs, RC giants, and red giant branch (RGB) giants based on their intrinsic colors of $[J-K_s]_0$. For each of these three sub-samples, two extinction maps were created using the median and 90\% percentile values of $A_{K_s}$ measured within each pixel. \citet{rjce2012} suggested that the 90\% percentile maps probably provide the most reliable estimates of the total integrated extinction due to the low fraction of foreground sources in the lines of sight. Considering the spatial distributions of different stellar populations, the maps produced by different sub-samples actually trace the dust column to different distances: $\sim$2-4 kpc for MS maps; $\sim$6-10 kpc for RC maps; and $\sim$15-24 kpc for RGB maps. On the other hand, the RC maps have the highest signal-to-noise ratio because the vast majority of sources in the data set are RC giants. Therefore, in this paper we only consider the 90\% percentile extinction maps produced with the RC giants.}

\subsection{\citet{rjce2019}}
\mzrevi{\citet{rjce2019} applied the RJCE method \citep[][see the description in Sect.~\ref{ap:rjceext}]{rjce2011} to the VVV aperture photometric catalog \citep{vvvdr1}, 2MASS, and {\it Spitzer} GLIMPSE data. They finally obtained the $A_{K_s}$ extinction maps with the spatial resolution of 1\arcmin. The maps cover $\sim$148 deg$^2$ in the Galactic disk, i.e., the area of 295\degr$\lesssim l \lesssim$~350\degr, $-$1.0\degr$\lesssim b \lesssim$~1.0\degr~(extending to $|b|\lesssim$~2.25\degr~at some longitudes). \citet{rjce2019} constructed the extinction maps with the all sources, MS stars, RC giants, and RGB giants as suggested by \citet{rjce2012}, respectively, but they only release the $A_{K_s}$ map produced by all stars.}

\subsection{\citet{schultheis2014}}\label{sect:schultheis}

\mzrevi{\citet{schultheis2014} presented a 3D extinction map with a spatial resolution of 6\arcmin~in the Galactic bulge by comparing the VVV photometric catalog \citep{vvvdr1} to the Galactic Besan\c{c}on model \citep{besancon2003,besancon2012}. Specifically, the sources from the VVV photometric catalog and the Besan\c{c}on model were located into the 6\arcmin~grid of the Galactic bulge area. In each cell, the extinction can be obtained by comparing the observed color to the simulated intrinsic color while the distance can be also calculated with the distance-color relation constructed with the simulated data. Repeat this calculation iteratively until the reddened simulated color can well mimic the observed color distribution. The final 3D extinction map can trace the dust column till 10 kpc, covering the whole VVV bulge area, i.e., $-$10\degr$<l<$~10\degr, $-$10\degr$<b<$~5\degr.}

\subsection{\citet{surot2020}}\label{sect:summarysurot}
\mzrevi{\citet{surot2020} presented a $E(J-K_s)$ color excess map in the Galactic bulge based on their VVV MW-BULGE-PSFPHOT catalogs \citep{vvvdaophot2019}. In each VVV tile, they selected the RC and RGB giants manually on the $J-K_s$ versus $K_s$ CMDs and then obtained the color map in a 610$\times$500 pixel grid. All 196 color maps were firstly self-calibrated using the overlap regions among all tiles. Then the color excess maps were obtained by applying the absolute calibration with the reference $E(J-K_s)$ map presented by \citet{beam2012,beam2018}. The final color excess maps covers the whole VVV bulge area, i.e., $|l|<$~10\degr, $-$10\degr$<b<$~5\degr, with the variant resolution of $\sim$2\arcmin-10\arcsec.}

\section{Glimpse of visual comparisons of our XPNICER extinction maps to the previous dust-based maps}\label{app:vcom}

\begin{figure*}
    \centering
    \includegraphics[width=1.0\linewidth]{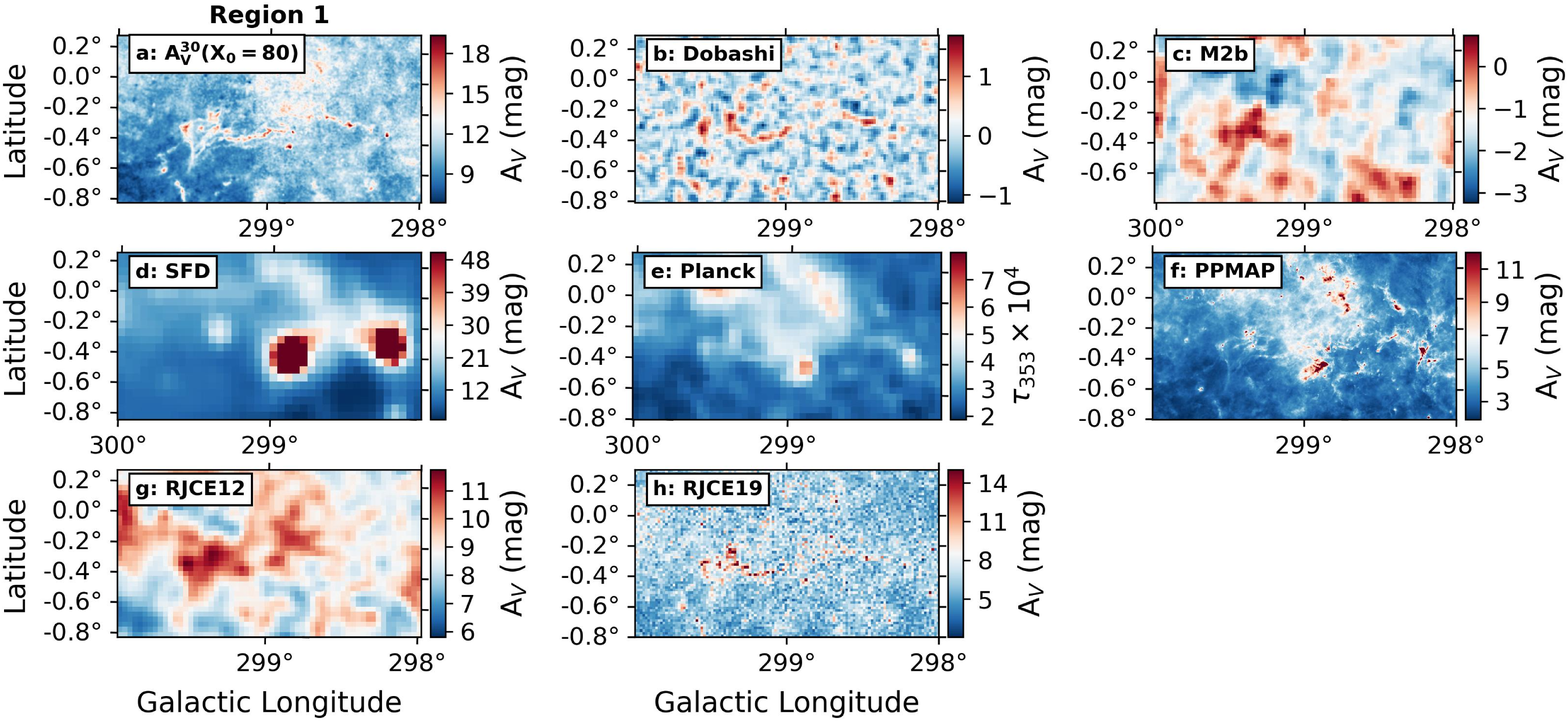}
    \caption{Zoom-in view of a $\sim$2\degr$\times$1\degr~region towards $l=$~299\degr~and $b=$~-0.3\degr. Others are the same as Fig.~\ref{fig:extmap_zoomin}.}
    \label{fig:reg_ap1}
\end{figure*}

\begin{figure*}
    \centering
    \includegraphics[width=1.0\linewidth]{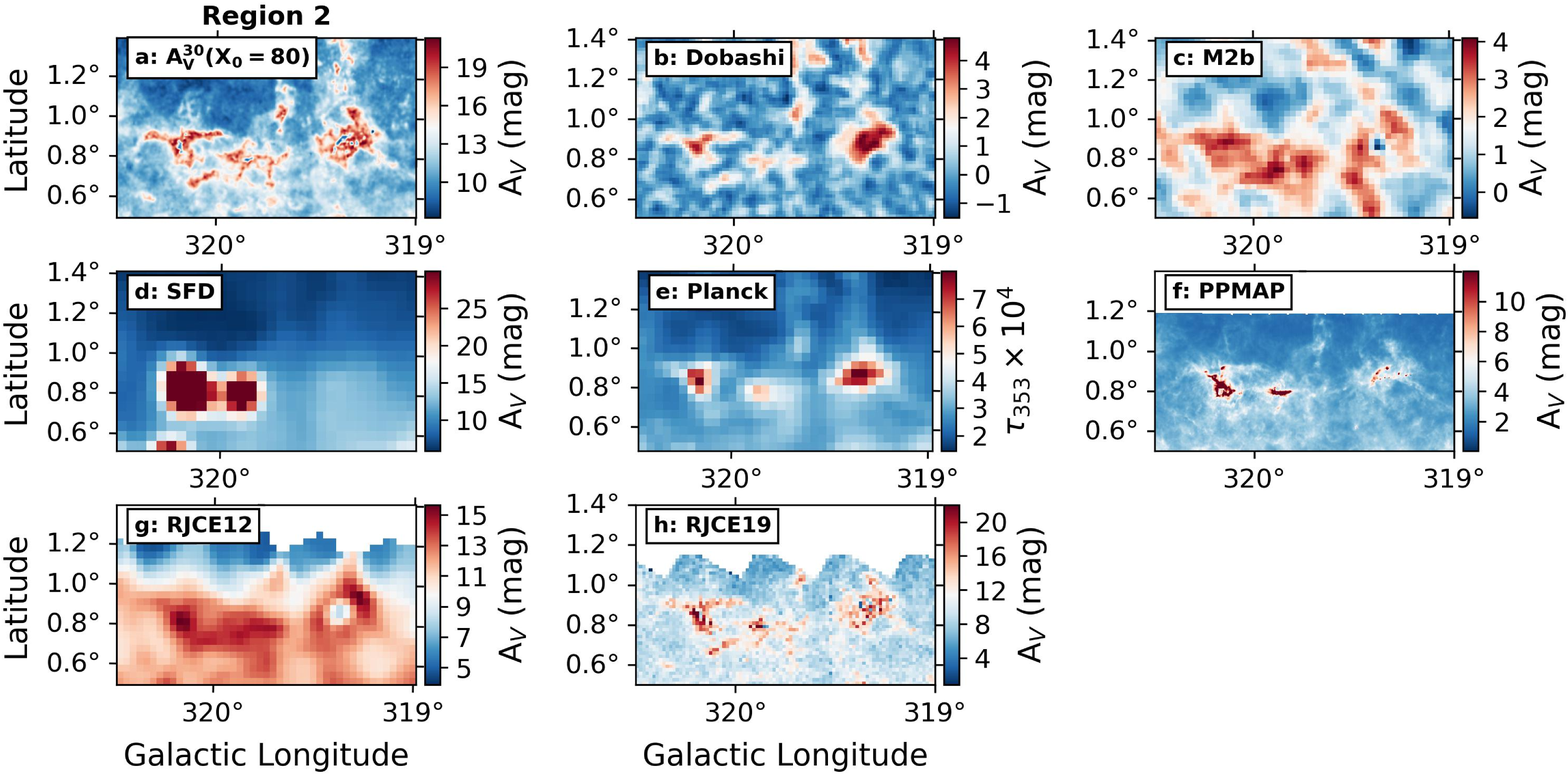}
    \caption{Zoom-in view of a $\sim$1.5\degr$\times$0.8\degr~region towards $l=$~319.7\degr~and $b=$~0.9\degr. Others are the same as Fig.~\ref{fig:extmap_zoomin}.}
    \label{fig:reg_ap2}
\end{figure*}

\begin{figure*}
    \centering
    \includegraphics[width=1.0\linewidth]{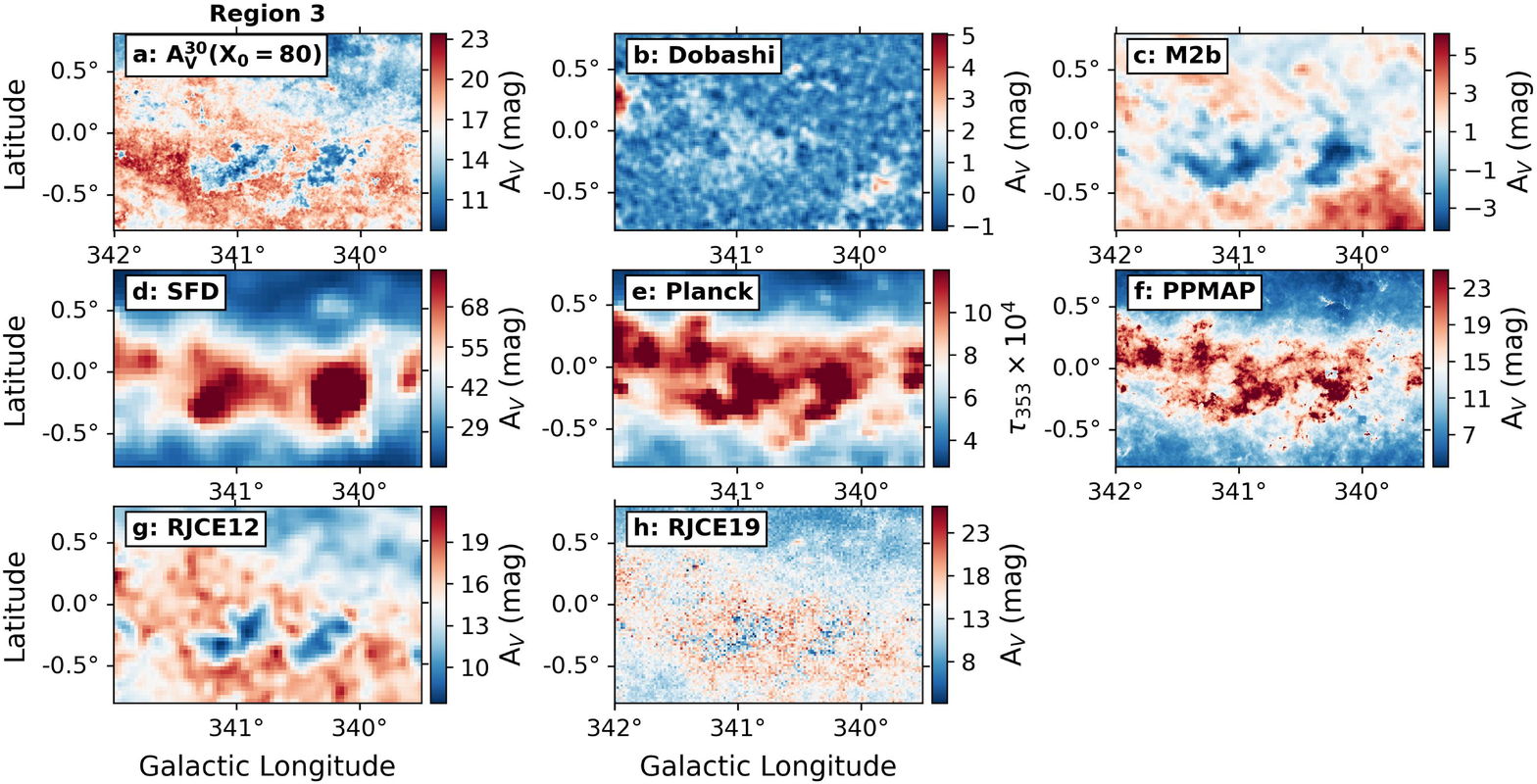}
    \caption{Zoom-in view of a $\sim$2.5\degr$\times$1.6\degr~region towards $l=$~340.75\degr~and $b=$~0.0\degr. Others are the same as Fig.~\ref{fig:extmap_zoomin}.}
    \label{fig:reg_ap3}
\end{figure*}

\begin{figure*}
    \centering
    \includegraphics[width=1.0\linewidth]{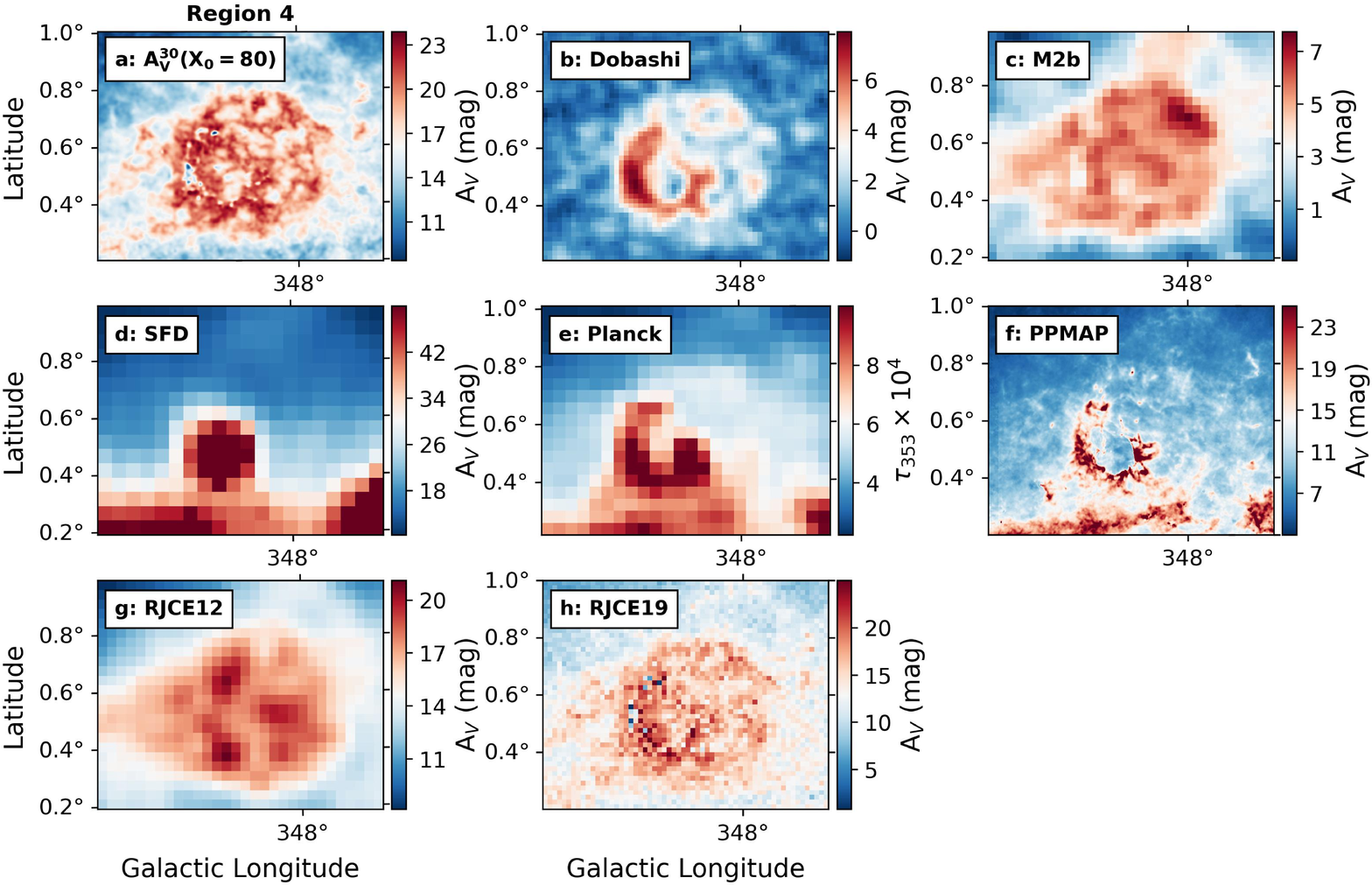}
    \caption{Zoom-in view of the region with RCW 120. Others are the same as Fig.~\ref{fig:extmap_zoomin}.}
    \label{fig:reg_ap4}
\end{figure*}

\begin{figure*}
    \centering
    \includegraphics[width=1.0\linewidth]{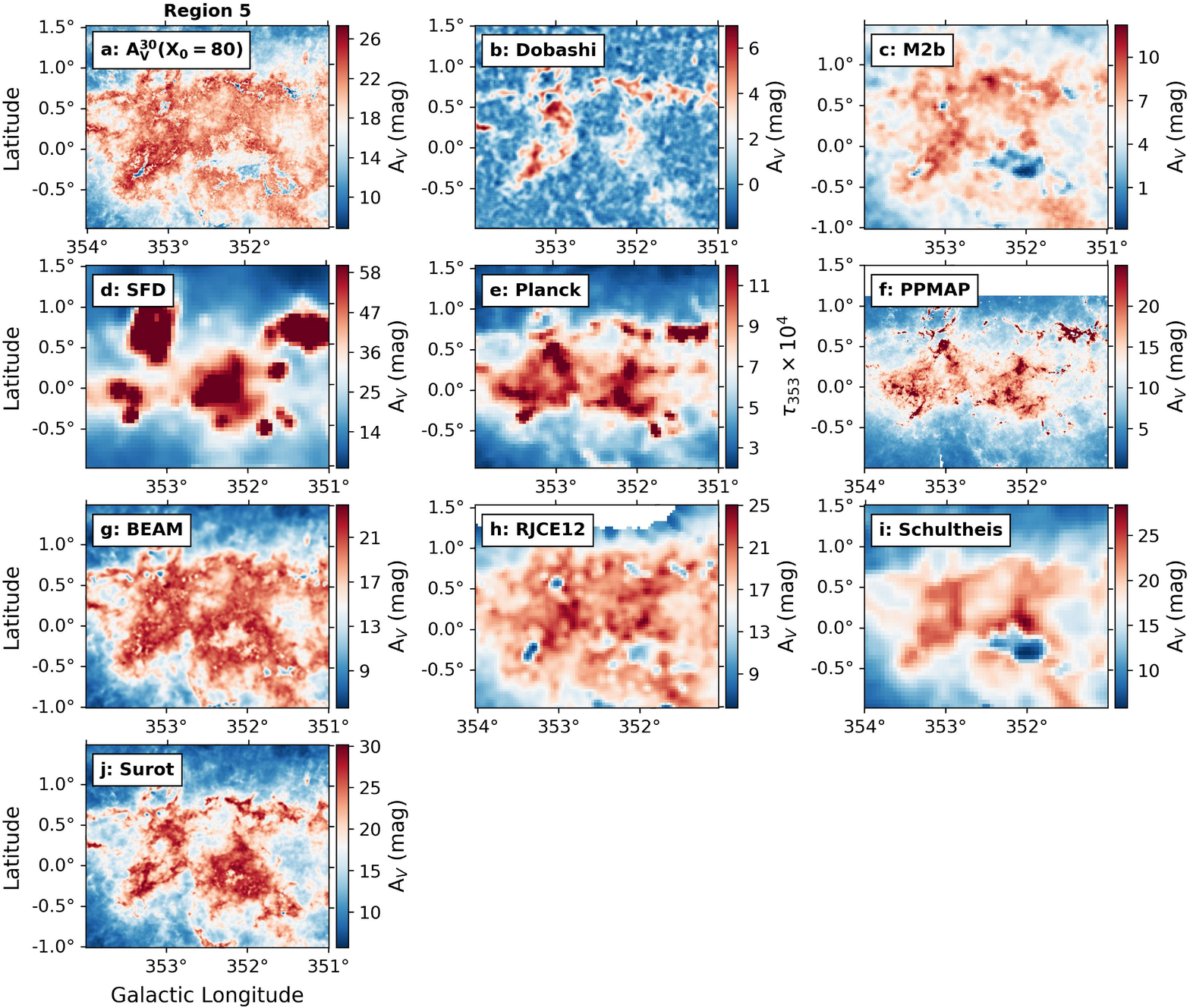}
    \caption{Zoom-in view of a $\sim$3\degr$\times$2.5\degr~region towards $l=$~352.5\degr~and $b=$~0.25\degr. Others are the same as Fig.~\ref{fig:extmap_zoomin}\mzrevi{, except the panels (g) is the BEAM extinction map by \citet{beam2012,beam2018}; (h) is the extinction map obtained with RJCE method by \citet{rjce2012}; (i) is the integrated 2D extinction map by \citet{schultheis2014}; and (j) is the extinction map by \citet{surot2020}.}}
    \label{fig:reg_ap5}
\end{figure*}

\begin{figure*}
    \centering
    \includegraphics[width=1.0\linewidth]{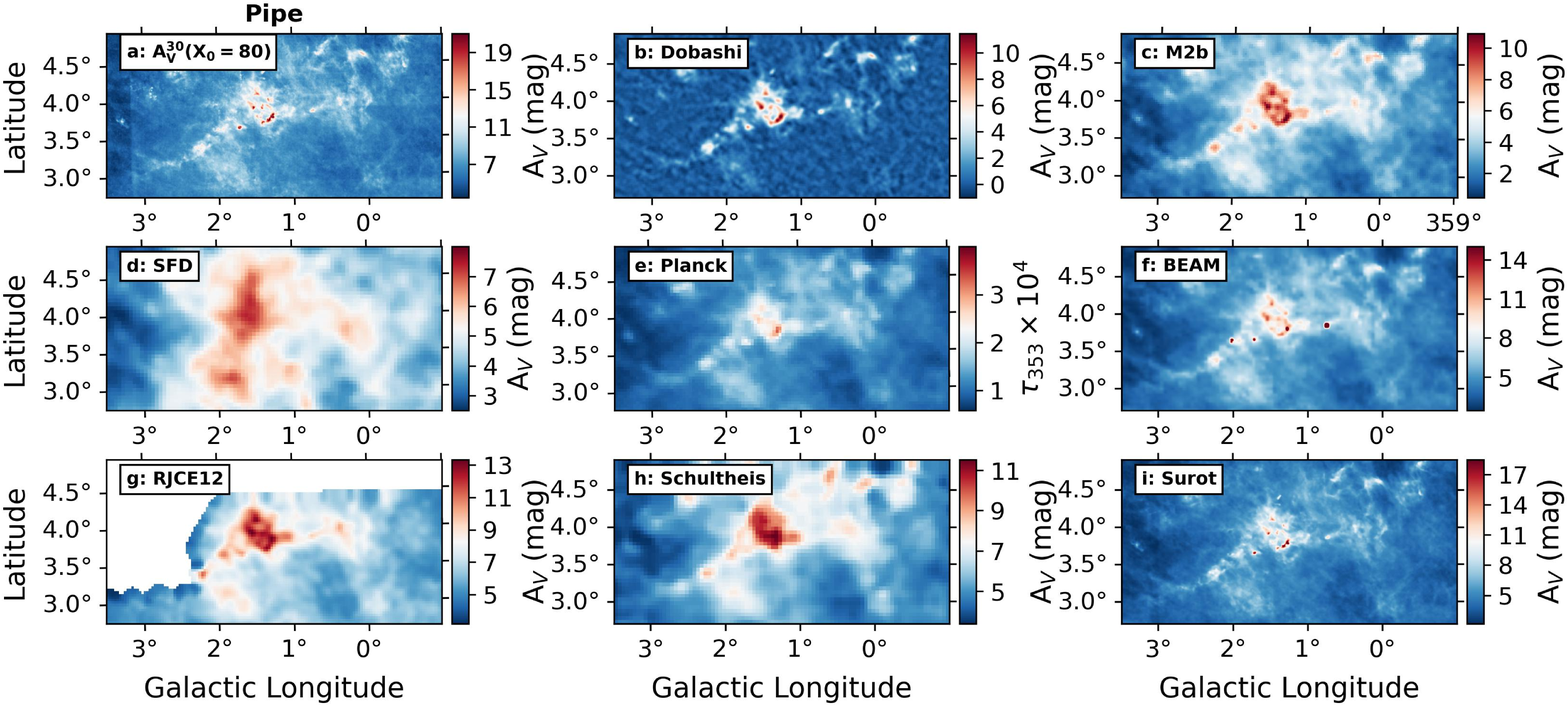}
    \caption{Zoom-in view of the region with the Pipe molecular cloud. Others are the same as Fig.~\ref{fig:extmap_zoomin}\mzrevi{, except the panels (f) is the BEAM extinction map by \citet{beam2012,beam2018}; (g) is the extinction map obtained with the RJCE method by \citet{rjce2012}; (h) is the integrated 2D extinction map by \citet{schultheis2014}; and (i) is the extinction map by \citet{surot2020}.}}
    \label{fig:reg_pipe}
\end{figure*}

\section{Background estimation for $A_V$(XPNICER)}\label{ap:backgroundest}
We used two methods to estimate the background of $A_V$(XPNICER) and obtained two backgrounds, i.e., $A_V$(BG1) and $A_V$(BG2).

$A_V$(BG1) was obtained by smoothing $A_V$(XPNICER) with a box kernel. We tried some different sizes of box kernels from 5\arcmin~to 3\degr~and found that a 25\arcmin~box kernel can minimize the $\chi^2$ that was defined as:
\begin{align*}
    \chi^2=\sum_{\mathrm{pixels}}\frac{[A_V\mathrm{(XPNICER)}-A_V\mathrm{(Dobashi)}]^2}{\sigma^2\mathrm{(XPNICER)}+\sigma^2\mathrm{(Dobashi)}+\sigma^2\mathrm{(BG1)}},
\end{align*}
where $\sigma$(XPNICER) and $\sigma$(Dobashi) were the uncertainty maps of $A_V$(XPNICER) and $A_V$(Dobashi), respectively, while $\sigma$(BG1) was the noise map of $A_V$(BG1).

We note that $A_V$(M2b) was obtained by assuming a diffuse dust model that was only related to the Galactic height. We used a similar model to fit $A_V$(XPNICER) as a slab. The model was defined as:
\begin{align*}
    A_V\mathrm{(BG2)}=c+A\times e^{-\frac{|b|}{h}},
\end{align*}
where $b$ was the Galactic latitude while $c$, $A$, and $h$ were free parameters. We fitted the median profile calculated over all Galactic longitudes of $A_V$(XPNICER) with the above model and obtained $c=$~4.38\,mag, $A=$~16.05\,mag, and $h=$~1.41\degr.

\section{Quantitative comparisons between our XPNICER map and previous dust-based maps in five sub-regions}\label{ap:compare2pre}

\begin{figure*}
    \centering
    \includegraphics[width=1.0\linewidth]{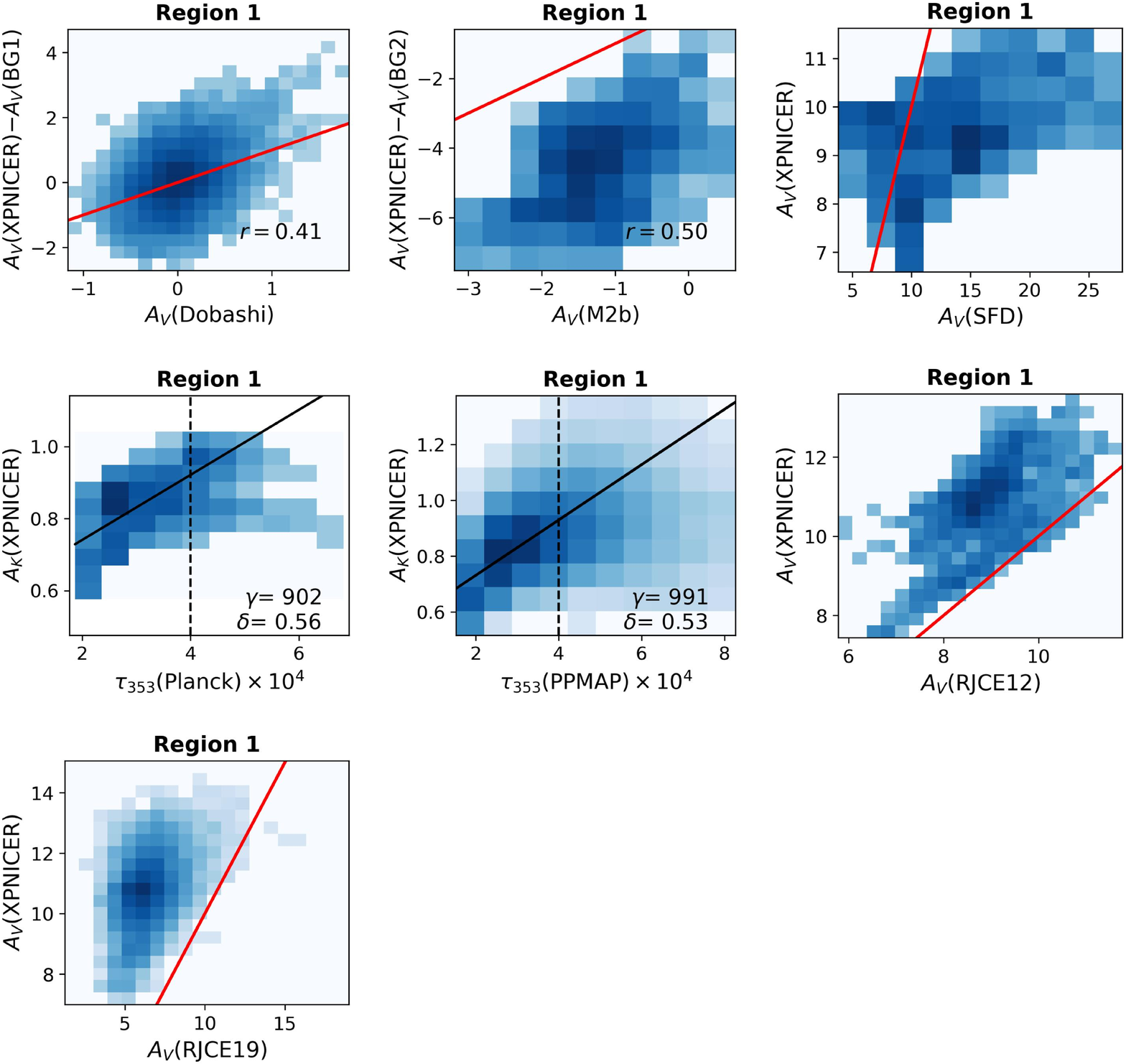}
    \caption{Pixel-to-pixel comparisons of our XPNICER extinction map with previous dust-based maps in \mzrevi{sub-region 1 (see Fig.~\ref{fig:reg_ap1})}, including the extinction maps from \citet[][$A_V$(Dobashi)]{dobashi2011} and \citet[][$A_V$(M2b)]{juvela16}, the SFD map \citep[][$A_V$(SFD)]{sfd1998}, the Planck dust map \citep[][$\tau_{\mathrm{353}}$(Planck)]{planck-dust-2014}, the Herschel PPMAPs \citep[][$\tau_{\mathrm{353}}$(PPMAP)]{ppmap2017}, \mzrevi{RJCE extinction map by \citet[][$A_V$(RJCE12)]{rjce2012}, and RJCE extinction map by \citet[][$A_V$(RJCE19)]{rjce2019}}. Others are the same as Fig.~\ref{fig:compare2pre}.}
    \label{fig:ap1com}
\end{figure*}

\begin{figure*}
    \centering
    \includegraphics[width=1.0\linewidth]{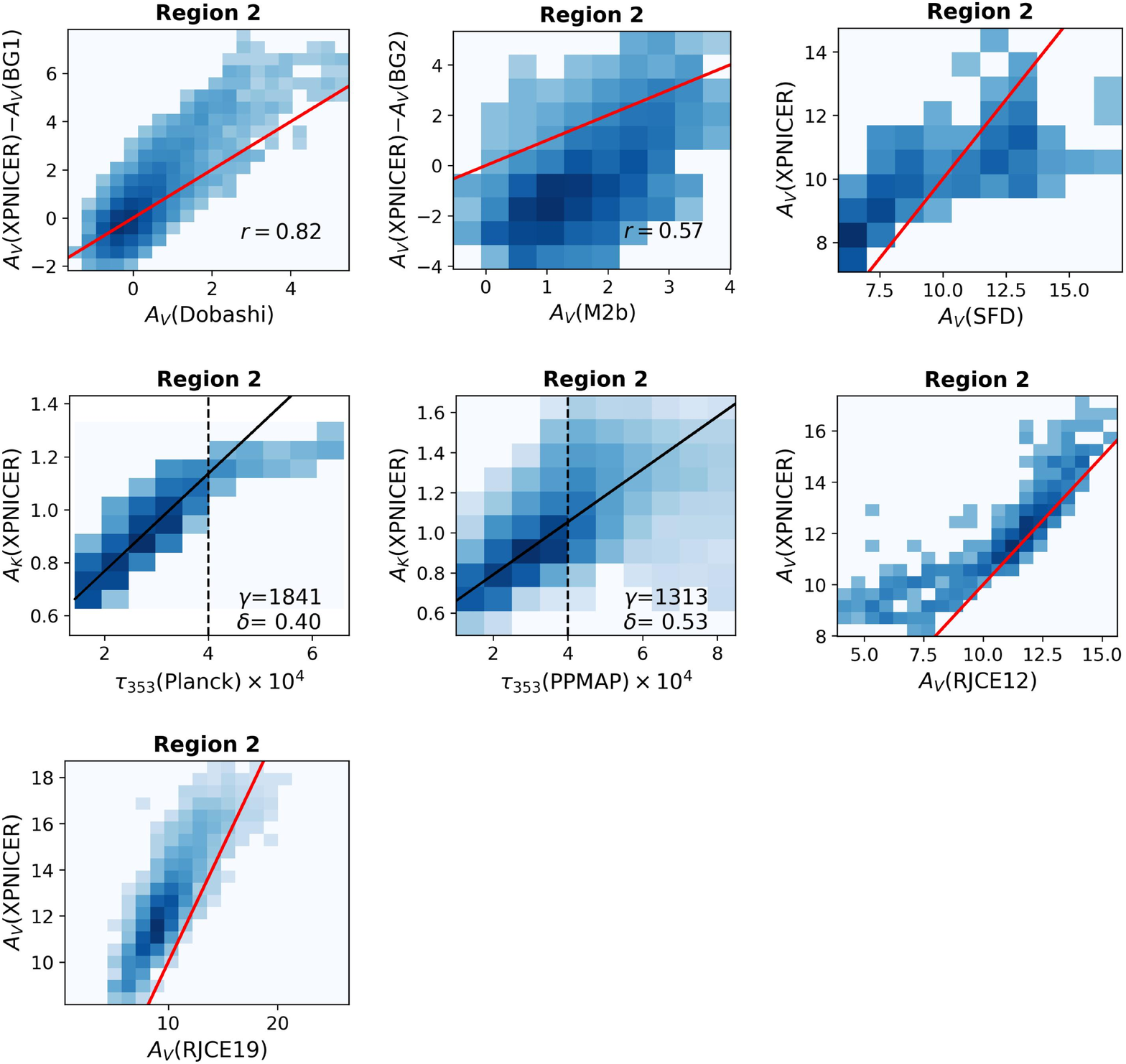}
    \caption{\mzrevi{Pixel-to-pixel comparisons of our XPNICER extinction map with previous dust-based maps in sub-region 2 (see Fig.~\ref{fig:reg_ap2}), including the extinction maps from \citet[][$A_V$(Dobashi)]{dobashi2011} and \citet[][$A_V$(M2b)]{juvela16}, the SFD map \citep[][$A_V$(SFD)]{sfd1998}, the Planck dust map \citep[][$\tau_{\mathrm{353}}$(Planck)]{planck-dust-2014}, the Herschel PPMAPs \citep[][$\tau_{\mathrm{353}}$(PPMAP)]{ppmap2017}, RJCE extinction map by \citet[][$A_V$(RJCE12)]{rjce2012}, and RJCE extinction map by \citet[][$A_V$(RJCE19)]{rjce2019}. Others are the same as Fig.~\ref{fig:compare2pre}.}}
    \label{fig:ap2com}
\end{figure*}

\begin{figure*}
    \centering
    \includegraphics[width=1.0\linewidth]{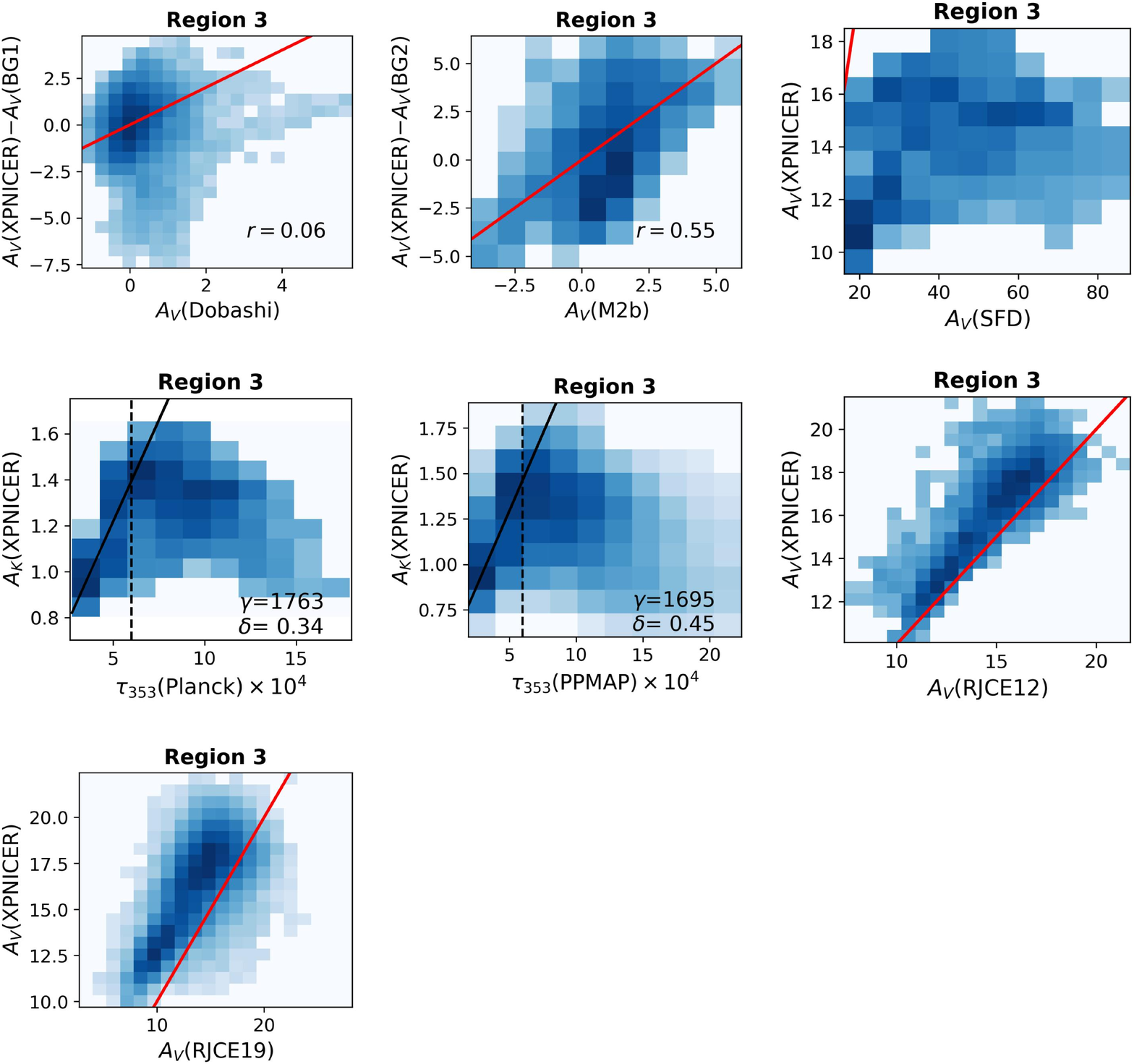}
    \caption{\mzrevi{Pixel-to-pixel comparisons of our XPNICER extinction map with previous dust-based maps in sub-region 3 (see Fig.~\ref{fig:reg_ap3}), including the extinction maps from \citet[][$A_V$(Dobashi)]{dobashi2011} and \citet[][$A_V$(M2b)]{juvela16}, the SFD map \citep[][$A_V$(SFD)]{sfd1998}, the Planck dust map \citep[][$\tau_{\mathrm{353}}$(Planck)]{planck-dust-2014}, the Herschel PPMAPs \citep[][$\tau_{\mathrm{353}}$(PPMAP)]{ppmap2017}, RJCE extinction map by \citet[][$A_V$(RJCE12)]{rjce2012}, and RJCE extinction map by \citet[][$A_V$(RJCE19)]{rjce2019}. Others are the same as Fig.~\ref{fig:compare2pre}.}}
    \label{fig:ap3com}
\end{figure*}

\begin{figure*}
    \centering
    \includegraphics[width=1.0\linewidth]{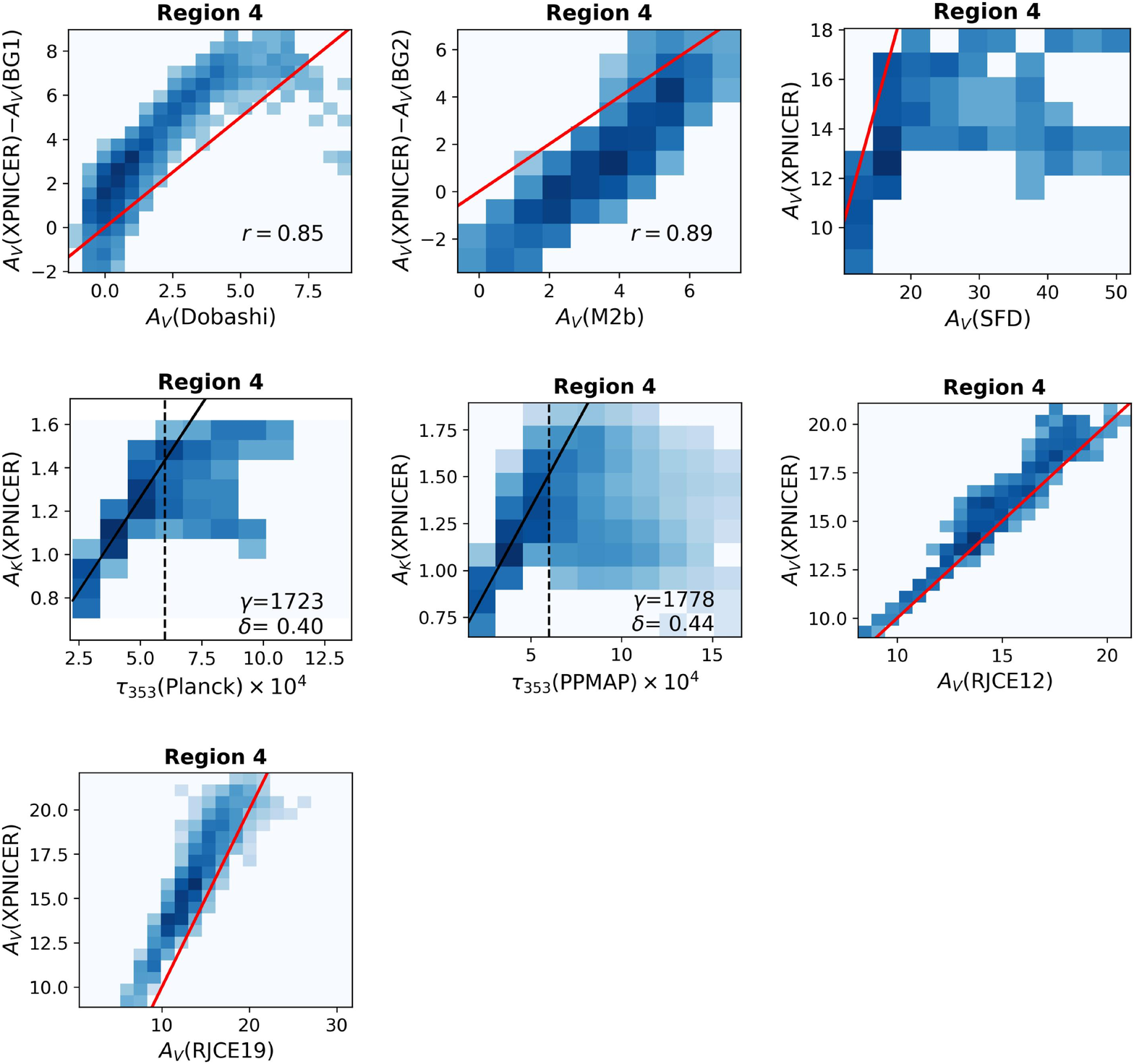}
    \caption{\mzrevi{Pixel-to-pixel comparisons of our XPNICER extinction map with previous dust-based maps in sub-region 4 (see Fig.~\ref{fig:reg_ap4}), including the extinction maps from \citet[][$A_V$(Dobashi)]{dobashi2011} and \citet[][$A_V$(M2b)]{juvela16}, the SFD map \citep[][$A_V$(SFD)]{sfd1998}, the Planck dust map \citep[][$\tau_{\mathrm{353}}$(Planck)]{planck-dust-2014}, the Herschel PPMAPs \citep[][$\tau_{\mathrm{353}}$(PPMAP)]{ppmap2017}, RJCE extinction map by \citet[][$A_V$(RJCE12)]{rjce2012}, and RJCE extinction map by \citet[][$A_V$(RJCE19)]{rjce2019}. Others are the same as Fig.~\ref{fig:compare2pre}.}}
    \label{fig:ap4com}
\end{figure*}

\begin{figure*}
    \centering
    \includegraphics[width=1.0\linewidth]{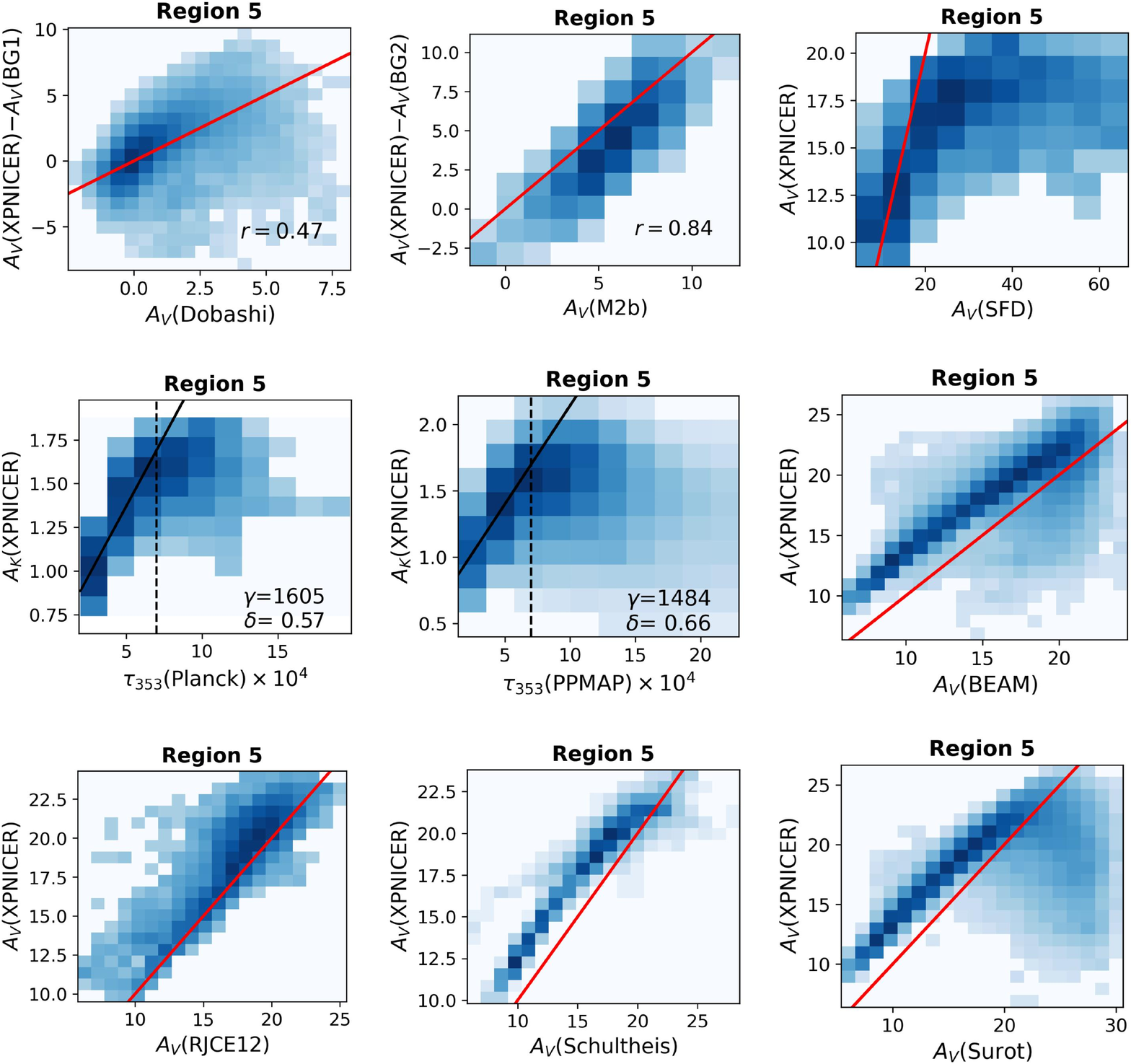}
    \caption{\mzrevi{Pixel-to-pixel comparisons of our XPNICER extinction map with previous dust-based maps in sub-region 5 (see Fig.~\ref{fig:reg_ap5}), including the extinction maps from \citet[][$A_V$(Dobashi)]{dobashi2011} and \citet[][$A_V$(M2b)]{juvela16}, the SFD map \citep[][$A_V$(SFD)]{sfd1998}, the Planck dust map \citep[][$\tau_{\mathrm{353}}$(Planck)]{planck-dust-2014}, the Herschel PPMAPs \citep[][$\tau_{\mathrm{353}}$(PPMAP)]{ppmap2017}, BEAM extinction map by \citet[][$A_V$(BEAM)]{beam2012,beam2018}, and RJCE extinction map by \citet[][$A_V$(RJCE12)]{rjce2012}, 2D extinction map integrated from the 3D dust map by \citet[][$A_V$(Schultheis)]{schultheis2014}, and extinction map by \citet[][$A_V$(Surot)]{surot2020}. Others are the same as Fig.~\ref{fig:compare2pre}.}}
    \label{fig:ap5com}
\end{figure*}

\section{Calibrating Herschel PPMAPs with Planck dust maps}\label{ap:calipp}

The archival Herschel data have been corrected by comparing to the Planck data \citep{planckdata2014}. The absolute flux calibration applied a constant-offset correction to the Herschel images, which assumed a average zero-level flux value over a given map \citep{bernard2010}. The Herschel PPMAPs were constructed with the Herschel images that were calibrated with the constant-offset correction.

However, \citet{av2017} pointed out that the constant-offset correction actually assumed that the absolute calibration was independent of angular scale. Unfortunately, the Herschel and Planck flux distribution could vary significantly within a given image, especially for the Hi-GAL images towards the Galactic mid-plane. \citet{av2017} improved a spatial dependence correction method based on Fourier transforms, which combined Planck model on large scales with the Herschel images on smaller scales. They compared these two different correction methods and found significant differences ($\gtrsim$~20\%) over $\sim$15\%~area of a given map at low column densities and high temperatures. \mztwo{The ratio of column densities corrected with their Fourier transforms method and constant-offset corrected column densities can be down to $<$0.3 and up to $>$1.1 at different spatial scales. Therefore, the zero-point of PPMAPs could be not uniform due to the inaccurate absolute flux calibration of the Herschel Hi-GAL images.}

\begin{figure*}
    \centering
    \includegraphics[width=1.0\linewidth]{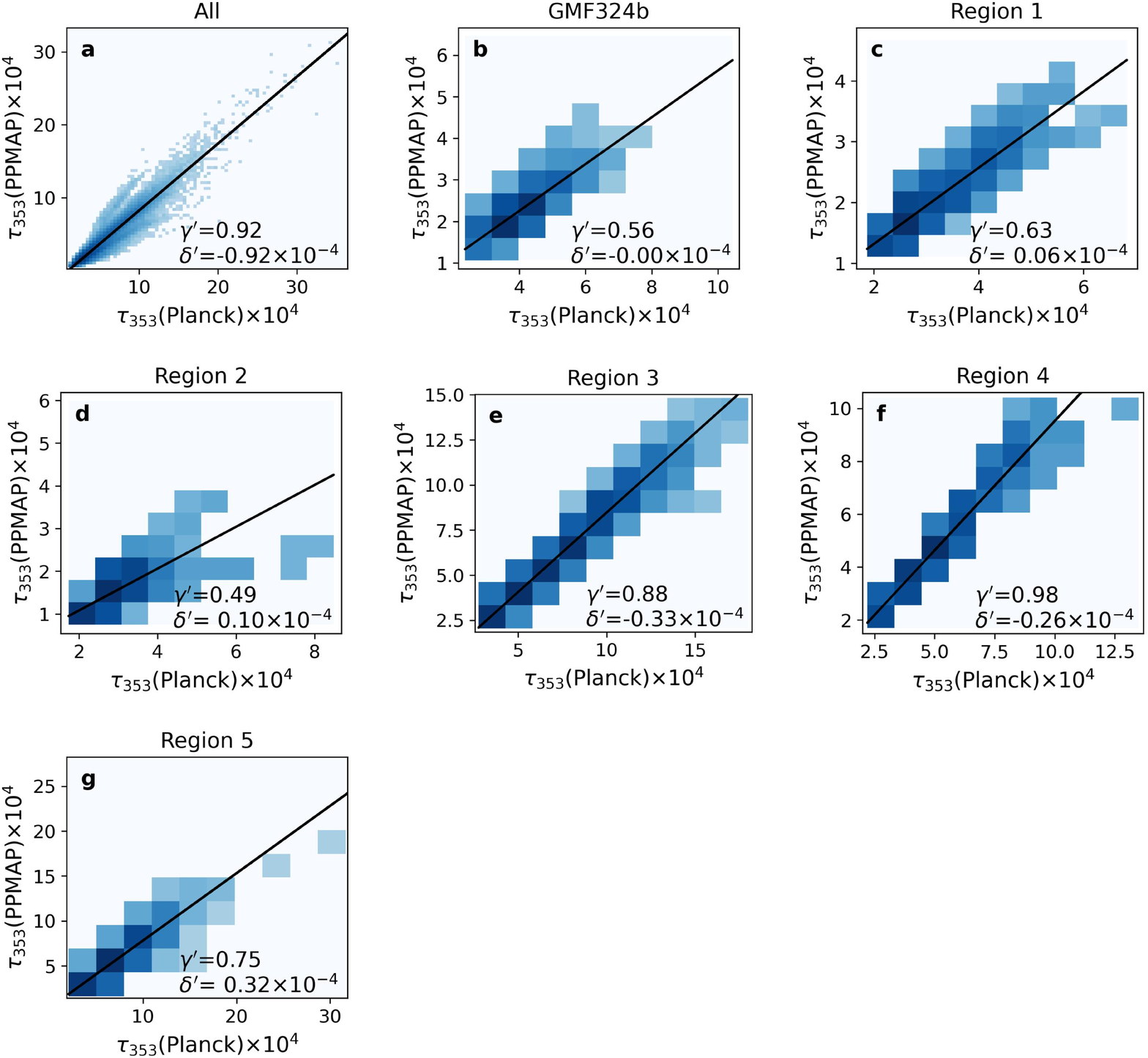}
    \caption{Pixel-to-pixel comparisons of the Planck dust maps and the Herschel PPMAPs in the whole VVV survey area and six sub-regions. The solid black lines show the linear fits with the slope of $\gamma^{\prime}$ and the intercept of $\delta^{\prime}$. The values of $\gamma^{\prime}$ and $\delta^{\prime}$ are also marked on each panel.}
    \label{fig:calipp}
\end{figure*}

We compared $\tau_{353}$(PPMAP) with the $\tau_{353}$(Planck) map (obtained in Sect.~\ref{sect:compare2planck}) at 5\arcmin~resolution. We limited the comparison in the VVV survey area (only $|b|\lesssim$~1 due to the Herschel Hi-GAL coverage). \mztwo{Figure~\ref{fig:calipp} shows the pixel-to-pixel relation between $\tau_{353}$(PPMAP) and $\tau_{353}$(Planck) for whole map and six sub-regions (see Figs.~\ref{fig:reg_ap1}-\ref{fig:reg_ap5}). There are linear relations that} can be fitted with the function of 
\begin{equation}
\tau_{353}\mathrm{(PPMAP)}=\gamma^{\prime}\tau_{353}\mathrm{(Planck)}+\delta^{\prime}.\label{eq:pppfit}
\end{equation}
\mztwo{The returned $\gamma^{\prime}$ is in the range of $\sim$0.5$-$1 while $\delta^{\prime}$ is in the range of $\sim$(-1$-$0.3)$\times 10^{-4}$. On average there is a approximating one-to-one relation between $\tau_{353}$(PPMAP) and $\tau_{353}$(Planck), but the ratio of $\tau_{353}$(PPMAP) to $\tau_{353}$(Planck) is quite different in different sub-regions, i.e., different spatial locations and scales. Therefore, $\tau_{353}$(PPMAP) and $\tau_{353}$(Planck) vary significantly at different angular scales. We consider this variance could result from the constant-offset absolute flux calibration of the Herschel images.} 

\mztwo{In principal, to eliminate the variance between $\tau_{353}$(PPMAP) and $\tau_{353}$(Planck) we need to re-calibrate the Herschel Hi-GAL images as suggested by \citet[][]{av2017} and then construct $\tau_{353}$(PPMAP) again with the method suggested by \citet[][]{ppmap2017}, which is obviously beyond the scope of this paper. To compare our XPNICER map and the PPMAPs, we simply calibrate $\tau_{353}$(PPMAP) with $\tau_{353}$(Planck) using the relations of Eq.\ref{eq:pppfit} for the whole map and the sub-regions. For example, Fig.~\ref{fig:calipp}b shows the relation between $\tau_{353}$(PPMAP) and $\tau_{353}$(Planck) in the region with GMF 324b (see Fig.~\ref{fig:extmap_zoomin}). The linear fitting returns $\gamma^{\prime}=$~0.56 and $\delta^{\prime}=$0.00. We calibrated $\tau_{353}$(PPMAP) of the region with GMF 324b by multiplying it by the factor of 1/0.56$\approx$1.8.} 


\bsp	
\label{lastpage}
\end{document}